\documentclass[
	aps,prd,
	reprint,
	showkeys,
	amsmath,amssymb,
	nofootinbib,
	floatfix,
        superscriptaddress,
	]{revtex4-2}
\usepackage[utf8]{inputenc}
\usepackage{aas_macros}
\pdfoutput=1
\usepackage{hyperref}

\usepackage[caption=false]{subfig}
\usepackage{graphicx}
\hypersetup{ 
			colorlinks=true,
			pdfauthor={Georgios Valogiannis and Cora Dvorkin},
			pdftitle={WST TBD},
			citecolor=blue
			}

\usepackage{graphicx}
\usepackage{dcolumn}
\usepackage{bm}

\begin{document}


\title{Towards unveiling the large-scale nature of gravity with the wavelet scattering transform}

\author{Georgios Valogiannis}
 \email{gvalogiannis@uchicago.edu}
 \affiliation{
 Department of Astronomy $\&$ Astrophysics, University of Chicago, Chicago, IL, 60637, USA\\
}
 \affiliation{
 Kavli Institute for Cosmological Physics, Chicago, IL, 60637, USA\\
}

\author{Francisco Villaescusa-Navarro}
 \affiliation{
 Center for Computational Astrophysics, New York, NY, 10010, USA\\
}
\author{Marco Baldi}
 \affiliation{
 Dipartimento di Fisica e Astronomia, Alma Mater Studiorum Universita di Bologna, via Piero Gobetti, 93/2, I-40129 Bologna, Italy\\
}
\begin{abstract}
We present the first application of the Wavelet Scattering Transform (WST) in order to constrain the nature of gravity using the three-dimensional (3D) large-scale structure of the universe. Utilizing the \textsc{Quijote-MG} N-body simulations, we can reliably model the 3D matter overdensity field for the f(R) Hu-Sawicki modified gravity (MG) model down to $k_{\rm max}=0.5$ h/Mpc. Combining these simulations with the \textsc{Quijote} $\nu$CDM collection, we then conduct a Fisher forecast of the marginalized constraints obtained on gravity using the WST coefficients and the matter power spectrum at redshift z=0. Our results demonstrate that the WST substantially improves upon the 1$\sigma$ error obtained on the parameter that captures deviations from standard General Relativity (GR), yielding a tenfold improvement compared to the corresponding matter power spectrum result. At the same time, the WST also enhances the precision on the $\Lambda$CDM parameters and the sum of neutrino masses, by factors of 1.2-3.4 compared to the matter power spectrum, respectively. Despite the overall reduction in the WST performance when we focus on larger scales, it still provides a relatively $4.5\times$ tighter 1$\sigma$ error for the MG parameter at
$k_{\rm max}=0.2$ h/Mpc, highlighting its great sensitivity to the underlying gravity theory. This first proof-of-concept study reaffirms the constraining properties of the WST technique and paves the way for exciting future applications in order to perform precise large-scale tests of gravity with the new generation of cutting-edge cosmological data. 
\end{abstract}

\maketitle

\section{Introduction}
The dramatically increasing quantity and quality of cosmological observations of the Large-Scale Structure (LSS) of the universe in the coming decade creates an exciting landscape for the exploration of the deepest mysteries in modern fundamental physics. The recent announcement of the first cosmology results \citep{2024arXiv240403000D,2024arXiv240403001D,2024arXiv240403002D} by the Dark Energy Spectroscopic Instrument (DESI) \citep{Levi:2013gra,2016DESI} marked the beginning of the Stage-IV era of precision cosmology, which will soon also include the corresponding analyses by the {\it Vera C. Rubin} Observatory Legacy Survey of Space and Time (LSST) \citep{Abell:2009aa,Abate:2012za}, {\it Euclid} \citep{Laureijs:2011gra}, the {\it Nancy Grace Roman} Space Telescope \citep{Spergel:2013tha} and SPHEREx \citep{2014arXiv1412.4872D}. By accurately tracing the distribution of galaxies in the 3-dimensional (3D) LSS of the universe, which sensitively depends on the interplay between its fundamental constituents and interactions, we are able to perform stringent tests of our assumptions about the physics of the early universe and other light relics \citep{Chen_2016,DePorzio:2020wcz, Xu:2021rwg}, the nature of dark matter \citep{LSSTDarkMatterGroup:2019mwo}, massive neutrinos \citep{LESGOURGUES2006307,Dvorkin:2019jgs}, the accelerated expansion of the universe \citep{Copeland:2006wr} and also the properties of gravity at large scales \citep{2016RPPh...79d6902K,Ishak:2018his,doi:10.1146/annurev-astro-091918-104423, Alam:2020jdv}.

Despite being the weakest out of the 4 fundamental interactions, gravity played a central role in shaping the LSS of the universe, driving the instabilities that converted the primordial fluctuations into the inhomogeneous pattern of galaxies observed at late times. As a result, the LSS is a valuable probe that enables precise cosmological tests of gravity. These tests do not come in a vacuum; General Relativity (GR), our best theory of classical gravity, has successfully passed a wide array of tests, ranging from laboratory and Solar System experiments \citep{2006LRR.....9....3W} all the way to the strong field regime \citep{2008LRR....11....9P} and gravitational wave (GW) observations \citep{2016PhRvL.116f1102A}. Its validity on the cosmic scales, however, relies upon an extrapolation of the above conclusions by many orders of magnitude, a fact that has recently also come into question in connection to the discovery of the accelerated expansion of the universe \citep{1999ApJ...517..565P,1998AJ....116.1009R} and the cosmological constant problem \citep{1989RvMP...61....1W} associated with its standard $\Lambda$ Cold Dark Matter ($\Lambda$CDM) explanation\footnote{It should also be noted that recent cosmological observations may be providing hints for a potential deviation from a background evolution consistent with a cosmological constant \citep{2024arXiv240403002D,DESI:2024aqx,DESI:2024kob}.}. As a result, the wealth of incoming cosmological data provides us with the timely opportunity to accurately test the large-scale nature of gravity and its potential links to cosmic acceleration \citep{2016RPPh...79d6902K,Ishak:2018his,doi:10.1146/annurev-astro-091918-104423, Alam:2020jdv}.  

Even though the LSS is in principle a great arena for testing fundamental physics, fully utilizing it is far from trivial in practice. The nonlinear evolution of the primordial Gaussian density field has caused the dispersion of its information content \citep{PhysRevLett.108.071301}, which was originally fully encoded by the power spectrum, into higher order correlations \citep{10.1093/mnras/stv961,10.1093/mnras/stw2679,BERNARDEAU20021,PhysRevD.105.043517,Chen:2021vba,Philcox:2021hbm,2024arXiv240407249P} that are expensive to extract and perhaps unable to even be informative in the deeply non-Gaussian regime \citep{PhysRevLett.108.071301}. On the particular front of testing gravity, these challenges are further exacerbated by the 'screening' mechanisms that various modified gravity (MG) theories typically invoke in order to restore their viability in the Solar System \citep{2010arXiv1011.5909K,2013arXiv1312.2006K}. Suitably tailored interaction terms in the scalar field potential basically manage to suppress the MG-induced 'fifth forces' in regions of high density and/or gravitational potential, where GR is recovered \citep{2004PhRvL..93q1104K,2008PhRvD..77d3524O,2010PhRvL.104w1301H,2009IJMPD..18.2147B,2011JHEP...08..108D,1972PhLB...39..393V}. Given, however, that it is mainly these high overdensities that dominate the signal in the 2-point function, what keeps such theories viable also makes them very challenging to detect using the power spectrum, even with the next generation of cosmological surveys.

Aiming to overcome these sets of challenges, a wide array of alternative techniques have been developed, with a primary focus on efficiently recovering the additional non-Gaussian information that lies beyond the power spectrum, in order to fully utilize the constraining potential of the wealth of upcoming cosmological data. As far as testing gravity is concerned, for example, transformations that ``mark'' the significance of low-density, unscreened, regions have demonstrated great promise for improving cosmological constraints \citep{White_2016,PhysRevD.97.023535,Alam:2020jdv,2018MNRAS.479.4824H,2018MNRAS.478.3627A,2024MNRAS.528.6631A}, matching and exceeding their performance in the context of conventional cosmologies \citep{Pisani:2019cvo,Massara_2015, 10.1093/mnras/stz1944, 10.1093/mnras/stv777, Hamaus_2015, Kreisch:2021xzq,Bonnaire:2021sie, 2023arXiv230205302R,Neyrinck_2009,PhysRevLett.108.071301,PhysRevLett.107.271301,White_2016,PhysRevD.97.023535,PhysRevLett.126.011301,2022arXiv220601709M,2024arXiv240404228M}. These are joined by a long list of related approaches to galaxy clustering, involving for instance the use of Minkowski functionals \citep{10.1046/j.1365-8711.1999.02912.x,10.1111/j.1365-2966.2012.21103.x,10.1093/mnras/stt1316,PhysRevLett.118.181301,2023arXiv230208162L,2024arXiv240313985Y}, density-split clustering \citep{10.1093/mnras/stab1654,2022arXiv220904310P,Bayer:2021iyb,Paillas2023:2309.16541,Cuesta-Lazaro2023:2309.16539}, proxy higher-order statistics \citep{PhysRevD.91.043530,Peel:2018aly,Dizgah_2020,Chakraborty:2022aok, 2023JCAP...03..045H,2024arXiv240113036C,2024arXiv240115074H}, k-nearest neighbors \citep{Banerjee:2020umh,10.1093/mnras/stab961}, 1-point cumulants \citep{Uhlemann:2019gni,2020PhRvD.102l3546J} or the minimum spanning tree \citep{Naidoo:2021dxz} (with a similar level of activity in the context of weak gravitational lensing (WL) \citep{2023arXiv230112890E}). On the other end of the interpretability spectrum, modern Artificial Intelligence (AI) seeks to capture the entire field-level information, without any compression, exhibiting very promising results in preliminary applications \citep{PhysRevD.97.103515,Peel:2018aei,Merten:2018bgr,Villaescusa_Navarro_2021,Perez:2022nlv,2023ApJ...952...69D,Dai:2023lcb,Sharma:2024pth}. Identifying the approach that delivers the optimal trade-off between performance, cost and interpretability for a given application is one of the most pressing open questions in the field of precision cosmology. 

A more balanced path between the above two fronts can be taken by the Wavelet Scattering Transform (WST) \citep{https://doi.org/10.1002/cpa.21413,6522407}, which is a novel summary statistic that subjects a target field to a cascade of successive convolutions with a set of localized wavelets, followed by a nonlinear modulus and a global averaging operation. Given the close resemblance between these basic WST operations and the ones performed by a Convolutional Neural Network (CNN) (convolution, nonlinearity, pooling), the former can be alternatively viewed as a fixed-kernel shallow neural net, delivering thus a more favorable trade-off between performance and interpretability \citep{Sifre_2013_CVPR,10.1214/14-AOS1276,6822556,PhysRevLett.108.071301}. As a result, its use for non-Gaussian information extraction has been increasingly growing in a broad range of physical applications \citep{cheng:2021xdw},  such as in astrophysics \citep{refId0, Saydjari_2021,refdust}, cosmology \citep{10.1093/mnras/staa3165,10.1093/mnras/stab2102,PhysRevD.105.103534,PhysRevD.106.103509,PhysRevD.109.103503,PhysRevD.102.103506,10.1093/mnras/stac977,2022arXiv220407646E,2023arXiv231015250R,10.1093/mnras/stac2662,10.1093/mnras/stac977,DES:2023qwe,Peron:2024xaw,2024arXiv240416085C} or molecular chemistry \citep{10.5555/3295222.3295400,doi:10.1063/1.5023798}. Notably, it has been recently successfully applied to galaxy clustering observations, enabling substantial improvements in the derived cosmological constraints compared to the conventional 2-point correlation function \citep{PhysRevD.106.103509,PhysRevD.109.103503}. 

Our previous applications \citep{PhysRevD.105.103534,PhysRevD.106.103509,PhysRevD.109.103503} were mainly focused on constraining the cosmological parameters of the standard $\Lambda$CDM scenario, with only a brief consideration of simple extensions to it in \citep{PhysRevD.109.103503}, due to the limitations imposed by the available simulations \citep{Villaescusa_Navarro_2020,2021Maksimova}. Motivated by these encouraging findings in the context of $\Lambda$CDM, in this work we move one step further and perform the first WST application to test a modified theory of gravity, the popular f(R) Hu-Sawicki model \citep{2007PhRvD..76f4004H}. Using a new set of state-of-the-art simulations for this cosmological scenario, that expand upon the publicly available \textsc{Quijote} suite \citep{Villaescusa_Navarro_2020,QujiMG}, we perform a WST Fisher forecast using the real space matter overdensity and compare against the constraints predicted by the standard matter power spectrum at redshift $z=0$. Our work lays the foundation for obtaining precise constraints on the nature of gravity using the WST technique. 

This paper is structured as follows: in \S\ref{sec:WST} we introduce the Wavelet Scattering Transform, in \S\ref{sec:MGmodel} the modified gravity model that we are going to focus on, while in \S\ref{sec:Simulations} we describe the simulations we use to make our predictions for it. We then proceed to lay out all the details of our Fisher forecast framework in \S\ref{sec:Fisher}, before presenting our results in \S\ref{sec:Results}. Finally, we summarize our conclusions in \S\ref{sec:Conclusions}. More technical details are discussed in Appendices \S\ref{app:Gauss},\S\ref{app:Stability} and \S\ref{app:Der}.

\section{Wavelet Scattering Transform}\label{sec:WST}

First proposed in the context of signal processing and computer vision applications, the Wavelet Scattering Transform (WST) \citep{https://doi.org/10.1002/cpa.21413,6522407} is a novel summary statistic that was intended to shed light on the performance of CNNs. As such, it can simultaneously also serve as a more interpretable alternative. Specifically, consider $I({\bf x})$ to be a physical field of interest defined in an arbitrary number of dimensions, and also $\psi_{j_1, l_1}(\bold{x})$ to be a localized wavelet, the spatial and angular support of which is characterized by the indices $j_1$ and $l_1$, respectively. If $\ast$ denotes a convolution, then the operation
\begin{equation}\label{eq:convmod}
S_1 \equiv \left\langle|I(\bold{x}) \ast \psi_{j_1, l_1}(\bold{x}) | \right\rangle,   
\end{equation}
will extract a global (through averaging) quantity $S_1$ which quantifies the degree of clustering around scales $j_1$ and $l_1$ (through the wavelet convolution). The successive repetition of the fundamental operation \eqref{eq:convmod} up to the $n^{th}$ order can then generate a cascade of such WST coefficients, $S_n$, given explicitly by the following hierarchy of expressions:
\begin{align}\label{eq:WSTcoeff:base}
 S_0 &= \langle |I(\bold{x})|\rangle, \nonumber \\
 S_1(j_1,l_1) &= \left\langle |I(\bold{x}) \ast \psi_{j_1, l_1}(\bold{x}) | \right\rangle, \\
 S_2(j_2,l_2,j_1,l_1) &= \left\langle |\left(|I(\bold{x}) \ast \psi_{j_1, l_1}(\bold{x}) |\right) \ast \psi_{j_2, l_2}(\bold{x}) |\right\rangle \nonumber,
\end{align}
up to $n=2$. Eq. \eqref{eq:WSTcoeff:base} is referred to as a $\textit{scattering network}$, the coefficients of which can summarize the strength of clustering in the input field when combined with a family of localized wavelets, which we will define below. In direct analogy to the familiar hierarchy of n-point correlation functions, it can be proven that a WST coefficient of order $n$ contains high-order information related to the correlation function of order up to $2^n$ \citep{https://doi.org/10.1002/cpa.21413,10.1214/14-AOS1276}. Thanks to its fundamental property of not elevating the input field to very high powers, however, it is more efficient in the highly non-Gaussian regime because it does not amplify the tail of the probability distribution, unlike the traditional moment expansion \citep{PhysRevLett.108.071301,cheng:2021xdw}. The same property leads to a greater degree of robustness and numerical stability, as we will also find in this analysis, while the compact wavelet basis in Eq. \eqref{eq:WSTcoeff:base} guarantees that the extracted information is not dispersed among a very high-dimensional data vector \citep{cheng:2021xdw}. Given that a CNN essentially subjects an input field to a convolution with a kernel, followed by a nonlinear operation, before summarizing with a global pooling operation, it can be immediately understood why the WST can be viewed as a more interpretable equivalent, that is, as a fixed-kernel (wavelet) shallow neural net lying at the middle-ground between a powerful yet ``black box'' CNN and the well-understood but limited power spectrum. This makes it a valuable novel summary statistic with a wide range of applications considered across the physical sciences \citep{cheng:2021xdw}.

If we slightly relax the requirement that the field enters the WST operation in a strictly linear fashion, as in Eq. \eqref{eq:WSTcoeff:base}, and allow it to be raised to a more generic power $q \neq 1$ instead, we end up with the generalized version
\begin{align} \label{eq:WSTcoeff:power}
 S_0 &= \langle |I(\bold{x})|^q\rangle, \nonumber \\
 S_1(j_1,l_1) &= \left\langle |I(\bold{x}) \ast \psi_{j_1, l_1}(\bold{x}) |^q \right\rangle, \\
 S_2(j_2,l_2,j_1,l_1) &= \left\langle |\left(|I(\bold{x}) \ast \psi_{j_1, l_1}(\bold{x}) |\right) \ast \psi_{j_2, l_2}(\bold{x}) |^q\right\rangle \nonumber,
\end{align}
which essentially provides the flexibility to highlight the overdense or underdense regions of the LSS (when working with the cosmic density field), by choosing powers $q>1$ or $q<1$, respectively. Given that cosmic voids are known to be sensitive probes of fundamental physics, such as the properties of massive neutrinos or gravity, the latter choice is of particular interest to cosmological applications, a fact that we already confirmed for the neutrino mass constraints in Ref. \citep{PhysRevD.105.103534}. As a result, we will adopt the same choice of Eq. \eqref{eq:WSTcoeff:power} in this modified gravity WST application as well, as we will further discuss below.

For a particular choice of a 3-dimensional (3D) field as input, which will be the case in our cosmological application, a convenient wavelet basis can be generated starting with a solid harmonic mother wavelet:
\begin{equation}\label{solid:sup}
\psi^{m}_{l}(\bold{x}) = \frac{1}{\left(2\pi\right)^{3/2}}e^{-|\bold{x}|^2 /2 \sigma^2}|\bold{x}|^l Y_l^m \left(\frac{\bold{x}}{|\bold{x}|}\right),
\end{equation}
that was first applied in the context of 3D molecular chemistry  \citep{10.5555/3295222.3295400,doi:10.1063/1.5023798}, but which we also adopted in our previous galaxy clustering applications \citep{PhysRevD.105.103534,PhysRevD.106.103509,PhysRevD.109.103503}. If $Y_l^m$ in Eq. \eqref{solid:sup} is a Laplacian spherical harmonic and $\sigma$ the Gaussian width in pixel units, then a dilation of the mother wavelet will give:
\begin{equation}\label{dil:sup}
\psi^{m}_{j, l}(\bold{x}) = 2^{-3 j}\psi^{m}_{l}(2^{-j} \bold{x}),
\end{equation}
which builds a family of wavelets spanning different dyadic scales, $2^{j}$, and angles $l$, after we sum over the second spherical harmonic index $m$. Adopting this configuration, the WST network eventually takes the following form:
\begin{align} \label{eq:WSTcoeff:sol}
 S_0 &= \langle |I(\bold{x})|^{q} \rangle, \nonumber \\
 S_1(j_1,l_1) &= \left\langle \left(\sum_{m=-l_1}^{m=l_1}|I(\bold{x}) \ast \psi^{m}_{j_1, l_1}(\bold{x}) |^{2}\right)^{\frac{q}{2}} \right\rangle, \\
 S_2(j_2,j_1,l_1) &= \left\langle \left(\sum_{m=-l_1}^{m=l_1}|U_1(j_1,l_1)(\bold{x}) \ast \psi^{m}_{j_2, l_1}(\bold{x}) |^{2}\right)^{\frac{q}{2}} \right\rangle \nonumber,
\end{align}
where
\begin{equation}
 U_1(j_1,l_1)(\bold{x}) = \left(\sum_{m=-l_1}^{m=l_1}|I(\bold{x}) \ast \psi^{m}_{j_1, l_1}(\bold{x}) |^{2}\right)^{\frac{1}{2}}.
\end{equation}
Aside from the solid harmonic wavelets that we adopt in this analysis, other applications have considered bump-steerable wavelets \citep{PhysRevD.102.103506,2022arXiv220407646E}, Morlet wavelets \citep{10.1093/mnras/staa3165,10.1093/mnras/stab2102} or wavelets with a preferred direction \citep{Saydjari_2021,2023arXiv231015250R}.

Furthermore, and staying consistent with previous applications \citep{10.1093/mnras/staa3165,10.5555/3295222.3295400,doi:10.1063/1.5023798,PhysRevD.105.103534,PhysRevD.106.103509}, we further reduce the dimensionality of the resulting WST data vector by discarding second order coefficients for $j_2<j_1$, since the $1^{st}$ order convolution over $j_1$ washes out all information below that scale, and also proceed with $l_2=l_1$, which provides a reasonable compromise between performance and computational cost.

This configuration leads to a total of 
\begin{equation}\label{Stotal}
S_0+S_1+S_2=1+(L+1)(J+2)(J+1)/2    
\end{equation}
WST coefficients up to $2^{nd}$ order, with the spatial and angular indices running in the following range
\begin{equation}
(j,l) \in([0,..,J-1,J],[0,..,L-1,L]),    
\end{equation}
respectively. For an input field resolved using NGRID cells on each dimension, the maximum spatial scale cannot exceed $J\le\log_2({\rm NGRID})$, given that the wavelet dilations scale as $2^j$.

Summarizing, the overall WST configuration is determined by the values of $J$, $L$, $q$ and $\sigma$, which the user is free to choose based on the particular application. We will justify how we pick their values for this particular analysis in the following sections. Finally, given an input field $I(\bold{x})$ of resolution $\rm NGRID^{3}$ \footnote{An additional layer of smoothing can be applied to the field in case we want to ensure that small-scale sensitivity is completely removed beyond a fixed cut-off, as we will explain below.}, the WST coefficients \eqref{Stotal} can be evaluated from Eq. \eqref{eq:WSTcoeff:sol}, using the public \textsc{Kymatio} package \citep{2018arXiv181211214A}\footnote{Available in \url{https://www.kymat.io/}}.

\section{Gravity model}\label{sec:MGmodel} 
Testing General Relativity against competing alternatives has a long history that traces all the way back to its original formulation; for an overview see \citep{2006LRR.....9....3W}. Ever since the discovery of the accelerated expansion of the universe \citep{1999ApJ...517..565P,1998AJ....116.1009R} and the subsequent dawn of precision cosmology, many such approaches have been naturally increasingly focusing on cosmological tests of gravity \citep{2016RPPh...79d6902K,Ishak:2018his,doi:10.1146/annurev-astro-091918-104423, Alam:2020jdv}. In order to evade the tight constraints already placed on gravity in the strong regime \citep{2008LRR....11....9P} and the vicinity of our Solar System \citep{2006LRR.....9....3W}, while at the same time being able to predict detectable signals at cosmic scales, alternative scenarios of this kind typically invoke a dynamical screening mechanism that restores their viability at smaller scales \citep{2010arXiv1011.5909K,2013arXiv1312.2006K,2004PhRvL..93q1104K,2008PhRvD..77d3524O,2010PhRvL.104w1301H,2009IJMPD..18.2147B,2011JHEP...08..108D,1972PhLB...39..393V}. 

One of the most popular such scenarios is the f(R) class of theories \citep{DeFelice:2010aj}, in which a suitably chosen function of the Ricci scalar, R, modifies the Einstein-frame expression of the Einstein-Hilbert action S, as follows:
\begin{equation}\label{actFr}
S=\int d^4x \sqrt{-g} \left[\frac{R+f(R)}{16 \pi G_{\rm N}} + \rm \mathcal{L}_m \right],
\end{equation}
where $G_{\rm N}$ is the Newtonian gravitational constant, $g$ the determinant of the metric tensor and $\rm \mathcal{L}_m $ the Lagrangian density of the matter sector. In the limit $\rm f(R)=0$ the standard GR expression is recovered, while non-zero values of the modifying function free up an additional degree of freedom, $f_{\rm R} = df(R)/dR$, that can potentially cause self-acceleration \citep{Carroll:2003wy}. In the weak-field regime and under the quasi-static approximation, the scalar field equations become:
\begin{equation}
\begin{aligned}\label{eq:modpoisson}
&\nabla^2 \Phi_{\rm N} = 4 \pi G_{\rm N} a^2 \delta \rho_{\rm m} - \frac{1}{2}\nabla^2 f_{\rm R}  , \\
&\nabla^2 f_{\rm R} =  -\frac{a^2}{3}\delta R - \frac{8 \pi G_{\rm N} a^2 }{3} \delta \rho_{\rm m} ,
\end{aligned}
\end{equation}
where $\delta \rho_{\rm m}$ and $\delta R$ are the matter density perturbations and the perturbation of the Ricci scalar at a scale factor $a$, respectively. Eqs. \eqref{eq:modpoisson} essentially describe how the ``scalaron'' $f_{\rm R}$ sources an additional fifth force, in addition to the standard Newtonian contribution described by the Poisson term.

Arguably the most popular member of this class, which is the one we are going to work with in this analysis, is the Hu-Sawicki model \citep{2007PhRvD..76f4004H}, with 
\begin{equation}\label{fRHu}
f(R)=-m^2\frac{c_1\left(R/m^2\right)^n}{c_2\left(R/m^2\right)^n+1},
\end{equation}
where $H_0$ denotes the Hubble constant, $\Omega_{\rm m}$ the fractional matter density evaluated today, $m=H_0\sqrt{\Omega_{\rm m}}$, and $c_1, c_2$ \& $n$ are free parameters of the model. When we restrict our attention to background histories exactly matching the $\Lambda$CDM one, with vacuum fractional energy density $\Omega_{\Lambda}$, we end up with
\begin{equation}\label{fr0}
\bar{f}_{R_0} =-n\frac{c_1}{c_2^2}\left(\frac{\Omega_{\rm m}}{3(\Omega_{\rm m}+\Omega_{\Lambda})}\right)^{n+1},
\end{equation}
which effectively reduces the number of free parameters to the pair $\bar{f}_{R_0}$ $\&$ $n$ that fully characterize the model in this case. Larger $\bar{f}_{R_0}$ absolute values lead to stronger deviations, with $\bar{f}_{R_0} \rightarrow 0$ corresponding to the GR limit. The popularity of this scenario stems from the fact that it realizes the chameleon screening mechanism \citep{2004PhRvL..93q1104K} (the other popular class being the Vainshtein mechanism \citep{1972PhLB...39..393V}); that is, the scalaron $f_{\rm R}$ acquires a non-zero mass, $m^2_{f_{\rm R}}=\frac{1}{3}f_{\rm RR}^{-1}$, which leads to an exponential Yukawa-type suppression of the fifth force. In high-density regions the field becomes very massive, heavily suppressing the deviations such that GR is recovered. As a result, it serves as a very useful toy model for testing modified gravity theories with a screening phenomenology. Last but not least, we note that one of its characteristic signatures is the scale-dependence of the growth factor even at the linear level, since the MG field has a non-zero background mass as well, given by:
\begin{equation}\label{Eq:mass}
m(a)=\left(\frac{1}{2|\bar{f}_{R_0}|}\right)^{\frac{1}{2}}\frac{\left(\Omega_{m} a^{-3}+4\Omega_{\Lambda}\right)^{1+\frac{n}{2}}}{\left(\Omega_{m}+4\Omega_{\Lambda}\right)^{\frac{n+1}{2}}},
\end{equation}
which leads to the following evolution of the linear growth factor, $D_m $:
\begin{equation}
\label{growthfact}
\ddot{D}_m + 2H\dot{D}_m = \frac{3}{2} \Omega_m(a)H^2 D_m\frac{G_{eff}(k,a)}{G_N},
\end{equation}
with 
\begin{equation}
\label{Geffective}
\frac{G_{eff}(k,a)}{G_N} = 1+ \frac{1}{3}\frac{k^2}{k^2 + a^2 m^2(a)}.
\end{equation}
We notice that the enhancement reaches a maximum value of $\frac{1}{3}$ at scales below the field Compton wavelength, $r \ll \lambda_c \sim 1/ a m(a)$. Conversely, the growth factor reduces to its standard GR form for large values of the background mass (equivalently, at scales $r \gg \lambda_c$).

\section{Simulations}\label{sec:Simulations} 
In order to be able to properly assess the extent to which higher order statistics such as the WST can allow us to extract the non-Gaussian information encoded in the cosmic density field, we need to accurately model structure formation down to the nonlinear scales, a step that can be reliably performed only through N-body simulations. In this work we utilize the suite of the \textsc{Quijote} simulations \citep{Villaescusa_Navarro_2020}\footnote{\url{https://quijote-simulations.readthedocs.io/}}, which have been ideally designed to enable Fisher forecasts and parameter inference in a simulation-based manner. These are N-body simulations run using the state-of-the-art TreePM code Gadget-III \citep{Springel:2005mi}, using a resolution of $512^3$ cold dark matter (CDM) particles placed inside a simulation cube with a side equal to 1.0 Gpc/h. The runs were initialized at a redshift $z_i=127$ using second order Lagrangian Perturbation Theory (2LPT), and evolved the dynamics all the way to $z=0$. The fiducial configuration consists of 15,000 random realizations run for a cosmology determined by the following parameter values: $\Omega_b = 0.049$, $\Omega_m = 0.3175$, $h = 0.6711$, $n_s = 0.9624$, $\sigma_8 = 0.834$, sum of the neutrino masses $M_{\nu} = 0.0$ eV and dark energy equation of state $w = -1$. Furthermore, in order to enable the numerical evaluation of derivatives needed for Fisher matrix calculations, additional simulations have been run for step-wise variations of each one of the base parameters, while keeping the rest fixed, and for 500 phase-matched realizations for each pair. For the neutrino mass case, where the fiducial value is $M_{\nu} = 0.0$ eV and cannot get negative values, 3 steps have been run for values $M^{+}_{\nu} = 0.10$ eV, $M^{++}_{\nu} = 0.20$ eV and $M^{+++}_{\nu} = 0.40$ eV, using $512^3$ neutrino particles alongside CDM, in order to enable the evaluation of high order derivatives using a forward-stepping scheme. Because these neutrino simulations were initialized using the first order, Zel'dovich Approximation (ZA), instead, 500 additional realizations have been run for the fiducial cosmology but with ZA initial conditions, in order to match this configuration. All the associated details are presented in Table 1 of Ref. \citep{Villaescusa_Navarro_2020} and the related discussion. 

In addition to the above collection of $\nu$CDM simulations, that we have already used in our previous WST application \citep{PhysRevD.105.103534}, in this work we employ a new set of \textsc{Quijote-MG} simulations \citep{QujiMG} \footnote{\url{https://quijote-simulations.readthedocs.io/en/latest/mg.html}}, which expand the previously described public suite in order to enable studies of modified gravity. The highly nonlinear nature of the screening mechanisms that MG models typically possess leads to slow convergence when attempting to integrate the scalar field equations \eqref{eq:modpoisson} with N-body simulations, significantly increasing the associated computational cost. Even though many approaches in the literature have attempted to circumvent this issue using a combination of analytical \citep{Valogiannis:2019xed,Valogiannis:2019nfz,Aviles:2020wme,Liu:2021weo,Rodriguez-Meza:2023rga}, emulation \citep{2021PhRvD.103l3525R,Bai:2024cgt} or hybrid techniques \citep{Valogiannis:2016ane,Wright:2022krq,Gupta:2024seu}, there has still been a scarcity, relative to standard models, of available mocks for simulation-based investigations of beyond-GR theories. Thanks to this new set of simulations, we are able to properly forecast constraints to gravity alongside other $\Lambda$CDM parameters while working down to the nonlinear regime. In particular, Quijote-MG contains N-body simulations for the Hu-Sawicki model introduced in Section \S\ref{sec:MGmodel}, that were performed using the code MG-Gadget \citep{Puchwein:2013lza}, which is a MG expansion of Gadget \citep{Springel:2005mi}. Among a total of 4,048 simulations, it contains 4 cases corresponding to increasing values of parameter $\bar{f}_{R_0}$ (equivalent to increasing degrees of deviation from GR), in a step-wise fashion, each one of which was initialized (at $z_i=127$) using ZA initial conditions for the same 500 realizations as the $\nu$CDM derivative grid described above. In increasing order of their corresponding $\bar{f}_{R_0}$ value, we label these steps as $f_{\rm R_p}$, $f_{\rm R_{pp}}$, $f_{\rm R_{ppp}}$ and $f_{\rm R_{pppp}}$, with $\bar{f}_{R_0}$ values:
\begin{align} \label{eq:frvals}
&\{f_{\rm R_p},f_{\rm R_{pp}},f_{\rm R_{ppp}},f_{\rm R_{pppp}}\} = \\
& \{-5\times 10^{-7},-5\times 10^{-6},-5\times 10^{-5},-5\times 10^{-4}\} \nonumber,
\end{align}
while the second Hu-Sawicki model parameter is always kept fixed to $n=1$. In the next section we will explain how these can be combined to evaluate numerical derivatives with respect to deviations from GR. The output of the simulations, including both dark matter particles and also halo catalogs, was saved at redshifts $z=3, 2, 1, 0.5, 0$. In addition to these $4\times500=$2,000 simulations available for the step-wise variations of $\bar{f}_{R_0}$, a second set of 2,048 runs was performed for simultaneous variations of the 6 base $\nu$CDM parameters $+$ $\bar{f}_{R_0}$, in a Sobol sequence. Readers interested in further details on these simulations are referred to Ref. \citep{QujiMG} as well as the public website. 

\begin{figure*}%
    \centering
    \subfloat{{\includegraphics[width=0.47\textwidth]{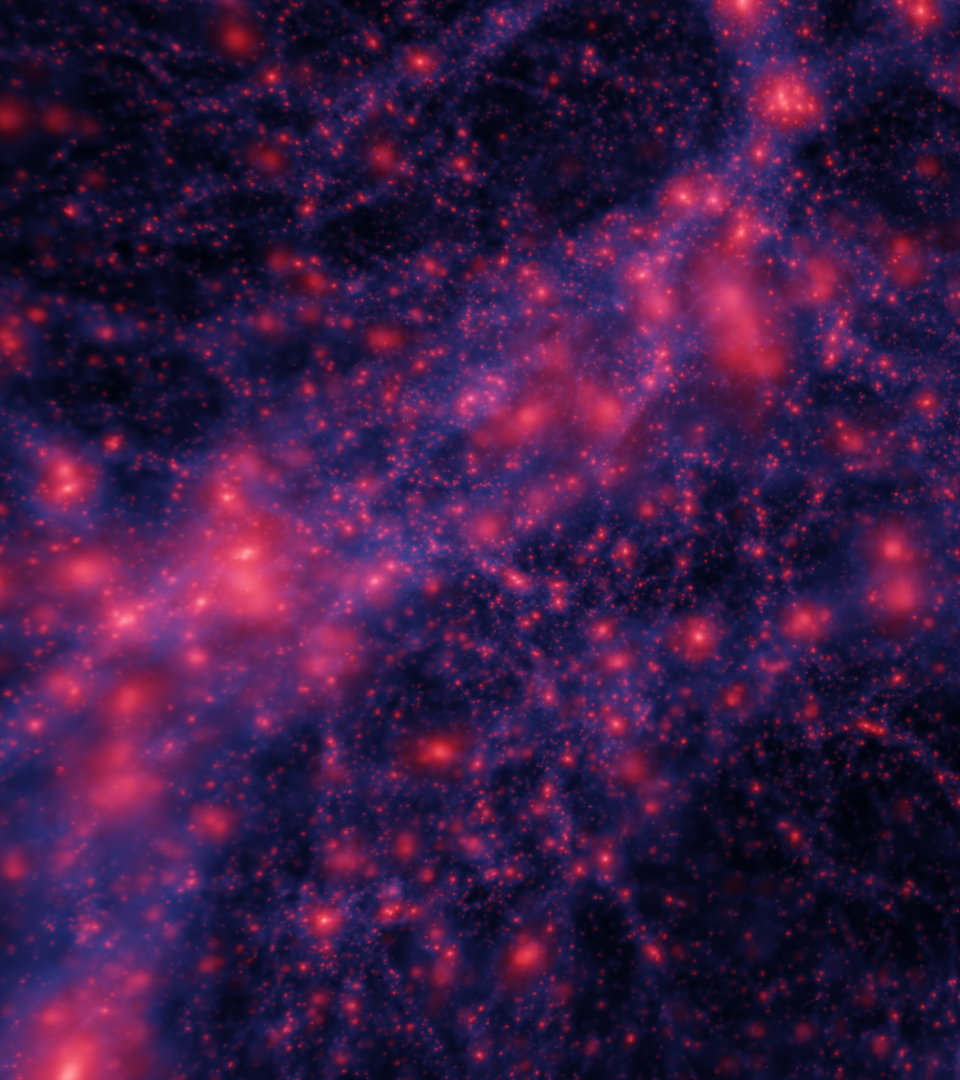} }}%
    \qquad
    \subfloat{{\includegraphics[width=0.47\textwidth]{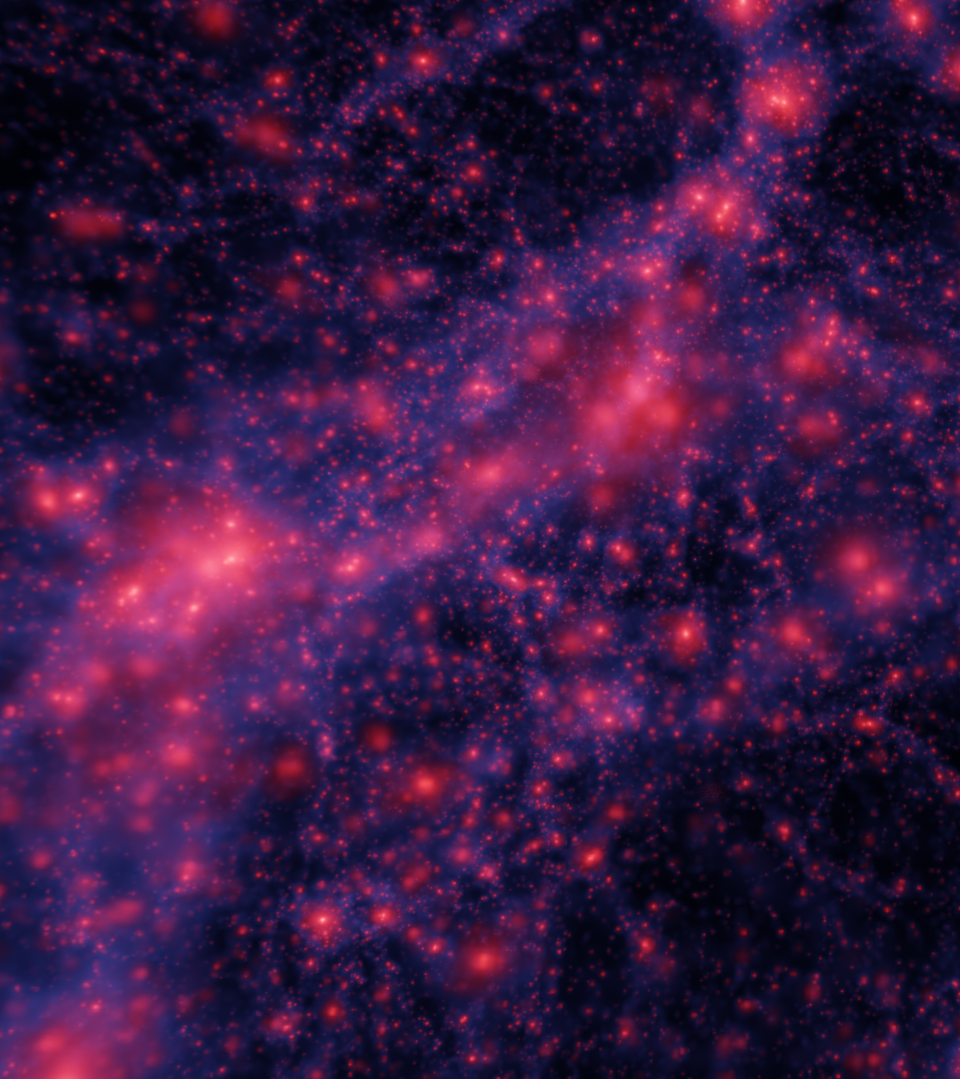} }}%
    \caption{Visualizations of the large-scale structure obtained by an instance of the GR simulation [left] and by the MG $f_{\rm R_{pppp}}$ simulation [right], both at $z=0$ and for the same set of initial conditions.}%
    \label{fig:1}%
\end{figure*}

\begin{figure}[ht]
\includegraphics[width=0.49\textwidth]{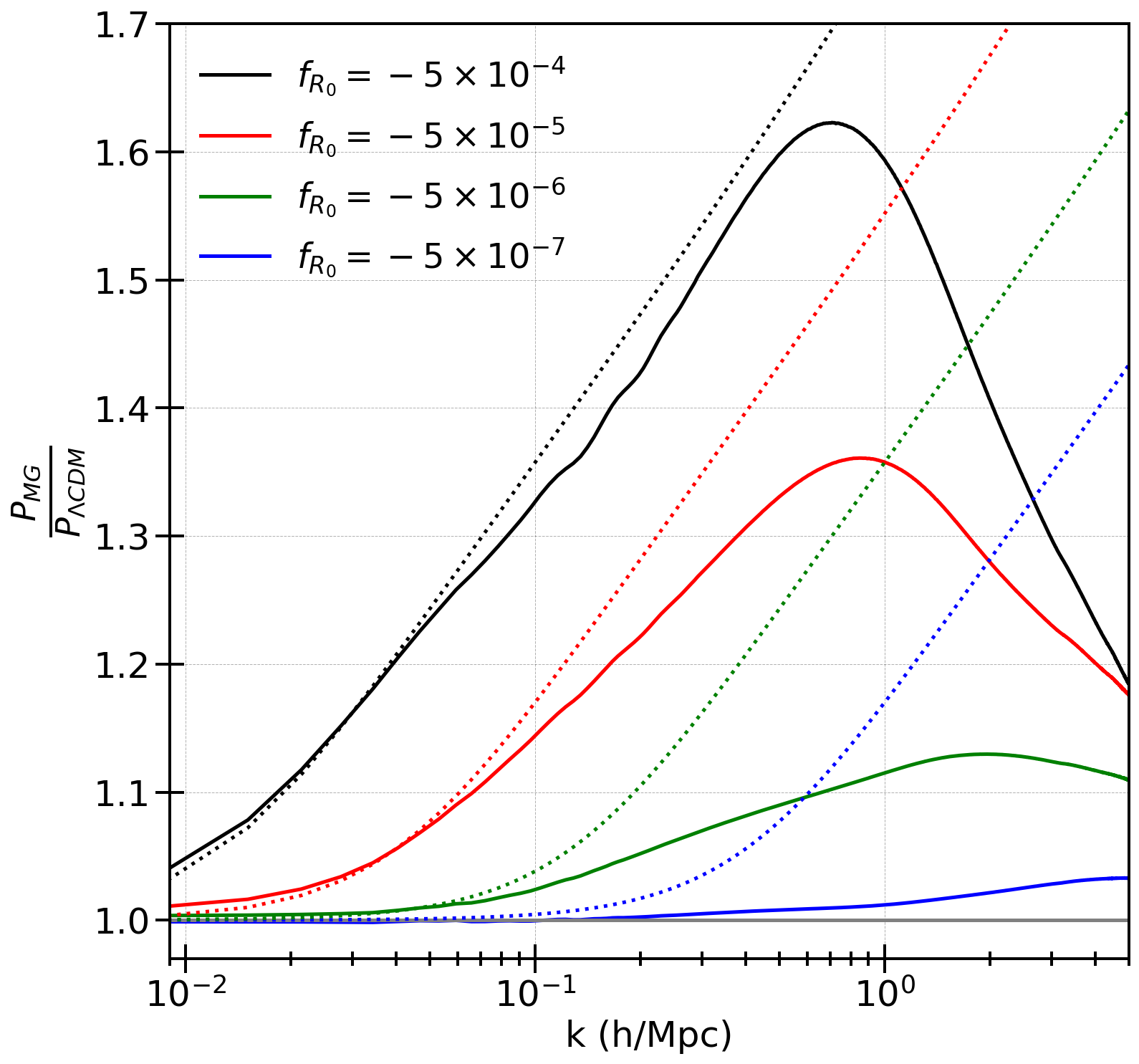}
\caption{\label{fig:epsart} The fractional deviation with respect to the $\Lambda$CDM real-space matter power spectrum is plotted for the 4 derivative steps in $\bar{f}_{R_0}$ \eqref{eq:frvals} of the Hu-Sawicki MG model, at $z=0$. The solid lines correspond to the results from the \textsc{Quijote} simulations while the dotted ones show the equivalent predictions from the linear theory prediction Eq.~\eqref{growthfact}.}
\end{figure}

\begin{figure}[ht]
\includegraphics[width=0.49\textwidth]{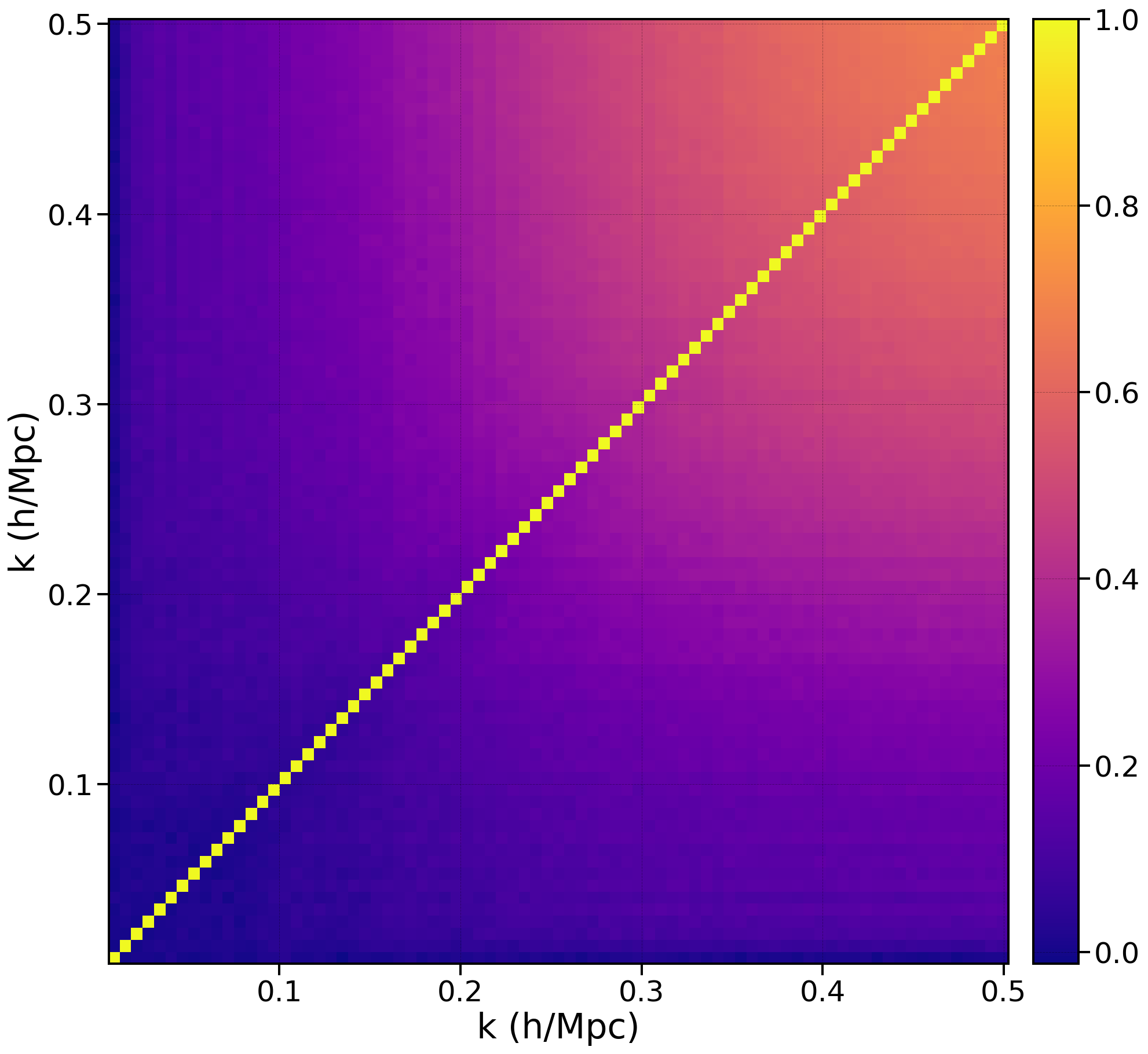}
\caption{\label{fig:Pkcov} Correlation matrix of the matter power spectrum monopole evaluated for the fiducial cosmology at redshift $z=0$.}
\end{figure}

\begin{figure}[ht]
\includegraphics[width=0.49\textwidth]{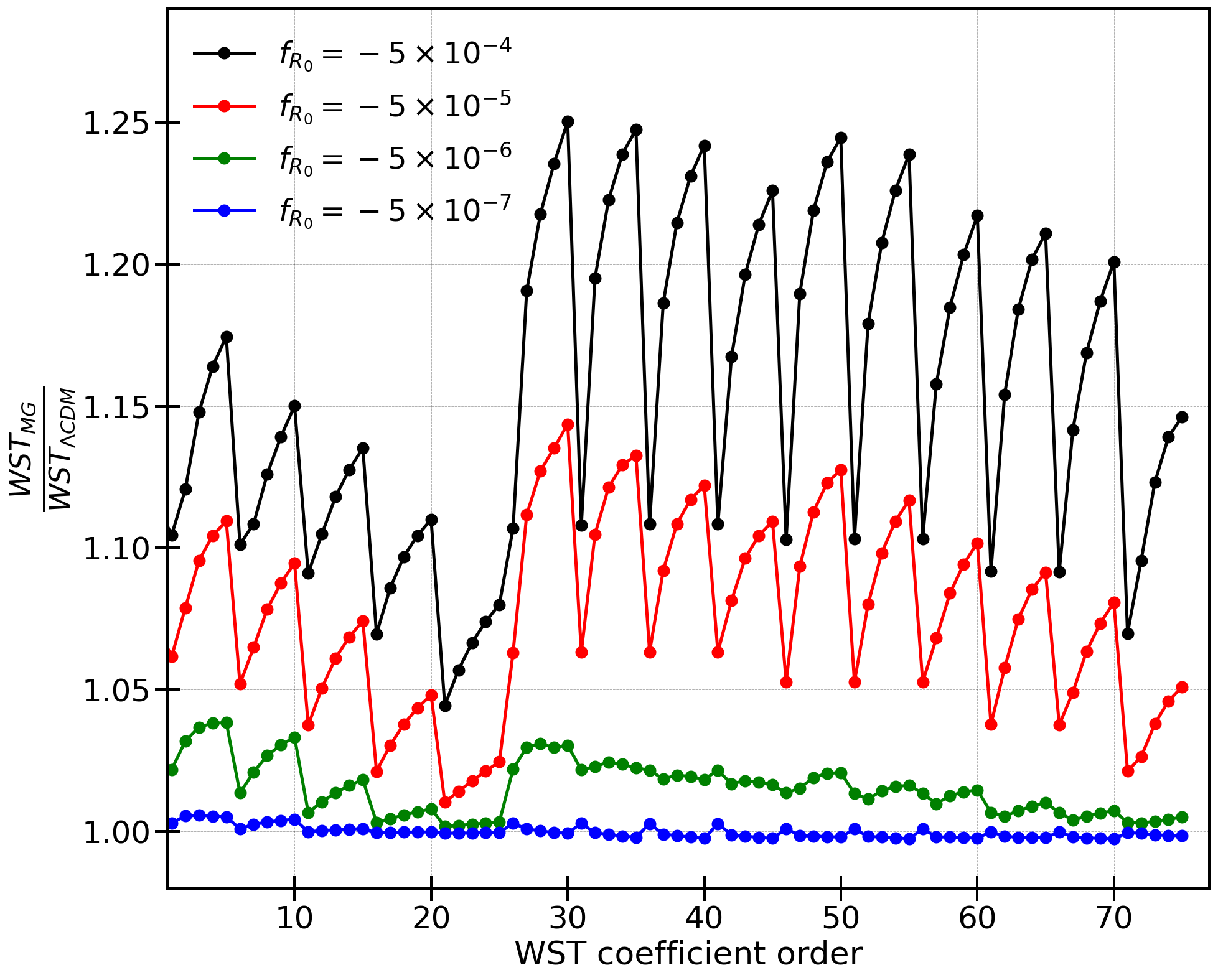}
\caption{\label{fig:WSTratio} Same as in Fig.~\ref{fig:epsart} but here shown for the data vector of 76 WST coefficients corresponding to our baseline configuration as defined in \S\ref{subsec:FisherWST}. The WST coefficients populate the data vector in order of increasing values of the $j_1$ and $l_1$ indices, with the $l_1$ index varied faster. As we look from left to right, the first 25 points comprise the $1^{st}$ order group of wavelets, before moving to the $2^{nd}$ order part.}
\end{figure}

\begin{figure}[ht]
\includegraphics[width=0.49\textwidth]{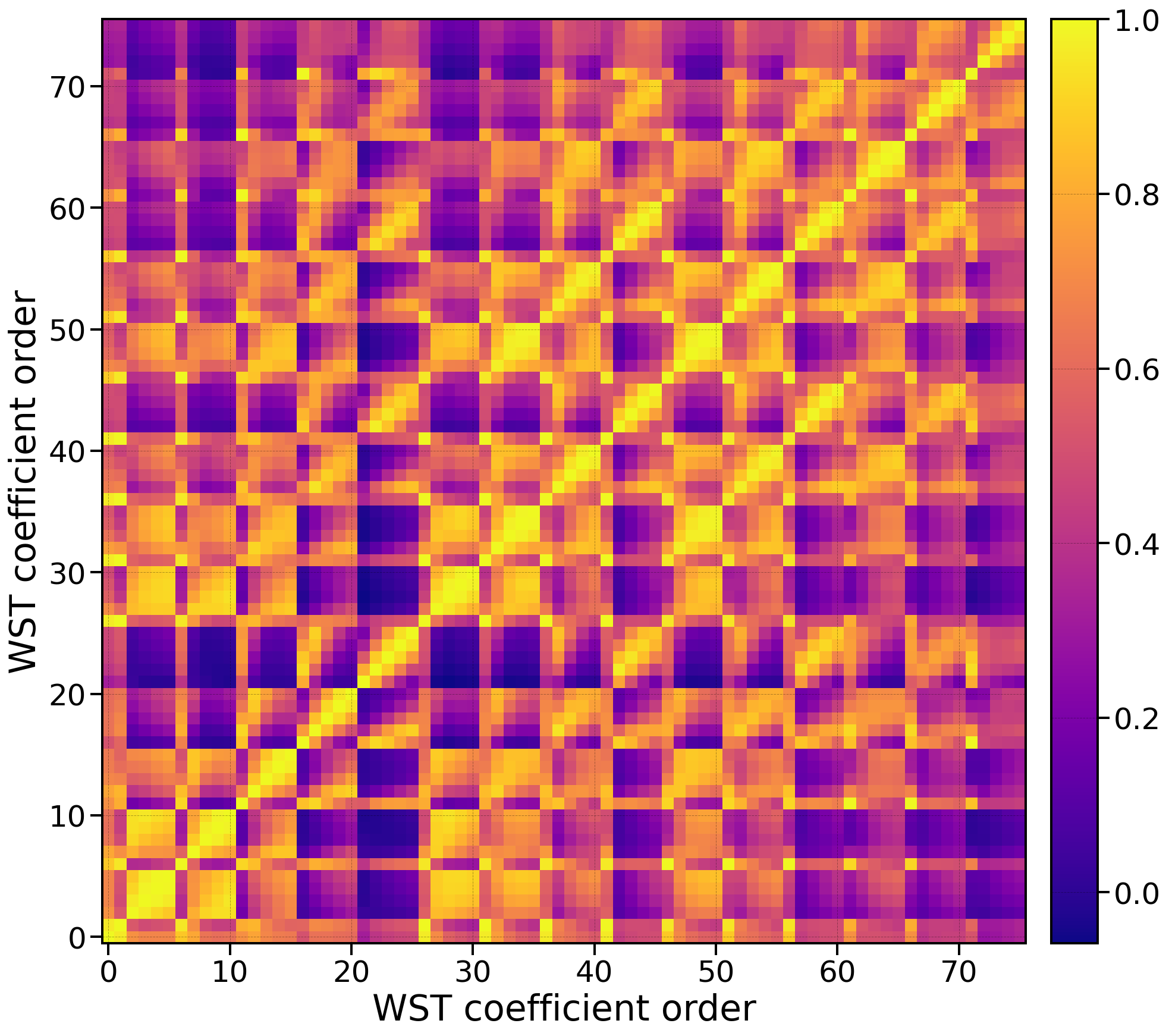}
\caption{\label{fig:WSTcov} Correlation matrix of all 76 coefficients of the WST data vector evaluated at the fiducial cosmology at $z=0$. The WST coefficients populate the data vector in order of increasing values of the $j_1$ and $l_1$ indices, with the $l_1$ index varied faster, just like in Fig.~\ref{fig:WSTratio}.}
\end{figure}

\section{Fisher forecast}\label{sec:Fisher}
The main focus of our analysis will be to forecast the constraining power of our summary statistics of interest, which, as we explain in detail below, will be the WST and the standard matter power spectrum. Under the assumption of a Gaussian likelihood, which we confirm in the Appendix \S\ref{app:Gauss}, this can be commonly achieved using the Fisher matrix formalism. According to it, if $O_m$ is a data vector of $m$  observations, which are considered functions of a set of cosmological parameters $\theta_{\alpha}$, then the Fisher matrix is defined as:
\begin{equation}\label{Fisher}
F_{\alpha \beta} = \frac{\partial O_i}{\partial \theta_\alpha} C^{-1}_{ij} \frac{\partial O^T_j}{\partial \theta_\beta}, 
\end{equation}
where $C_{ij}$ denotes the covariance matrix of $O_m$. The Fisher matrix expression of Eq. \eqref{Fisher} assumes that the covariance matrix is independent of the cosmology, which has been found to be a good approximation \citep{refId0Car,Tegmark_1997}. According to the Kramer-Rao inequality, it then follows that the marginalized $1\sigma$ errors on each one of the parameters $\theta_{\alpha}$ is always $\sigma_{\alpha} \geq \sqrt{\left(F^{-1}\right)_{\alpha\alpha}}$, providing thus a lower bound. 

In this work we evaluate the ingredients of the Fisher matrix \eqref{Fisher}, which in turn gives the $1\sigma$ errors on the parameters through $\sigma_{\alpha} = \sqrt{\left(F^{-1}\right)_{\alpha\alpha}}$, using the Quijote simulations described in \S\ref{sec:Simulations}. In particular, for the covariance matrix estimation we use the 15,000 realizations available for the fiducial cosmology to get:
\begin{equation}\label{eq:covmat}
C = \frac{1}{\rm N_{\rm r} - 1}\sum_{\rm k=1}^{\rm N_{\rm r}} \left(O^k_m-\bar{O}_m\right)\left(O^k_m-\bar{O}_m\right)^{\rm T},
\end{equation} 
where $\bar{O}_m$ is the mean value of the observable evaluated over the $N_{\rm r}=$15,000 realizations and $O^k_m$ its corresponding value for the k-th realization. Furthermore, upon inversion of the covariance matrix in Eq. \eqref{Fisher}, we apply the standard de-biasing Hartlap factor \citep{refId22}:
\begin{equation}\label{Hartlap}
\hat{C}^{-1}= \frac{N_{\rm r}-N_{d}-2}{N_{\rm r}-1}C^{-1}, 
\end{equation}
where $N_{d}$ is the dimensionality of the observable data vector $O_m$. For the vectors of WST coefficients and matter power spectrum that we consider in this work, for which, as we will explain shortly, $N_{d}<100$, the large number of available realizations $N_{\rm r}=$15,000 guarantees that this factor is practically very close to 1. 

The numerical derivatives w.r.t. the $\Lambda$CDM parameters $\{\Omega_m, \Omega_b, h, n_s, \sigma_8\}$ can be evaluated using the central difference expression:
\begin{equation}\label{central}
\frac{\partial O_{\rm m}}{\partial \theta}=\frac{O_{\rm m}(\theta+\delta \theta)-O_{\rm m}(\theta-\delta \theta)}{2 \delta \theta}, 
\end{equation}
with corresponding step values $\{0.02,0.004,0.04,0.04,0.03\}$ around the fiducial cosmology, as can be seen in Table I of \citep{Villaescusa_Navarro_2020}. For the neutrino mass $M_{\nu}$ we follow our previous works \citep{Villaescusa_Navarro_2020,PhysRevD.105.103534} and work with the highest order forward difference derivative version using all 3 available steps, that is:
\begin{equation}\label{Mnuder}
\frac{\partial O}{\partial M_{\nu}}=\frac{O(M^{+++}_{\nu})-12 O(M^{++}_{\nu})+32 O(M^{+}_{\nu})-21 O(0)}{12 M^{+}_{\nu}}, 
\end{equation}
where the subscript in $O_{\rm m}$ above was omitted for compactness. We also clarify that, since we are interested in 3D galaxy clustering applications, we will only work with the density field traced by cold dark matter and baryons in the presence of massive neutrinos, $\delta_{cb} = \delta_{CDM}+\delta_b$ (since galaxies are biased tracers of it), rather than the total field $\delta_m = \delta_{CDM}+\delta_b+\delta_{\nu}$,  which can only be probed by weak lensing \citep{Bayer:2021kwg}. 

Last but not least, for the Hu-Sawicki model extension, there are 4 available steps away from GR, distributed equally in a logarithmic scale, as seen in Eq. \eqref{eq:frvals}. In order to map them to equal steps in a linear scale, which is what the derivative expressions assume, we first perform the following variable transformation: 
\begin{equation}\label{eq:Yarr}
Y=\left(\bar{f}_{R_0}\right)^{\log_{10}2},
\end{equation}
which maps to 4 equal steps in the linear space, as follows:
\begin{align} \label{eq:Yarrsteps}
&\{Y_{\rm p},Y_{\rm pp},Y_{\rm ppp},Y_{\rm pppp}\} = \\
& \{-0.0127, -0.0254, -0.0507, -0.101\} \nonumber,
\end{align}
where $Y_{\rm pppp}=2Y_{\rm ppp}=4Y_{\rm pp}=8Y_{\rm p}$ and the fiducial value for the $\Lambda$CDM cosmology is again $Y=0$. Using these steps, one can in principle construct multiple versions of the derivatives, including a higher order version than Eq. \eqref{Mnuder}. However, and as we explain in detail in the Appendix \S\ref{app:Der}., we chose not to use the smallest step $Y_{\rm p}$, because this value corresponds to a very small deviation from the GR limit in our summary statistics, and the observed signal might be masked by numerical noise at large scales. We instead work with $Y_{\rm pppp},Y_{\rm ppp},Y_{\rm pp}$, which can be combined, similarly to the neutrino mass case, to give the prediction for the derivative away from the $Y=0$ $\Lambda$CDM limit:
\begin{equation}\label{fRder}
\frac{\partial O}{\partial Y}=\frac{O(Y_{\rm pppp})-12 O(Y_{\rm ppp})+32 O(Y_{\rm pp})-21 O(0)}{12 Y_{\rm pp}}. 
\end{equation}
We emphasize that we have checked and confirmed that our conclusions are nevertheless robust against the choice of the version used to evaluate the derivative w.r.t. Y, as explained in the Appendix \S\ref{app:Der}. 

To summarize, we use the above methods to evaluate the numerical derivatives of the observable $O_{\rm m}$ (in Eq. \eqref{Fisher}) with regards to the 7 cosmological parameters of interest:
\begin{equation}
\theta_{\alpha}=\{\Omega_m, \Omega_b, H_0, n_s, \sigma_8, M_{\nu},Y\},
\end{equation}
averaging over the 500 random realizations available for each parameter. Given that the massive neutrino and Hu-Sawicki cosmology simulations were initialized using ZA initial conditions, we use the ZA fiducial simulations for the evaluation of the corresponding $\Lambda$CDM limit, $O(0)$, in their derivatives, in Eqs. \eqref{Mnuder} $\&$ \eqref{fRder}. In the Appendix \S\ref{app:Stability} we carefully check and confirm the numerical stability of our results as a function of the available number of realizations for the evaluation of the derivatives and the covariance matrices of our statistics.

We should note at this point, that a given Fisher constraint on Y can in principle be subsequently translated to a constraint on $\bar{f}_{R_0}$, which is the parameter most commonly quoted in the literature (for example, see Ref.~\citep{Lombriser:2014dua}). Indeed, using the standard error propagation formula for the power law Eq.~\eqref{eq:Yarr}) gives:
\begin{equation}\label{Eq:Errprop}
\sigma_{\bar{f}_{R_0}}=\frac{\sigma_Y}{\log_{10} 2}\left(\frac{\bar{f}_{R_0}}{Y}\right).
\end{equation}
However, for our fiducial values, $Y=\bar{f}_{R_0}=0$, the fraction in Eq.\eqref{Eq:Errprop} becomes singular, preventing us from applying the transformation in this particular case. As a result we will quote constraints on Y in this analysis. 

Having laid out the details of our Fisher framework, we finally proceed to describe the exact configuration of the summary statistics that we will apply it to:
\subsection{WST}\label{subsec:FisherWST}
In order to evaluate the WST coefficients, we first begin with the real-space 3D matter overdensity field at $z=0$, 
\begin{equation}\label{eq:deltaef}
\delta_m(\vec{x}) = \frac{\rho_m(\vec{x})}{\bar{\rho}_m} -1,
\end{equation}
which we extract from each one of the 24,000 Quijote realizations we use for the covariance matrix and the derivatives. In particular, we make use of the public package \textsc{Pylians}\citep{Pylians} \footnote{\url{https://pylians3.readthedocs.io/en/master/index.html}} to evaluate the overdensity field \eqref{eq:deltaef} on a grid of $256^3$ resolution using the Piecewise-Cubic Spline (PCS) interpolation scheme. These fields are then fed as input into the WST Eqs. \eqref{eq:WSTcoeff:sol}, which are evaluated using \textsc{Kymatio} \citep{2018arXiv181211214A}. Staying consistent with our previous works \citep{PhysRevD.105.103534,PhysRevD.106.103509,PhysRevD.109.103503}, we adopt a baseline WST configuration for $J=L=4$ and $q=\sigma=0.8$, which results in a vector of $N_{d}=76$ $S_0+S_1+S_2$ coefficients up to $2^{nd}$ order, and $N_{d}=26$ coefficients up to first order. Last but not least, in order to better facilitate an ``apples vs apples'' comparison against the power spectrum that is commonly evaluated at a fixed $k_{\rm max}$, we further implement an additional pre-processing step and apply a sharp k-cut filter to the field \eqref{eq:deltaef} before the WST evaluation. This step guarantees the removal of any residual small-scale sensitivity that might be coming from the tails of the Gaussian-like wavelets that the WST convolves with, as was also performed in \citep{PhysRevD.109.103503} (see a relevant discussion in Appendix E of that work). For our baseline WST case, we will restrict ourselves to a conservative $k_{\rm max}=0.5$ h/Mpc, in order to make sure we stay within the regime in which we can reliably trust the validity of our simulations. 

\begin{figure*}[ht]
\includegraphics[width=0.99\textwidth]{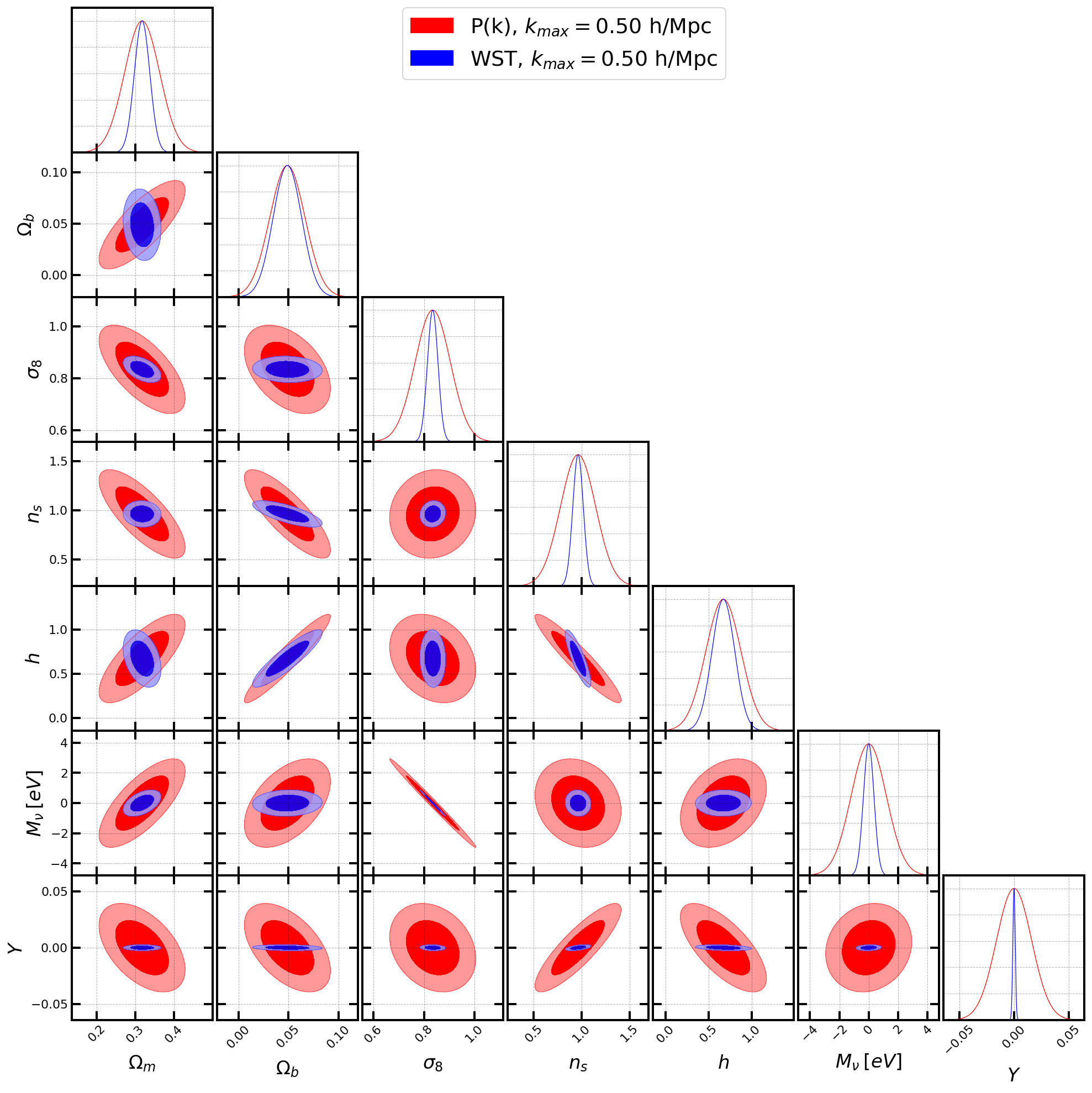}
\caption{\label{fig:epsart2} 1$\sigma$ and  2$\sigma$ contours of the cosmological parameters as obtained from the Fisher forecast using the real-space matter power spectrum (red) and the WST coefficients (blue) at $z=0$.}
\end{figure*}

\subsection{Power spectrum}
In order to have a benchmark that can allow us to assess the extent to which the WST can improve upon the constraints to gravity obtained using conventional statistics, we also perform forecasts for the real-space matter power spectrum at $z=0$, given by \footnote{The angular brackets in Eqs. \eqref{eq:Pk0} and also \eqref{eq:WSTcoeff:power}-\eqref{eq:WSTcoeff:sol} denote ensemble average}: 
\begin{equation}\label{eq:Pk0}
P_m(k) = \langle |\tilde{\delta}_m(k)|^2 \rangle,
\end{equation}
which we evaluate in 79 equally-spaced logarithmic bins and $1024^3$ grid resolution down to $k_{\rm max}=0.5$ h/Mpc, using \textsc{Pylians} \citep{Pylians}. Given that we only work in real-space, where the field \eqref{eq:deltaef} is isotropic, we do not need to consider higher order ($\ell > 0$) multipoles of the power spectrum, as they vanish. 

         \begin{table}[t!]
    	\begin{tabular}{ | p{9em} | p{3.7em} |p{3.7em}   || p{3.7em} |p{3.7em}|}
		\hline
		{Statistic} & \multicolumn{2}{c||}{P(k)} & \multicolumn{2}{c|}{WST}
		\\ \hline
    	$k_{\rm max} \ [h/Mpc]$ &  $0.2$ & $0.5$&  $0.2$& $0.5$	
		\\ \hline \hline
		$\sigma(\Omega_m)$
		& 0.079 & 0.046 
		& 0.077 &  0.020 
		\\ \hline
		$\sigma(\Omega_b)$
		& 0.034 &  0.017 
		& 0.033 &  0.014 
		\\ \hline
		$\sigma(\sigma_8)$
		& 0.080 &  0.069 
		& 0.080 &  0.021 
		\\ \hline
		$\sigma(n_s)$
		& 0.415 &  0.184 
		& 0.356 &  0.054 
		\\ \hline
		$\sigma(h)$ 
		& 0.432 &  0.204 
		& 0.40 & 0.132 
		\\ \hline
		$\sigma(M_\nu$) [eV]
		& 1.310 &   1.197 
		& 1.270 &  0.357	 
            \\ \hline
		$\sigma(Y)$ 
		& 0.025 &   0.016 
		& 0.0054 &  0.0012	  
		\\ \hline
		\end{tabular}
			\caption{Marginalized $1\sigma$ errors on all cosmological parameters obtained from the matter $P(k)$ and the baseline WST configuration at $z=0$. The results are reported using both $k_{\rm max}=0.2$ h/Mpc and also $k_{\rm max}=0.5$ h/Mpc.
			}
    	\label{tab1}
	\end{table}

\section{Results}\label{sec:Results}
We begin this section by visualizing the characteristic effects of the Hu-Sawicki model on the large-scale structure formation, through a side-by-side comparison between the outputs of the largest deviation MG simulation, $f_{\rm R_{pppp}}$, and the corresponding GR case at $z=0$, in Fig. \ref{fig:1}. The visual differences in the clustering pattern caused by the effect of the MG fifth force are apparent, even through this simple comparison. We then proceed to examine our familiar case of the matter power spectrum. In Fig.~\ref{fig:epsart} we plot its fractional deviation with respect to the fiducial $\Lambda$CDM prediction (often referred to as the ``boost") for the 4 derivative steps \eqref{eq:frvals}, evaluated at $z=0$. The linear theory predictions from Eq. \eqref{growthfact} (dotted lines) illustrate the effect of the Compton mass of the scalaron at the background level \eqref{Eq:mass}, below which GR is recovered and the ratio $\frac{P_{MG}}{P_{\Lambda CDM}}$ thus approaches unity at large scales. Increasing absolute values of $\bar{f}_{R_0}$ lead to smaller values of the Compton mass and more pronounced deviations w.r.t. the $\Lambda$CDM power spectrum, which (for the linear theory prediction) monotonically increase as we look into larger Fourier modes (smaller scales). The nonlinear effect of the chameleon screening becomes clearly apparent when we look at the corresponding simulated predictions (solid lines), which become progressively more suppressed compared to their linear theory counterparts as we focus on smaller scales in Fig.~\ref{fig:epsart}. For decreasing absolute values of $\bar{f}_{R_0}$, this suppression becomes increasingly more pronounced, a reflection of a higher degree of chameleon screening. These results reflect the characteristic phenomenology of the Hu-Sawicki model, which has been extensively studied in the literature (for example see an overview in Ref.~\citep{Alam:2020jdv}). Looking into the correlation matrix of the matter power spectrum, in Fig.~\ref{fig:Pkcov}, it exhibits a well-known structure that is perfectly diagonal at linear scales, but progressively includes non-diagonal contributions as we focus on the nonlinear regime \citep{Villaescusa_Navarro_2020}. 

In Fig.~\ref{fig:WSTratio} we then repeat the same plotting exercise for the data vector of the 76 WST coefficients of our baseline configuration up to second order, that was defined in \S\ref{subsec:FisherWST}. Even though the result is unsurprisingly harder to interpret than the familiar power spectrum case, we notice that it does reflect the same basic trends: the fractional deviation of the WST summary statistic w.r.t. its $\Lambda$CDM prediction becomes increasingly more pronounced for larger absolute values of $\bar{f}_{R_0}$, just like in the power spectrum case of Fig.~\ref{fig:epsart}. In a similar fashion, the amplitude of the deviation is relatively larger for the groups of coefficients that probe the smallest scales, when compared to the ones capturing the largest scales at the same order (i.e. from left to right in Fig.~\ref{fig:WSTratio}). When we finally inspect the correlation matrix of the WST coefficients, that is shown in Fig.~\ref{fig:WSTcov}, we observe the same patterns as in our previous applications \citep{PhysRevD.105.103534,PhysRevD.106.103509,PhysRevD.109.103503}: starting with the $1^{st}$ order group of wavelets (that is, until index 26), we notice the existence of strong correlations between nearby scales and angles (close to the diagonal), which progressively weaken and eventually vanish as we look into the coupling between the smallest and the largest wavelet scales. Similar patterns permeate into the $2^{nd}$ order group of wavelets and their correlations with the corresponding scales at the $1^{st}$ order level.

Moving on to the discussion of our cosmological results, in Fig.~\ref{fig:epsart2} we plot the 1$\sigma$ and 2$\sigma$ Fisher ellipses obtained on all 7 cosmological parameters of interest using the matter power spectrum and the baseline WST configuration at $z=0$. At the same time, the values of the marginalized  1$\sigma$ errors are explicitly listed in Table~\ref{tab1}. We clarify again at this point that both predictions include contributions from scales down to at most $k_{\rm max}=0.5$ h/Mpc, which serves as a conservative limit. Looking into the results, and starting from the MG parameter Y, we immediately observe that the WST delivers an impressive improvement on the value of the 1$\sigma$ constraint, which is $\sim10\times$ tighter than the corresponding prediction from the matter power spectrum. The parameter that exhibits the second best improvement using the WST is the sum of the neutrino masses, with an 1$\sigma$ error that is tighter by a factor of $\sim3.4$ compared to the corresponding power spectrum value, in line with our previous findings in \citep{PhysRevD.105.103534}. Furthermore, using the WST coefficients improves the constraints obtained on the rest of the $\Lambda$CDM parameters by a factor in the range $1.2-3.3$, as seen in Table~\ref{tab1}, also in broad agreement with \citep{PhysRevD.105.103534}. We note that a joint forecast using P(k)+WST does not improve the 1$\sigma$ errors on $M_{\nu}$ and $Y$ by more than $20\%$ compared to the WST-only analysis, which is why we do not explicitly show the results for this combination. We also report that the 1$\sigma$ errors obtained on the 6 $\nu$CDM parameters increase by up to $90\%$ for the power spectrum and by up to $20 \%$ for the WST when we marginalize over the new MG parameter Y, as compared to the pure GR forecast that we performed in \citep{PhysRevD.105.103534} with these \textsc{Quijote} simulations. In the specific case of the WST, the implementation of a sharp k-filter on top of the Gaussian smoothing has led to an additional slight increase in the errors compared to the values reported in \citep{PhysRevD.105.103534}. We emphasize again at this point that the numerical stability of these results as a function of the available number of realizations for the evaluation of the Fisher matrix has been checked and confirmed in the Appendix \S\ref{app:Stability}.

As we already clarified in \S\ref{sec:Fisher}, since our fiducial cosmology is the one governed by GR gravity, with a value of $Y=0$, we cannot propagate our 1$\sigma$ errors on Y to the typically quoted parameter $\bar{f}_{R_0}$, since the transformation \eqref{Eq:Errprop} becomes singular in this case. If we however assume a cosmology-independent Fisher matrix for a small perturbation around the fiducial cosmology, for example a scenario with $\bar{f}_{R_0}=-10^{-6}$, we would get 1$\sigma$ errors $\sigma_{\bar{f}_{R_0}}=2.55\times 10^{-7}$ from the WST and $\sigma_{\bar{f}_{R_0}}=3.4\times 10^{-6}$ using the matter power spectrum, allowing us to connect our analysis to the commonly quoted Hu-Sawicki constraints when analyzing cosmological and astrophysical observational probes \citep{Lombriser:2014dua}. Such a comparison reveals that these constraints are competitive with the current tightest bounds coming from galactic structure \citep[e.g.][]{Desmond:2020gzn}, which might however be relying on simple assumptions to model the screening effect in galactic settings \citep{{Burrage:2023eol}}.

\begin{figure*}[ht]
\includegraphics[width=0.99\textwidth]{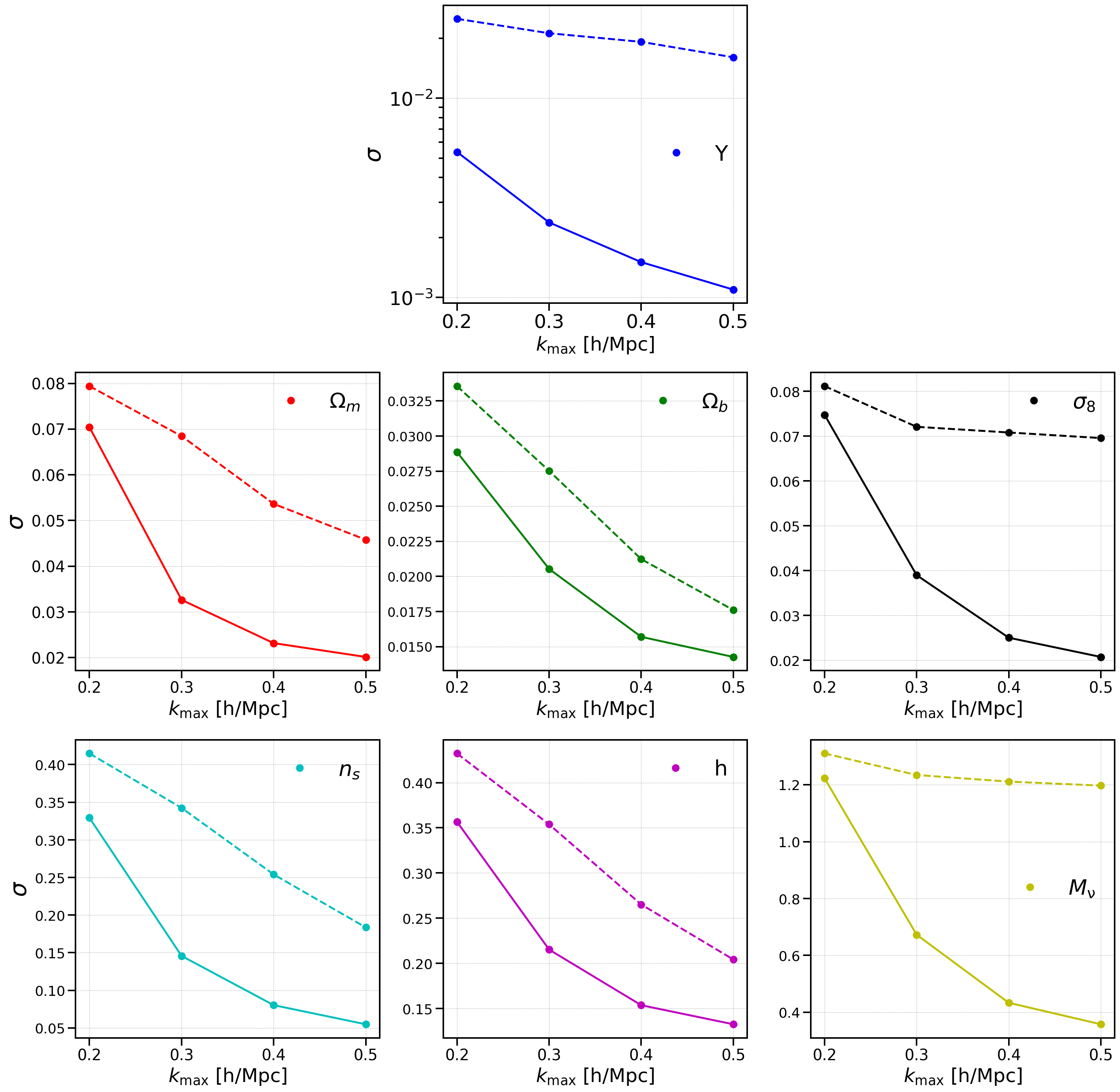}
\caption{\label{fig:sigmavskmax} The marginalized 1$\sigma$ errors obtained on all 7 parameters are plotted as a function of the $k_{\rm max}$ value used for the evaluation of the matter power spectrum (dashed lines) and the WST coefficients (solid lines).}
\end{figure*}

\begin{figure*}[ht]
\includegraphics[width=0.99\textwidth]{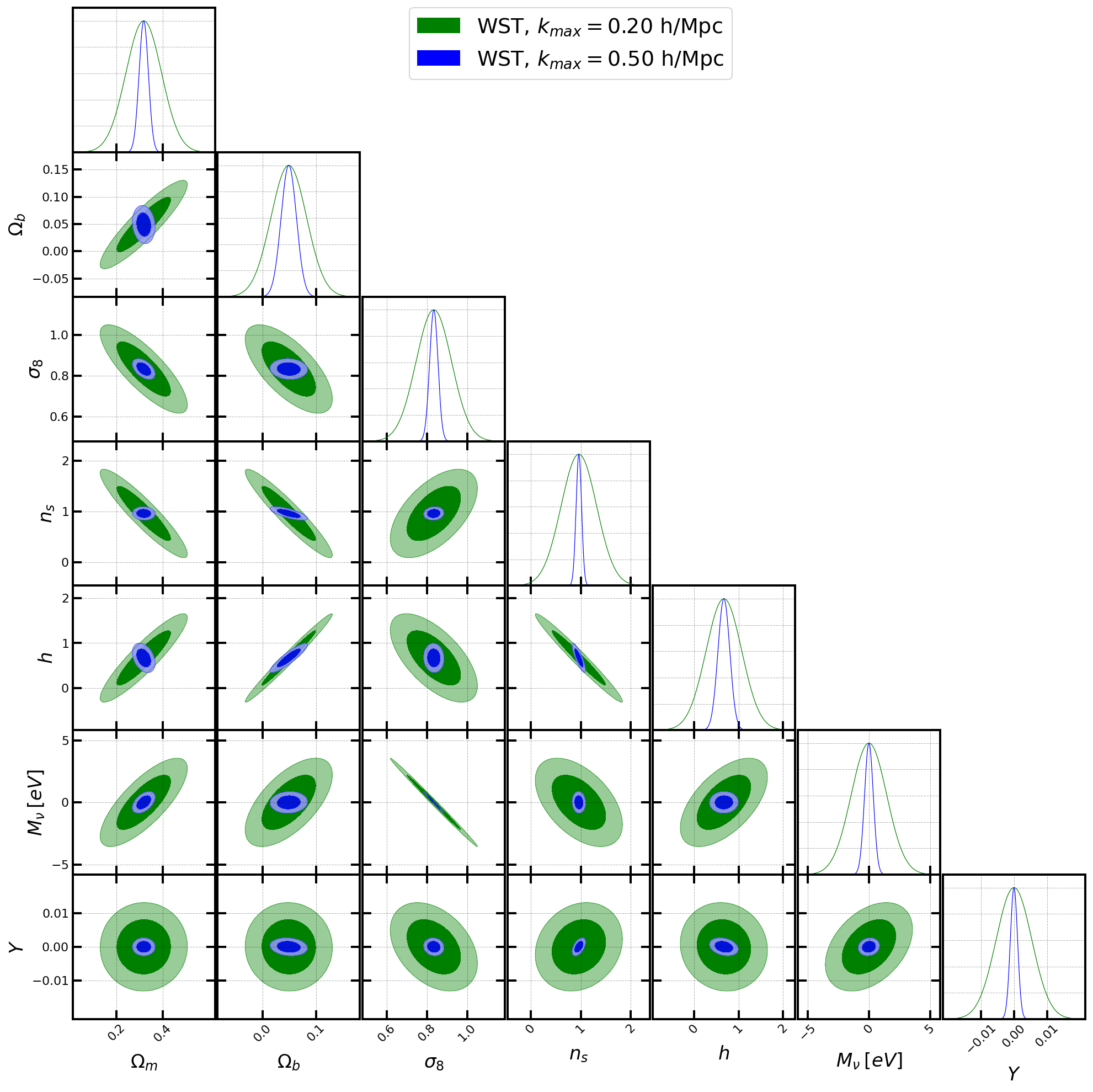}
\caption{\label{fig:epsart4} 1$\sigma$ and 2$\sigma$ Fisher contours obtained on the cosmological parameters using the WST estimator evaluated up to a maximum wavenumber $k_{\rm max}=0.2$ h/Mpc (green) and $k_{\rm max}=0.5$ h/Mpc (blue), at $z=0$.}
\end{figure*}

\begin{figure*}[ht]
\includegraphics[width=0.99\textwidth]{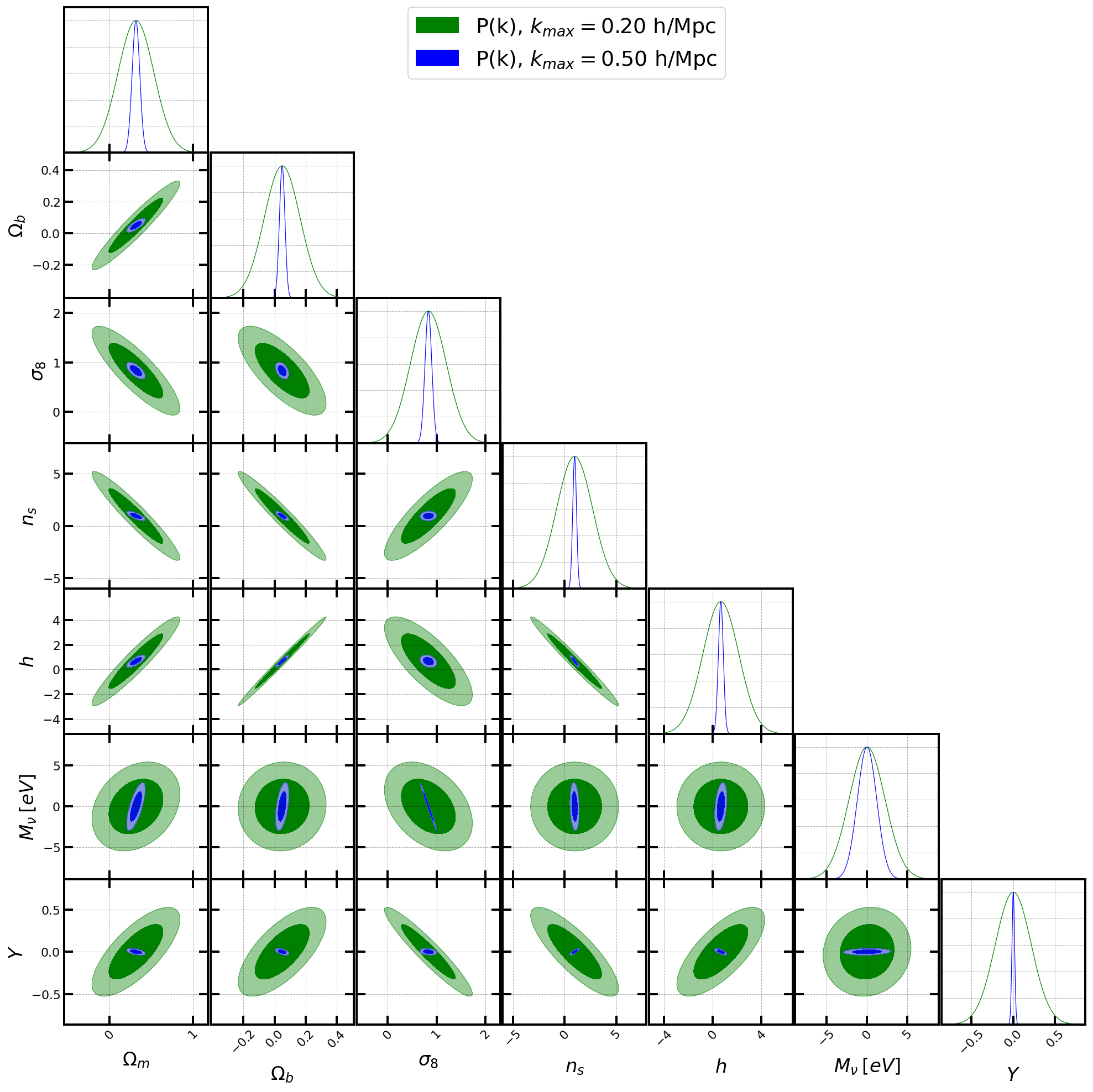}
\caption{\label{fig:epsartPk} The $k_{\rm max}$ comparison of Fig.~\ref{fig:epsart4} is repeated for the matter power spectrum at $z=0$.}
\end{figure*}

Even though the main focus of this application is to evaluate the WST's ability to extract non-Gaussian information that lies at the smallest scales, it is also useful to examine how our results change if we restrict our focus on larger scales instead, through the application of sharp k-filters with a smaller $k_{\rm max}$ \footnote{We note that this effect can also be investigated by adopting different combinations of the Gaussian smoothing width, $\sigma$, and field resolution, $N_{\rm grid}$, that enter the WST evaluation, as we explained in \S\ref{subsec:FisherWST}. For the exploratory purposes of this work, however, the comparison is more straightforward if we stick to the same WST settings and only change the $k_{\rm max}$ value of the top-hat filter.}. Indeed, in Fig.~\ref{fig:sigmavskmax} we plot the successive degradation of the marginalized 1$\sigma$ errors on the cosmological parameters obtained from both summary statistics as we impose more conservative scale cuts, up to a $k_{\rm max}=0.2$ h/Mpc. At the same time, we also compare the Fisher ellipses obtained using the smallest and the largest scale cut for the WST in Fig.~\ref{fig:epsart4} and for the power spectrum in Fig.~\ref{fig:epsartPk}. In Fig.~\ref{fig:sigmavskmax} we notice that, as expected, the values of the 1$\sigma$ constraints obtained from both estimators increase as we progressively remove valuable small-scale information. For most parameters, the slope of this increase is steeper for the WST, such that the overall net improvement w.r.t. the power spectrum is smaller at $k_{\rm max}=0.2$ h/Mpc compared to the baseline $k_{\rm max}=0.5$ h/Mpc case. This is not surprising given that most of the constraining power is, in principle, expected to reside in the nonlinear scales. As these get progressively removed, the two results will eventually converge in the linear regime that contains only Gaussian information. Furthermore, the coarse logarithmic binning of the WST configuration we adopted in this work has not been fully optimized for the extraction of large-scale information. Nevertheless, the WST is still able to provide an overall improvement in the errors on all the parameters even at $k_{\rm max}=0.2$ h/Mpc, which is actually equal to a factor of $\sim4.5$ for the MG parameter Y (and up to $15 \%$ for the rest of the parameters). This result seems to be consistent with recent findings hinting at the existence of substantially more information available beyond the power spectrum, even at the quasi-linear regime \citep{Nguyen:2024yth}. Lastly, we note that when we attempt to filter out scales $k_{\rm max}<0.2$ h/Mpc, our WST predictions become substantially more noisy, leading to insufficient levels of numerical convergence of our Fisher predictions, which we cannot trust and do not report as a result. More simulations would be needed to investigate these scales with this WST configuration, which we defer to future work. 

\begin{figure*}[ht]
\includegraphics[width=0.99\textwidth]{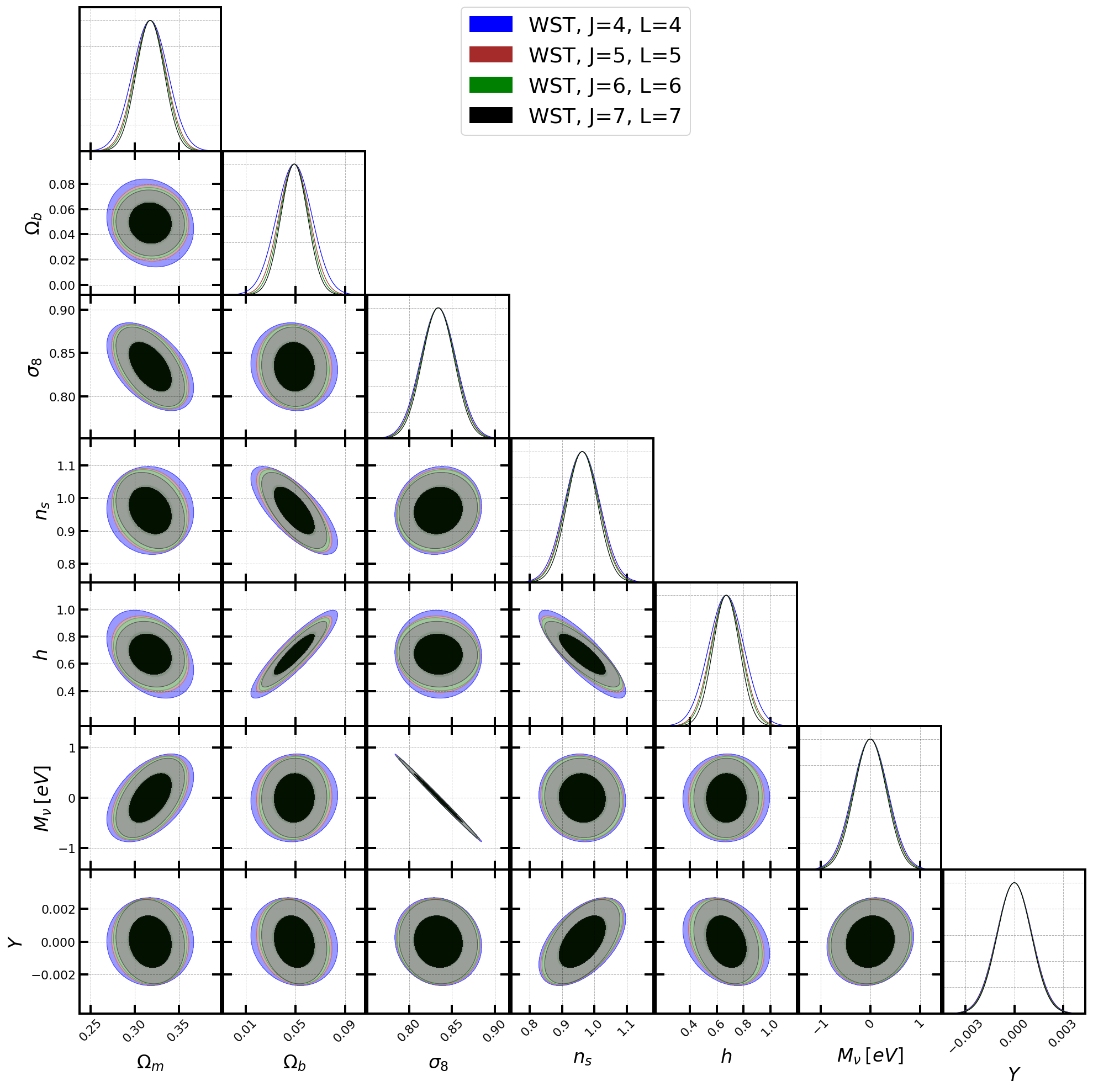}
\caption{\label{fig:epsart5} Dependence of the 1$\sigma$ and 2$\sigma$ Fisher ellipses obtained on the cosmological parameters as a function of the number of spatial J and angular L configurations included in the WST estimator, at $z=0$ and for $k_{\rm max}=0.5$ h/Mpc.}
\end{figure*}

As we explained in \S\ref{subsec:FisherWST}, our analysis adopted a baseline WST configuration that uses $J=L=4$ spatial and angular scales, which we have found to deliver a satisfactory trade-off between performance and computational cost. In Fig.~\ref{fig:epsart5} we proceed to explicitly show how the inclusion of larger scales, up to $J=L=7$, can further improve our constraints. We find that, for the maximum case using $J=L=7$, the 1$\sigma$ errors do not improve by more than $30 \%$ compared to the baseline case, with the corresponding improvement on the MG parameter Y being equal to $\sim 5 \%$ only. Given that the main focus of this work was to derive constraints on the MG model, and also that including these additional scales makes the WST evaluation slower by a factor of $\sim 3$, we did not consider this to be a worthwhile trade-off, but we urge users to reconsider the optimal WST configuration for each individual application. 

We finish this section with a brief interpretation of these results. As we already pointed out in Ref.~\citep{PhysRevD.105.103534} and also explained in \S\ref{sec:WST}, the higher-order nature of the WST  allows it to capture non-Gaussian information and as a result to improve upon the performance of the standard power spectrum. In addition, raising the modulus to powers $q<1$ in \eqref{eq:WSTcoeff:sol} allows it to effectively upweight the significance of cosmic voids (lower over-density regions), which are known to be valuable probes of fundamental information \citep{Pisani:2019cvo} related to e.g. the nature of massive neutrinos or theories of gravity. In Ref.~\citep{PhysRevD.105.103534} we showed that by realizing these properties in a novel way the WST was able to significantly improve upon the standard constraints to the neutrino mass, with parallels to the performance of the marked power spectrum \citep{PhysRevLett.126.011301}. Given that many such techniques were originally proposed \citep{White_2016} and subsequently explored \citep{PhysRevD.97.023535,Alam:2020jdv,2018MNRAS.479.4824H,2018MNRAS.478.3627A,2024MNRAS.528.6631A} in the context of testing MG models with a density-dependent screening mechanism, the significant improvement on the MG constraints by the WST is thus not a surprise. In fact, in light of these results we would expect other higher-order statistics of this kind to also be able to improve upon the standard MG constraints, which is something that we are planning to explore with these simulations in the future. We however note, at this point, that our reported results are very optimistic as we have worked with the 3D matter overdensity field, which is not observable by realistic galaxy surveys. Marginalizing over the uncertainties introduced by galaxy physics and also accounting for survey systematics are expected to substantially weaken these constraints, tasks that we are going to undertake in future work. Such a forecast was, for example, performed in Ref.~\citep{Liu:2021weo}, working however only with quasi-linear scales using perturbation theory. Nevertheless, this first proof-of-concept application paves the way for the application of the WST technique in order to obtain competitive constraints on extended theories of gravity and other non-standard scenarios with future LSS data, going beyond \citep{PhysRevD.109.103503}. For a recent similar first exploration in the context of primordial non-Gaussianity, see Ref.~\citep{Peron:2024xaw}. 

\section{Conclusions}\label{sec:Conclusions}
In this work we move beyond previous $\Lambda$CDM applications \citep{PhysRevD.105.103534,PhysRevD.106.103509,PhysRevD.109.103503} and perform the first exploration in the direction of applying the Wavelet Scattering Transform in order to constrain the nature of gravity using the 3-dimensional Large-Scale Structure of the universe. 

Using the new suite of the \textsc{Quijote-MG} N-body simulations \citep{QujiMG}, in particular, we are able to reliably predict structure formation for the popular Hu-Sawicki screened MG scenario down to the nonlinear regime. In combination with the preceding \textsc{Quijote} $\nu$CDM collection, we then proceed to perform a simulation-based Fisher forecast of the marginalized constraints to gravity obtained using the WST coefficients and the matter power spectrum at redshift $z=0$. The WST statistic is found to deliver an impressive improvement in the marginalized 1$\sigma$ error obtained on the parameter characterizing the deviations from standard GR, which is tighter by a factor of $\sim 10$ compared to the corresponding prediction of the regular matter power spectrum. At the same time it also substantially improves upon the power spectrum constraints obtained on the rest of the $\Lambda$CDM parameters and the sum of the neutrino masses, by $1.2-3.3\times$ and $3.4\times$, respectively, in full agreement with our previous work \citep{PhysRevD.105.103534}. Even though the WST constraints and their relative improvement over the power spectrum inevitably degrade as we progressively restrict our focus on larger scales, we find that the WST is still able to deliver an error that is relatively $\sim 4.5 \times$ tighter for $k_{\rm max}=0.2$ h/Mpc, highlighting its great sensitivity to the nature of the underlying gravity model. We carefully check and confirm the robustness of our Fisher analysis and then proceed to briefly explain how the inherent properties exhibited by the WST estimator allow it to be particularly informative as far as cosmological tests of gravity are concerned.

Moving forward, a series of improvements will be necessary before the above result can actually be exploited using real observational data, following the previous successful WST applications in the context of $\Lambda$CDM \citep{PhysRevD.106.103509,PhysRevD.109.103503}. First of all, we only worked with the underlying matter density field, which is not observable by surveys of the 3D LSS. In follow-up studies we will accordingly revisit this analysis using biased tracers, simulated dark matter halos and realistic galaxy mocks obtained for the same MG model. Furthermore, the effect of anisotropic redshift-space distortions (RSD) of spectroscopic galaxy observations along the line-of-sight will need to be accounted for in our model, together with additional systematics related to a specific galaxy survey. Since RSD is known to be particularly sensitive to the underlying gravity theory, adopting anisotropic wavelets \citep{2021arXiv210411244S,2023arXiv231015250R} can form a powerful basis for future tests of gravity with spectroscopic data. Marginalizing over these effects is of course expected to lead to more realistic constraints over the optimistic ones reported in this dark matter-only application. Given the very promising findings from this first exploration in the context of gravity theories beyond GR, it would also be very interesting to explore the extent to which suitably tailored WST bases could be employed in order to test for other types of MG screening mechanisms, such as, for example, the Vainshtein mechanism or more general classes encompassing the broad Horndeski family \citep{2008PhRvD..77d3524O,2010PhRvL.104w1301H,2009IJMPD..18.2147B,2011JHEP...08..108D,1972PhLB...39..393V,Horndeski:1974wa}. The ever increasing interest in developing efficient and accurate simulation methods for more general theories of gravity \citep{Wright:2022krq,Gupta:2024seu} will naturally enable such simulation-based endeavors. Parallel to all these improvements on the modeling front, lastly, in future work we plan to go beyond the limits imposed by the Fisher method and revisit this analysis using likelihood-free inference.

This first proof-of-concept study reaffirms the constraining properties of the WST technique and lays the foundation for exciting future applications in order to perform precise large-scale tests of gravity with the new generation of cutting-edge cosmological data.

\begin{acknowledgments}
We would like to thank Sihao Cheng, Cora Dvorkin, Joshua Frieman, Wayne Hu, Austin Joyce and Chris Wilson for useful discussions over the course of this work.

GV acknowledges the support of the Eric and Wendy Schmidt AI in Science Postdoctoral Fellowship at the University of Chicago, a Schmidt Sciences program.
\end{acknowledgments}

\appendix

\section{Gaussianity of the Likelihood}\label{app:Gauss}
The Fisher matrix formalism \eqref{Fisher} that we employ for the forecasts presented in this work relies upon the assumption that the summary statistic of interest follows a Gaussian probability distribution. This approximation is accurate when the binning used to evaluate the statistic is wide enough that a sufficient number of modes contributes to its evaluation at each bin, such that the Central Limit Theorem kicks in and the observable becomes Gaussianized. This is a well-known fact in the case of the power spectrum, that needs to also be confirmed before performing likelihood analyses using alternative estimators. Following our previous application \citep{PhysRevD.109.103503}, we test and confirm this to be the case for the WST coefficients using the 15,000 \textsc{Quijote} realizations available for the fiducial cosmology, as explained in \S\ref{sec:Simulations}. Specifically, if $\bold{O}_i$ is the estimator value evaluated at the $i^{\rm th}$ realization, $\bar{\bold{O}}$ the mean value over all realizations and $C$ its covariance matrix from Eq. \eqref{eq:covmat}, then the 15,000 realizations will follow a $\chi^2$ distribution, given by:
\begin{equation}\label{chisq}
\chi^2_i = \left[\bold{O}_i-\bar{\bold{O}}\right]^{\rm T} C^{-1}\left[\bold{O}_i-\bar{\bold{O}}\right].
\end{equation}

\begin{figure}[ht!]
\centering 
\includegraphics[width=0.49\textwidth]{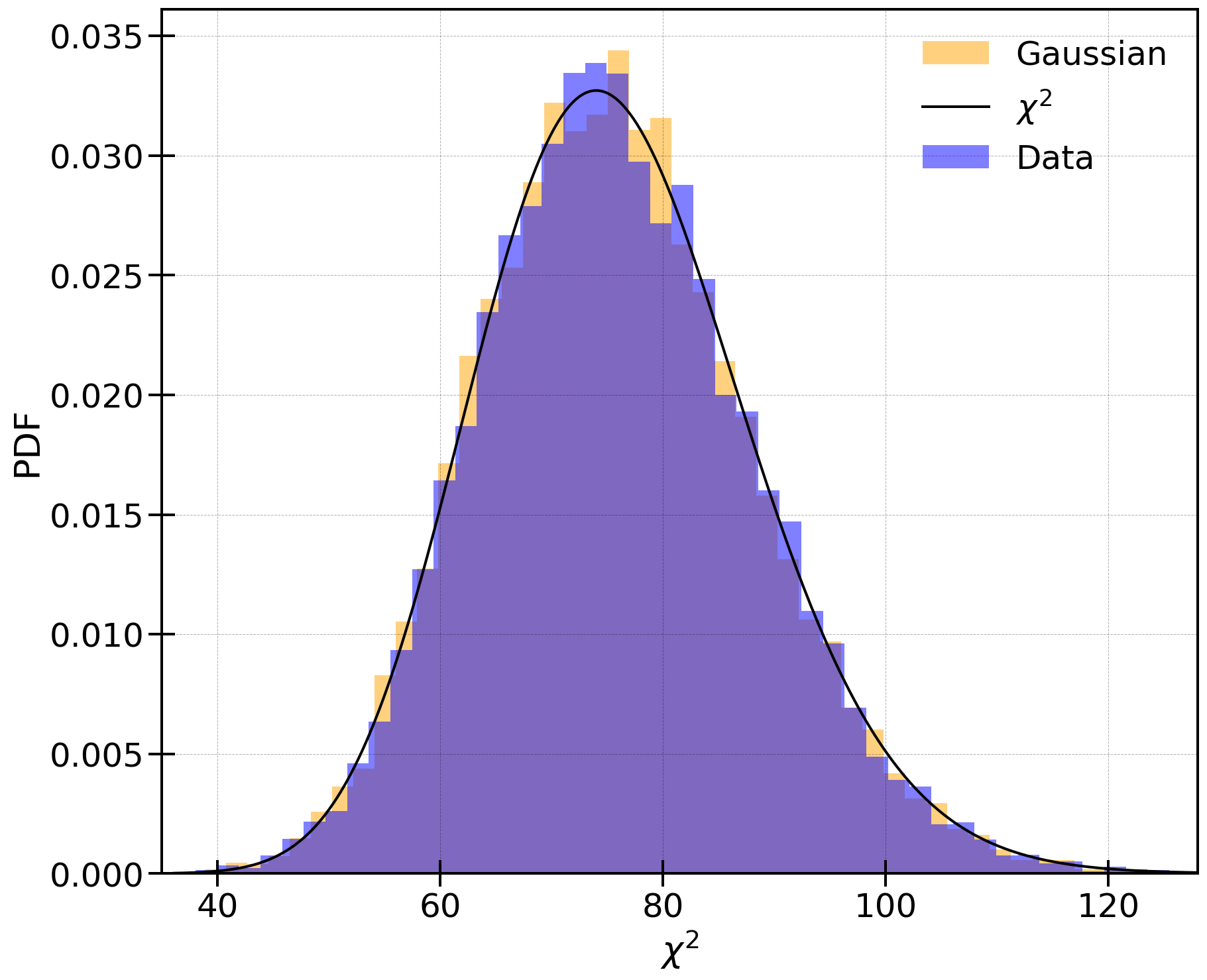}
\caption{The probability density function (PDF) of the $\chi^2$ distribution of the WST coefficients as obtained from the 15,000 realizations of the \textsc{Quijote} simulations at $z=0$ (blue) is plotted alongside a theoretical $\chi^2$ distribution with $N_{\bold{d}}=76$ degrees of freedom (black line) and samples drawn from a Gaussian distribution with the same mean and covariance (orange). The WST estimator does not exhibit any significant deviations from a Gaussian distribution.}
\label{Fig:Gausstest}
\end{figure}
\begin{figure}[ht!]
\centering 
\includegraphics[width=0.49\textwidth]{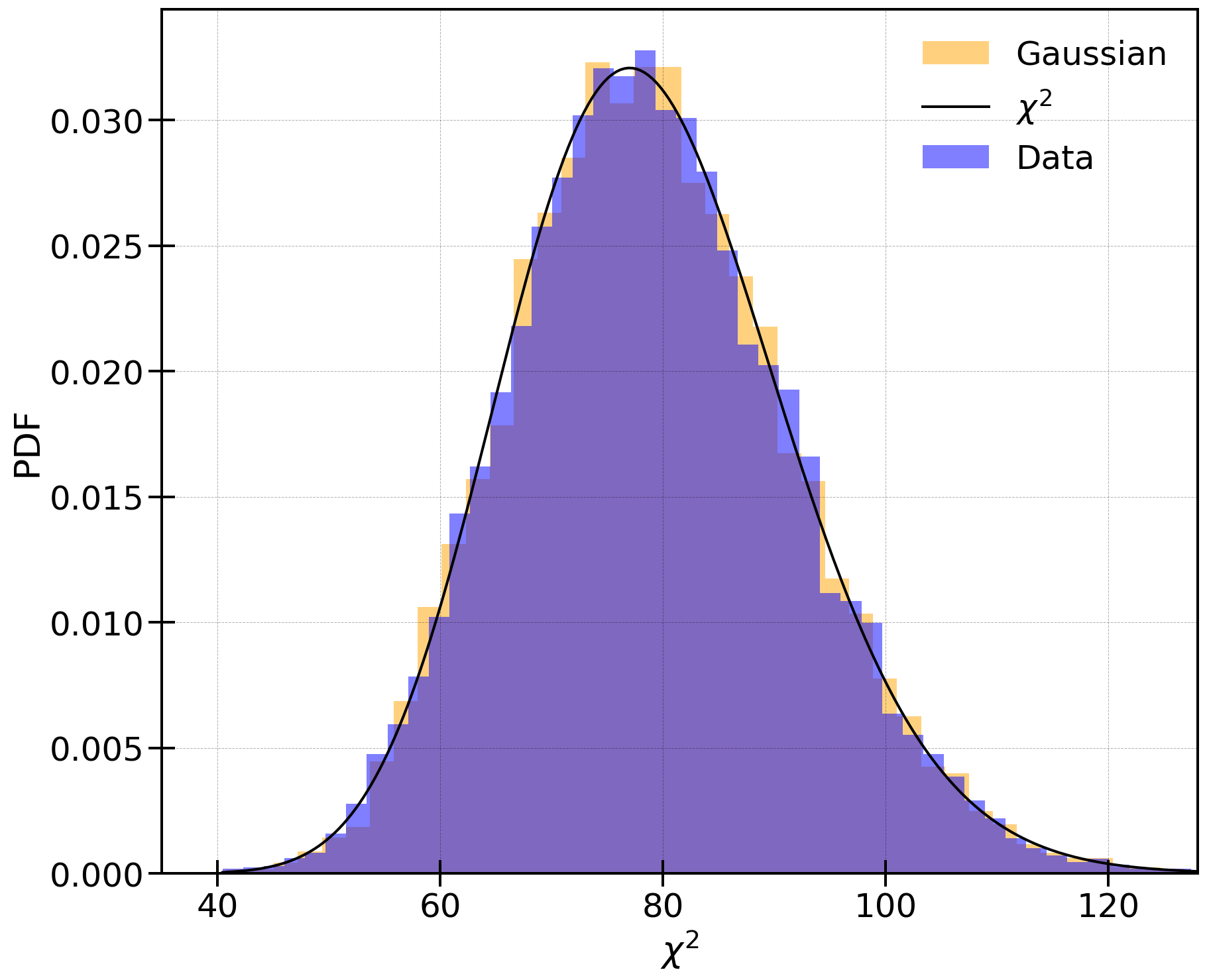}
\caption{The same $\chi^2$ analysis as in Fig. \ref{Fig:Gausstest} is repeated for the matter power spectrum monopole}
\label{Fig:Gausstestxi}
\end{figure}
For a summary statistic that actually follows a Gaussian distribution, the probability density function (pdf) of the $\chi^2$ values given by Eq. \eqref{chisq} will match the theoretical prediction for a distribution with a number of degrees of freedom equal to the dimensionality of the data vector. In Fig.~\ref{Fig:Gausstest} we plot this comparison for the WST coefficients, with $N_{\bold{d}}=76$ degrees of freedom for our base configuration, finding no apparent discrepancy between the two predictions and also a distribution of samples randomly drawn from a Gaussian with the same mean and covariance. Performing the equivalent comparison for the standard power spectrum, in Fig.~\ref{Fig:Gausstestxi}, reaffirms the known Gaussianity of the power spectrum and highlights the very similar levels of consistency between the two summary statistics considered in this application. These findings are in line with our previous WST application to the galaxy density field \citep{PhysRevD.109.103503}, but also to weak lensing maps \citep{10.1093/mnras/stab2102}, as well as with similar findings when considering other alternative summary statistics in the literature \citep{2021MNRAS.508.3125F,2022arXiv220904310P,2023Yuan}. For further discussion on how to handle cases that significantly deviate from Gaussianity in the context of Fisher forecasting or simulation-based inference, we refer readers to \cite{Park:2022hzj} or \citep{2023arXiv231015250R}, respectively.

\begin{figure}[ht]
\includegraphics[width=0.49\textwidth]{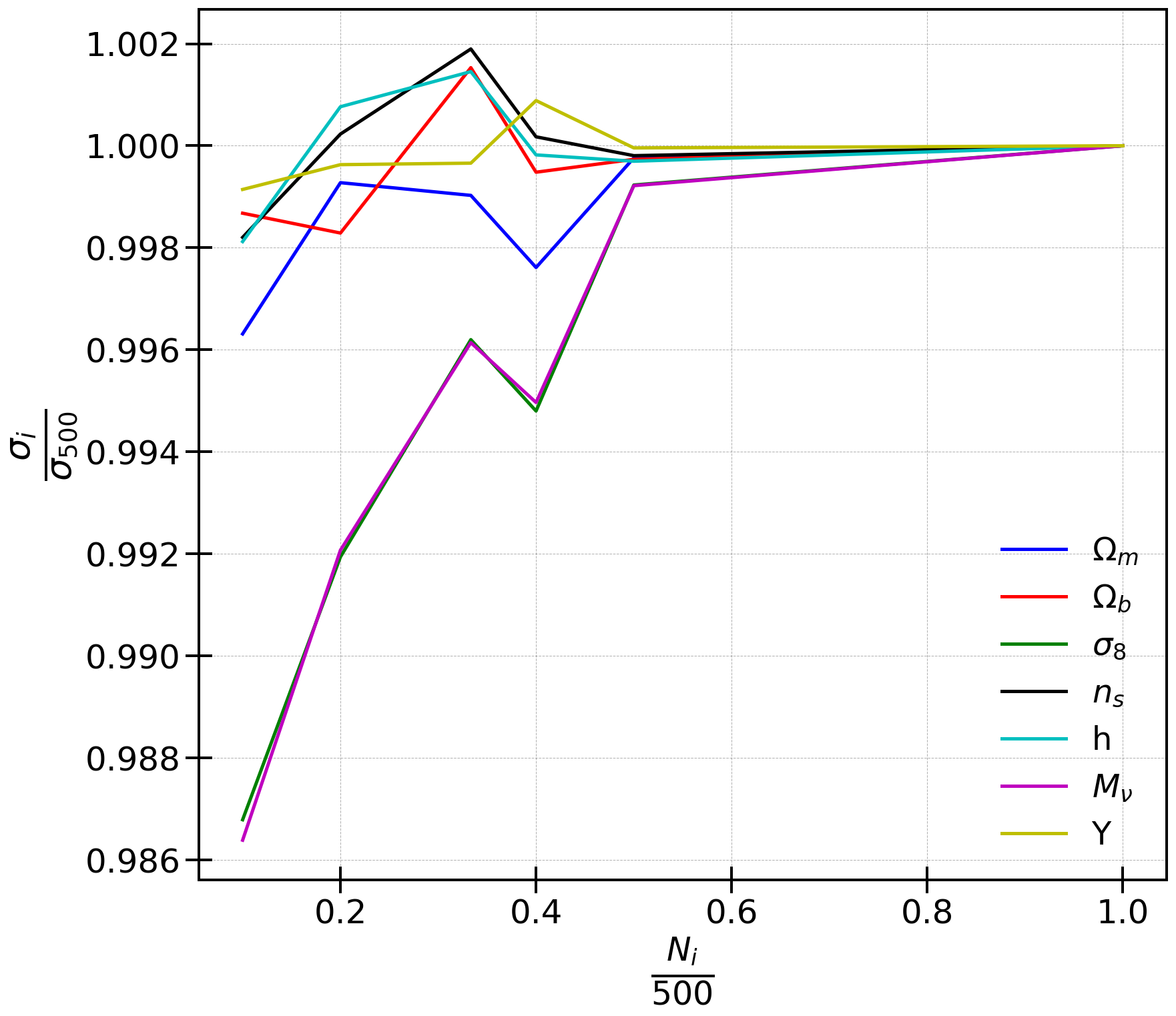}
\caption{\label{fig:WSTstability} Numerical convergence analysis of the 1$\sigma$ errors obtained on the cosmological parameters by the WST as a function of the 500 realizations used to evaluate the numerical derivatives. This example corresponds to the WST base configuration defined in \S\ref{sec:Fisher} and used in Fig.~\ref{fig:epsart2}.}
\end{figure}

\begin{figure}[ht]
\includegraphics[width=0.49\textwidth]{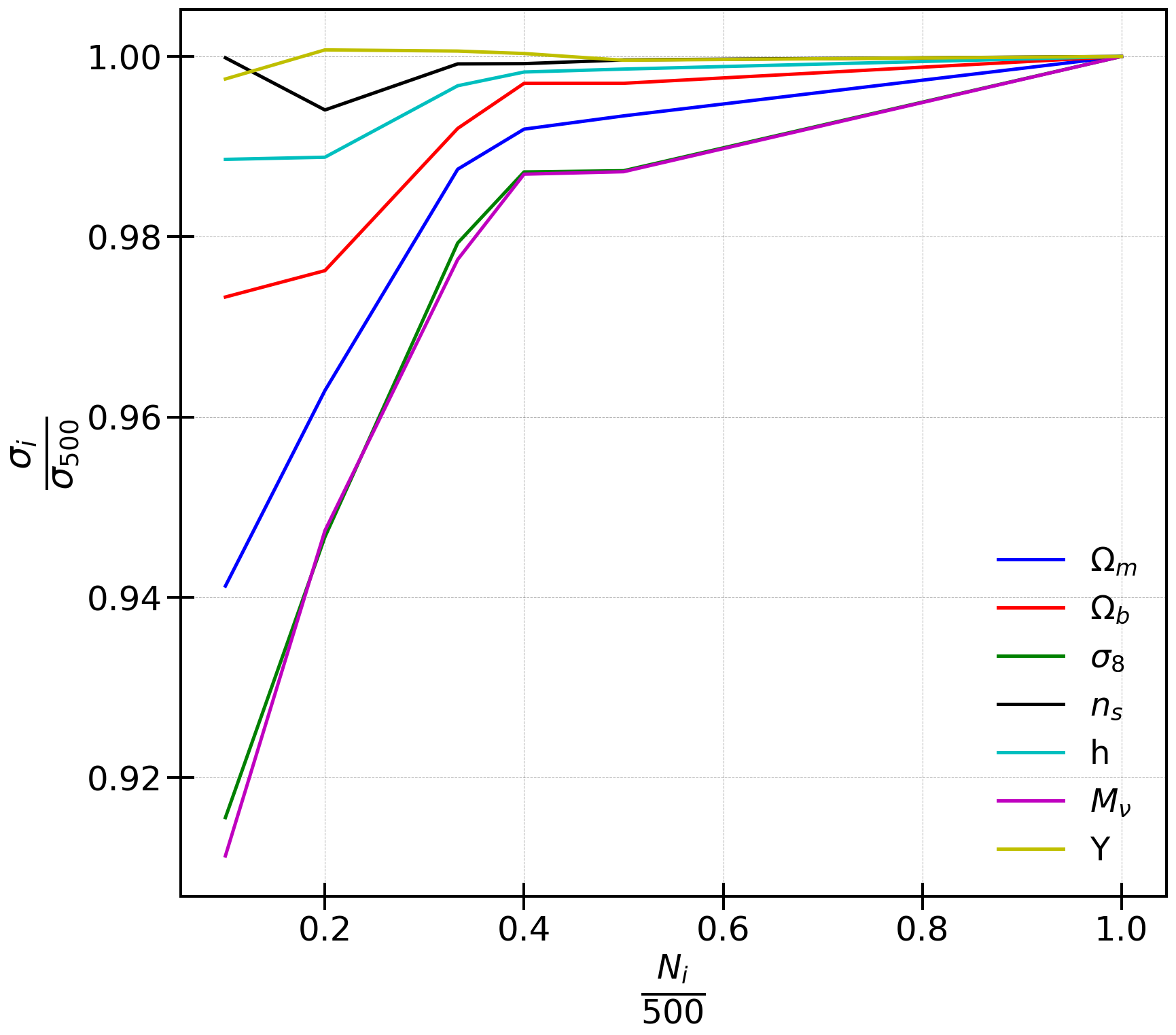}
\caption{\label{fig:Pkstability} Same analysis as in Fig.~\ref{fig:WSTstability} but here shown for the matter power spectrum.}
\end{figure}

\section{Numerical stability analysis}\label{app:Stability}
Even though the Fisher matrix technique is a valuable tool in a variety of cosmological applications, it can lead to biased constraints when using simulations to evaluate its ingredients. In particular, when the number of available random realizations is not sufficiently large to provide a smooth, noise-free, prediction, the residual Monte Carlo noise in the observable data vector (particularly the derivatives) can lead to an overestimation of the Fisher matrix and, in turn, the prediction of artificially tight 1$\sigma$ errors after its inversion \citep{Coulton:2022rir,Coulton:2023sfu, Wilson:2024nhm}. In this appendix we explicitly make sure that this is not the case in our forecasts. Specifically, in Fig.~\ref{fig:WSTstability} we investigate how the 1$\sigma$ errors obtained from the WST Fisher matrix in the base configuration vary as a function of the 500 random realizations that are available for the evaluation of the numerical derivatives, finding a remarkable level of numerical stability for all the parameters, including the new modified gravity case; the predictions are converged at a level better than $1\%$ even when using only $20\%$ of the 500 total realizations. When using half of the available realizations, the predictions do not fluctuate by more than $0.2\%$ compared to the ones from the full set, demonstrating the prominent numerical stability of the WST  estimator. It is also worth noting that it seems to converge even faster than the standard power spectrum case, which is shown in Fig.~\ref{fig:Pkstability}, and which is converged at the $\sim 1\%$ level when using half of the available realizations for the numerical evaluation of the derivatives. These findings are in line with our previous Fisher applications using the WST \citep{PhysRevD.105.103534,PhysRevD.106.103509}. We also note that we have double-checked the numerical stability of our results using the code by Ref.~\citep{Coulton:2023sfu}\footnote{\url{https://github.com/wcoulton/CompressedFisher/tree/main}}, finding no detected biases. We caution, nevertheless, that these conclusions hold for this particular WST configuration applied to the matter field, only. Using the WST coefficients evaluated at different scales and angles, and also when raising the modulus of the field in different powers $q$ in Eq. \eqref{eq:WSTcoeff:sol}, can lead to slower convergence. Similarly, the halo and galaxy fields tend to be inherently noisier than the matter case, leading to significant challenges when attempting to perform simulation-based forecasts \citep{Wilson:2024nhm}. In such cases, compression techniques can be employed to reduce the noise and obtain unbiased estimates \citep{Coulton:2023sfu}. In fact, we note that we did repeat our analysis using these compressed variants as well, finding entirely consistent results with the standard forecast, given the high degree of convergence already reported above. For a more thorough investigation into the impact of numerical noise on simulation-based Fisher forecasts, we refer readers to \citep{Wilson:2024nhm}.

 Lastly, we similarly confirmed that the 15,000 realizations available at the fiducial cosmology are more than sufficient for the evaluation of a well-converged covariance matrix. Even if we use half of them (7,500) to evaluate the covariance matrix from Eq. \eqref{eq:covmat}, the Fisher predictions for the WST do not change by more than 0.5 $\%$ for any of the parameters. We choose not to explicitly include these plots here as well, for brevity, since this is a result already discussed in \citep{PhysRevD.105.103534}, which used the exact same set of fiducial simulations for the evaluation of the covariance matrix.

 \begin{figure*}[ht]
\includegraphics[width=0.99\textwidth]{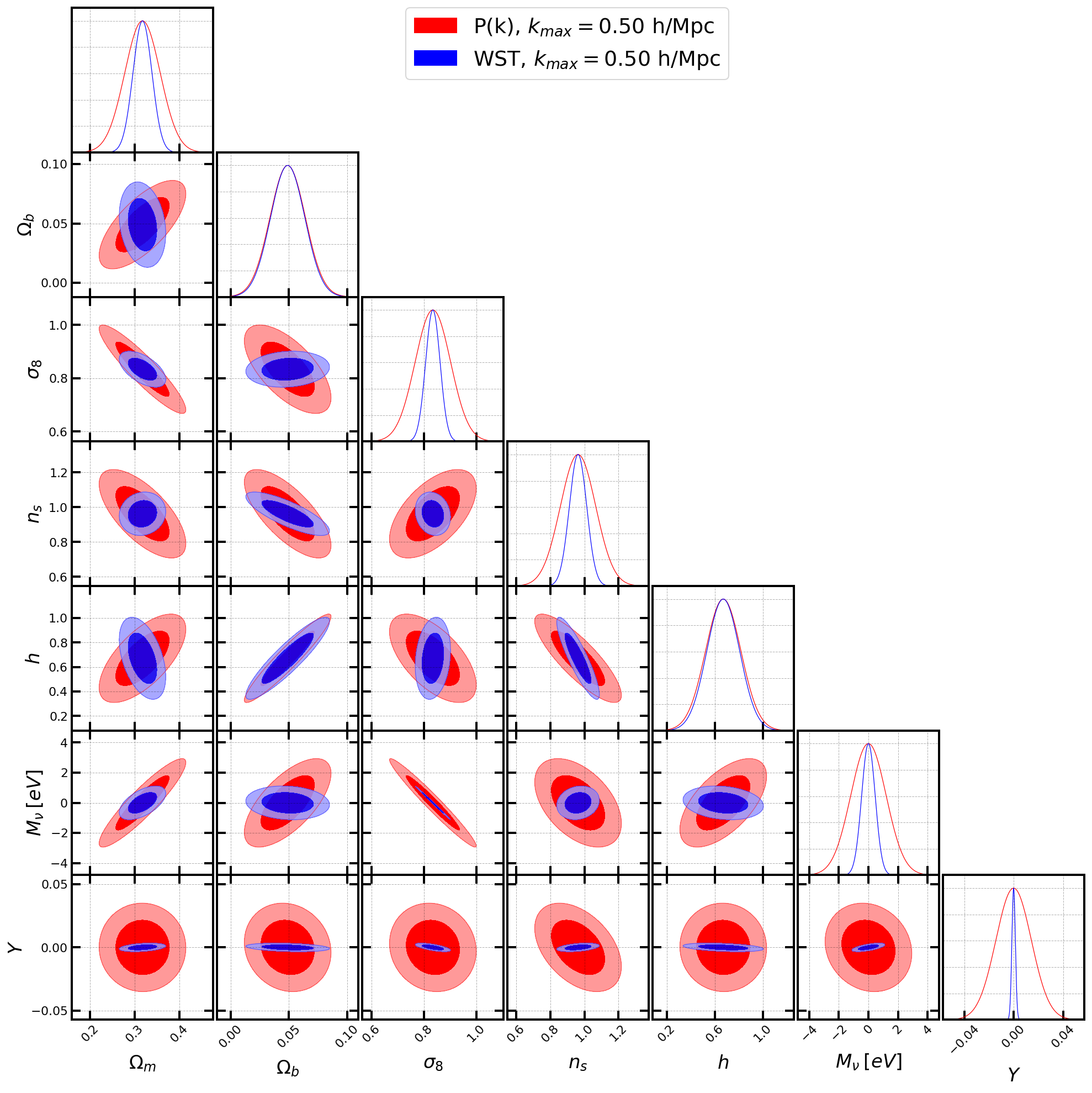}
\caption{\label{fig:derfrp} Same as in Fig. \ref{fig:epsart2} but using Eq. \eqref{app:fRder4} instead of \eqref{fRder} for the evaluation of the derivative w.r.t. the MG parameter Y for both summary statistics.}
\end{figure*}
 
\section{Dependence on derivative formula}\label{app:Der}
When multiple steps are available for the evaluation of the numerical derivatives, as is the case for the MG parameter Y (seen in \eqref{eq:Yarrsteps}) (but also for the neutrino mass $M_{\nu}$), we have more than one options to evaluate the derivatives using a forward stepping scheme. Starting with the smallest step, $Y_p$, for example, we can get the following versions:
\begin{widetext}
\begin{align} 
\frac{\partial O}{\partial Y} &= \frac{O(Y_{\rm p})-O(0)}{Y_{\rm p}}, \label{app:fRder1}\\ 
\frac{\partial O}{\partial Y} &= \frac{-O(Y_{\rm pp})+4 O(Y_{\rm p})-3 O(0)}{2 Y_{\rm p}}, \label{app:fRder2}\\ 
\frac{\partial O}{\partial Y} &= \frac{O(Y_{\rm ppp})-12 O(Y_{\rm pp})+32 O(Y_{\rm p})-21 O(0)}{12 Y_{\rm p}}, \label{app:fRder3}\\ 
\frac{\partial O}{\partial Y} &= \frac{-O(Y_{\rm pppp})+28 O(Y_{\rm ppp})-224 O(Y_{\rm pp})+512 O(Y_{\rm p})-315 O(0)}{168 Y_{\rm p}} \label{app:fRder4} 
\end{align}
\end{widetext}
in increasing order of evaluation. If the difference between a given summary statistic evaluated at the fiducial cosmology (i.e. at $Y=0$) and at $Y_p$ is large enough such that it is not masked by numerical noise fluctuations, then Eq. \eqref{app:fRder4} above will in principle provide the most precise estimation of the derivative. In our case, however, and as we explained in the main text, the smallest step $Y_p$ corresponds to a highly-screened instance of the Hu-Sawicki model that typically gives very small deviations from $\Lambda$CDM. If we look at the power spectrum case in Fig. \ref{fig:epsart}, for example, we find that the fractional deviation from the $\Lambda$CDM prediction never reaches $1 \%$ for $Y_p$, even at the smallest scales we consider, $k_{\rm max}=0.5$ h/Mpc. For $k<0.1$ h/Mpc, the differences between the two predictions are so small that they are inevitably masked by the numerical noise. In order to prevent this fact from contaminating our predictions, we chose to skip $Y_p$ and only work with the other 3 available larger steps, $Y_{\rm pp},Y_{\rm ppp},Y_{\rm pppp}$, as we explained in \S\ref{sec:Fisher} of the main text. With these 3 steps at hand, Eq. \eqref{app:fRder4} is no longer possible so we then use the next most precise version, Eq. \eqref{app:fRder3} for $Y_{\rm pp}=Y_p$ and so on, which coincides with  Eq. \eqref{fRder}.

Even though we adopted this above choice out of an abundance of caution, we emphasize at this point that the conclusions of our analysis are robust against the derivative version. In fact, the same levels of improvement are found even if we use Eq. \eqref{app:fRder4} (or any of its lower order variants with $Y_p$). This can be seen in Fig. \ref{fig:derfrp}, where we repeat the comparison of Fig.~\ref{fig:epsart2} for the two estimators, but this time using Eq. \eqref{app:fRder4} instead for the evaluation of the MG derivative. The WST is once again found to improve upon the power spectrum 1$\sigma$ error prediction for Y by a factor of $\sim 10 \times$, just like in the main analysis portrayed in Fig.~\ref{fig:epsart2}. As a matter of fact, we also found this to hold true when experimenting with forecasts using all the other possible variations of Eqs. \eqref{app:fRder1}-\eqref{app:fRder4}. Even though the individual predictions for the 1-$\sigma$ error on Y may fluctuate a bit when using different versions of the derivative, the trend is similar for both the power spectrum and the WST predictions, such that the relative improvement is always robust against the particular version of the derivative. Lastly, the different predictions also exhibit similar high levels of numerical stability as the one presented in Fig. \ref{fig:WSTstability}, even when using the smallest step $Y_p$. These results mirror our previous findings w.r.t. the neutrino derivative and its variants \citep{PhysRevD.105.103534}.

As a result of all the above, we conclude that our main conclusions wrt to the improvement delivered by the WST relative to the power spectrum are robust against the derivative choice for the MG parameter Y. Nevertheless, we caution users to be aware of these nuances and make use of the most appropriate combination of Eqs. \eqref{app:fRder1}-\eqref{app:fRder4} according to the particular application. For further discussion see Ref.~\citep{Wilson:2024nhm}.

\newpage
\bibliographystyle{apsrev4-2_16.bst}

\begin{thebibliography}{155}%
\makeatletter
\providecommand \@ifxundefined [1]{%
 \@ifx{#1\undefined}
}%
\providecommand \@ifnum [1]{%
 \ifnum #1\expandafter \@firstoftwo
 \else \expandafter \@secondoftwo
 \fi
}%
\providecommand \@ifx [1]{%
 \ifx #1\expandafter \@firstoftwo
 \else \expandafter \@secondoftwo
 \fi
}%
\providecommand \natexlab [1]{#1}%
\providecommand \enquote  [1]{``#1''}%
\providecommand \bibnamefont  [1]{#1}%
\providecommand \bibfnamefont [1]{#1}%
\providecommand \citenamefont [1]{#1}%
\providecommand \href@noop [0]{\@secondoftwo}%
\providecommand \href [0]{\begingroup \@sanitize@url \@href}%
\providecommand \@href[1]{\@@startlink{#1}\@@href}%
\providecommand \@@href[1]{\endgroup#1\@@endlink}%
\providecommand \@sanitize@url [0]{\catcode `\\12\catcode `\$12\catcode
  `\&12\catcode `\#12\catcode `\^12\catcode `\_12\catcode `\%12\relax}%
\providecommand \@@startlink[1]{}%
\providecommand \@@endlink[0]{}%
\providecommand \url  [0]{\begingroup\@sanitize@url \@url }%
\providecommand \@url [1]{\endgroup\@href {#1}{\urlprefix }}%
\providecommand \urlprefix  [0]{URL }%
\providecommand \Eprint [0]{\href }%
\providecommand \doibase [0]{https://doi.org/}%
\providecommand \selectlanguage [0]{\@gobble}%
\providecommand \bibinfo  [0]{\@secondoftwo}%
\providecommand \bibfield  [0]{\@secondoftwo}%
\providecommand \translation [1]{[#1]}%
\providecommand \BibitemOpen [0]{}%
\providecommand \bibitemStop [0]{}%
\providecommand \bibitemNoStop [0]{.\EOS\space}%
\providecommand \EOS [0]{\spacefactor3000\relax}%
\providecommand \BibitemShut  [1]{\csname bibitem#1\endcsname}%
\let\auto@bib@innerbib\@empty
\bibitem [{\citenamefont {{DESI Collaboration}}\ \emph
  {et~al.}(2024{\natexlab{a}})\citenamefont {{DESI Collaboration}, {Adame},
  {Aguilar}, {Ahlen}, {Alam}, {Alexander}, {Alvarez}, {Alves}, {Anand},
  {Andrade} et~al.}}]{2024arXiv240403000D}%
  \BibitemOpen
  \bibfield  {author} {\bibinfo {author} {\bibnamefont {{DESI Collaboration}}},
  \bibinfo {author} {\bibfnamefont {A.~G.}\ \bibnamefont {{Adame}}}, \bibinfo
  {author} {\bibfnamefont {J.}~\bibnamefont {{Aguilar}}}, \bibinfo {author}
  {\bibfnamefont {S.}~\bibnamefont {{Ahlen}}}, \bibinfo {author} {\bibfnamefont
  {S.}~\bibnamefont {{Alam}}}, \bibinfo {author} {\bibfnamefont {D.~M.}\
  \bibnamefont {{Alexander}}}, \bibinfo {author} {\bibfnamefont
  {M.}~\bibnamefont {{Alvarez}}}, \bibinfo {author} {\bibfnamefont
  {O.}~\bibnamefont {{Alves}}}, \bibinfo {author} {\bibfnamefont
  {A.}~\bibnamefont {{Anand}}}, \bibinfo {author} {\bibfnamefont
  {U.}~\bibnamefont {{Andrade}}}, \bibinfo {author} {\bibfnamefont
  {E.}~\bibnamefont {{Armengaud}}}, \bibinfo {author} {\bibfnamefont
  {S.}~\bibnamefont {{Avila}}}, \bibinfo {author} {\bibfnamefont
  {A.}~\bibnamefont {{Aviles}}}, \bibinfo {author} {\bibfnamefont
  {H.}~\bibnamefont {{Awan}}}, \bibinfo {author} {\bibfnamefont
  {S.}~\bibnamefont {{Bailey}}}, \bibinfo {author} {\bibfnamefont
  {C.}~\bibnamefont {{Baltay}}}, \bibnamefont {et~al.},\ }\bibfield  {title}
  {\bibinfo {title} {{DESI 2024 III: Baryon Acoustic Oscillations from Galaxies
  and Quasars}},\ }\href {https://doi.org/10.48550/arXiv.2404.03000} {\bibfield
   {journal} {\bibinfo  {journal} {arXiv e-prints}\ ,\ \bibinfo {eid}
  {arXiv:2404.03000}} (\bibinfo {year} {2024}{\natexlab{a}})},\ \Eprint
  {https://arxiv.org/abs/2404.03000} {arXiv:2404.03000 [astro-ph.CO]}
  \BibitemShut {NoStop}%
\bibitem [{\citenamefont {{DESI Collaboration}}\ \emph
  {et~al.}(2024{\natexlab{b}})\citenamefont {{DESI Collaboration}, {Adame},
  {Aguilar}, {Ahlen}, {Alam}, {Alexander}, {Alvarez}, {Alves}, {Anand},
  {Andrade} et~al.}}]{2024arXiv240403001D}%
  \BibitemOpen
  \bibfield  {author} {\bibinfo {author} {\bibnamefont {{DESI Collaboration}}},
  \bibinfo {author} {\bibfnamefont {A.~G.}\ \bibnamefont {{Adame}}}, \bibinfo
  {author} {\bibfnamefont {J.}~\bibnamefont {{Aguilar}}}, \bibinfo {author}
  {\bibfnamefont {S.}~\bibnamefont {{Ahlen}}}, \bibinfo {author} {\bibfnamefont
  {S.}~\bibnamefont {{Alam}}}, \bibinfo {author} {\bibfnamefont {D.~M.}\
  \bibnamefont {{Alexander}}}, \bibinfo {author} {\bibfnamefont
  {M.}~\bibnamefont {{Alvarez}}}, \bibinfo {author} {\bibfnamefont
  {O.}~\bibnamefont {{Alves}}}, \bibinfo {author} {\bibfnamefont
  {A.}~\bibnamefont {{Anand}}}, \bibinfo {author} {\bibfnamefont
  {U.}~\bibnamefont {{Andrade}}}, \bibinfo {author} {\bibfnamefont
  {E.}~\bibnamefont {{Armengaud}}}, \bibinfo {author} {\bibfnamefont
  {S.}~\bibnamefont {{Avila}}}, \bibinfo {author} {\bibfnamefont
  {A.}~\bibnamefont {{Aviles}}}, \bibinfo {author} {\bibfnamefont
  {H.}~\bibnamefont {{Awan}}}, \bibinfo {author} {\bibfnamefont
  {S.}~\bibnamefont {{Bailey}}}, \bibinfo {author} {\bibfnamefont
  {C.}~\bibnamefont {{Baltay}}}, \bibnamefont {et~al.},\ }\bibfield  {title}
  {\bibinfo {title} {{DESI 2024 IV: Baryon Acoustic Oscillations from the Lyman
  Alpha Forest}},\ }\href {https://doi.org/10.48550/arXiv.2404.03001}
  {\bibfield  {journal} {\bibinfo  {journal} {arXiv e-prints}\ ,\ \bibinfo
  {eid} {arXiv:2404.03001}} (\bibinfo {year} {2024}{\natexlab{b}})},\ \Eprint
  {https://arxiv.org/abs/2404.03001} {arXiv:2404.03001 [astro-ph.CO]}
  \BibitemShut {NoStop}%
\bibitem [{\citenamefont {{DESI Collaboration}}\ \emph
  {et~al.}(2024{\natexlab{c}})\citenamefont {{DESI Collaboration}, {Adame},
  {Aguilar}, {Ahlen}, {Alam}, {Alexander}, {Alvarez}, {Alves}, {Anand},
  {Andrade} et~al.}}]{2024arXiv240403002D}%
  \BibitemOpen
  \bibfield  {author} {\bibinfo {author} {\bibnamefont {{DESI Collaboration}}},
  \bibinfo {author} {\bibfnamefont {A.~G.}\ \bibnamefont {{Adame}}}, \bibinfo
  {author} {\bibfnamefont {J.}~\bibnamefont {{Aguilar}}}, \bibinfo {author}
  {\bibfnamefont {S.}~\bibnamefont {{Ahlen}}}, \bibinfo {author} {\bibfnamefont
  {S.}~\bibnamefont {{Alam}}}, \bibinfo {author} {\bibfnamefont {D.~M.}\
  \bibnamefont {{Alexander}}}, \bibinfo {author} {\bibfnamefont
  {M.}~\bibnamefont {{Alvarez}}}, \bibinfo {author} {\bibfnamefont
  {O.}~\bibnamefont {{Alves}}}, \bibinfo {author} {\bibfnamefont
  {A.}~\bibnamefont {{Anand}}}, \bibinfo {author} {\bibfnamefont
  {U.}~\bibnamefont {{Andrade}}}, \bibinfo {author} {\bibfnamefont
  {E.}~\bibnamefont {{Armengaud}}}, \bibinfo {author} {\bibfnamefont
  {S.}~\bibnamefont {{Avila}}}, \bibinfo {author} {\bibfnamefont
  {A.}~\bibnamefont {{Aviles}}}, \bibinfo {author} {\bibfnamefont
  {H.}~\bibnamefont {{Awan}}}, \bibinfo {author} {\bibfnamefont
  {B.}~\bibnamefont {{Bahr-Kalus}}}, \bibinfo {author} {\bibfnamefont
  {S.}~\bibnamefont {{Bailey}}}, \bibnamefont {et~al.},\ }\bibfield  {title}
  {\bibinfo {title} {{DESI 2024 VI: Cosmological Constraints from the
  Measurements of Baryon Acoustic Oscillations}},\ }\href
  {https://doi.org/10.48550/arXiv.2404.03002} {\bibfield  {journal} {\bibinfo
  {journal} {arXiv e-prints}\ ,\ \bibinfo {eid} {arXiv:2404.03002}} (\bibinfo
  {year} {2024}{\natexlab{c}})},\ \Eprint {https://arxiv.org/abs/2404.03002}
  {arXiv:2404.03002 [astro-ph.CO]} \BibitemShut {NoStop}%
\bibitem [{\citenamefont {Levi}\ \emph {et~al.}(2013)\citenamefont {Levi
  et~al.}}]{Levi:2013gra}%
  \BibitemOpen
  \bibfield  {author} {\bibinfo {author} {\bibfnamefont {M.}~\bibnamefont
  {Levi}} \bibnamefont {et~al.} (\bibinfo {collaboration} {DESI
  collaboration}),\ }\bibfield  {title} {\bibinfo {title} {{The DESI
  Experiment, a whitepaper for Snowmass 2013}},\ }\href@noop {} {\  (\bibinfo
  {year} {2013})},\ \Eprint {https://arxiv.org/abs/1308.0847} {arXiv:1308.0847
  [astro-ph.CO]} \BibitemShut {NoStop}%
\bibitem [{\citenamefont {{DESI Collaboration}}\ \emph
  {et~al.}(2016)\citenamefont {{DESI Collaboration}, {Aghamousa}, {Aguilar},
  {Ahlen}, {Alam}, {Allen}, {Allende Prieto}, {Annis}, {Bailey}, {Balland}
  et~al.}}]{2016DESI}%
  \BibitemOpen
  \bibfield  {author} {\bibinfo {author} {\bibnamefont {{DESI Collaboration}}},
  \bibinfo {author} {\bibfnamefont {A.}~\bibnamefont {{Aghamousa}}}, \bibinfo
  {author} {\bibfnamefont {J.}~\bibnamefont {{Aguilar}}}, \bibinfo {author}
  {\bibfnamefont {S.}~\bibnamefont {{Ahlen}}}, \bibinfo {author} {\bibfnamefont
  {S.}~\bibnamefont {{Alam}}}, \bibinfo {author} {\bibfnamefont {L.~E.}\
  \bibnamefont {{Allen}}}, \bibinfo {author} {\bibfnamefont {C.}~\bibnamefont
  {{Allende Prieto}}}, \bibinfo {author} {\bibfnamefont {J.}~\bibnamefont
  {{Annis}}}, \bibinfo {author} {\bibfnamefont {S.}~\bibnamefont {{Bailey}}},
  \bibinfo {author} {\bibfnamefont {C.}~\bibnamefont {{Balland}}}, \bibinfo
  {author} {\bibfnamefont {O.}~\bibnamefont {{Ballester}}}, \bibinfo {author}
  {\bibfnamefont {C.}~\bibnamefont {{Baltay}}}, \bibinfo {author}
  {\bibfnamefont {L.}~\bibnamefont {{Beaufore}}}, \bibinfo {author}
  {\bibfnamefont {C.}~\bibnamefont {{Bebek}}}, \bibinfo {author} {\bibfnamefont
  {T.~C.}\ \bibnamefont {{Beers}}}, \bibinfo {author} {\bibfnamefont {E.~F.}\
  \bibnamefont {{Bell}}}, \bibnamefont {et~al.},\ }\bibfield  {title} {\bibinfo
  {title} {{The DESI Experiment Part I: Science,Targeting, and Survey
  Design}},\ }\href@noop {} {\bibfield  {journal} {\bibinfo  {journal} {arXiv
  e-prints}\ ,\ \bibinfo {eid} {arXiv:1611.00036}} (\bibinfo {year} {2016})},\
  \Eprint {https://arxiv.org/abs/1611.00036} {arXiv:1611.00036 [astro-ph.IM]}
  \BibitemShut {NoStop}%
\bibitem [{\citenamefont {Abell}\ \emph {et~al.}(2009)\citenamefont {Abell
  et~al.}}]{Abell:2009aa}%
  \BibitemOpen
  \bibfield  {author} {\bibinfo {author} {\bibfnamefont {P.~A.}\ \bibnamefont
  {Abell}} \bibnamefont {et~al.} (\bibinfo {collaboration} {LSST Science
  Collaborations, LSST Project}),\ }\bibfield  {title} {\bibinfo {title} {{LSST
  Science Book, Version 2.0}},\ }\href@noop {} {\  (\bibinfo {year} {2009})},\
  \Eprint {https://arxiv.org/abs/0912.0201} {arXiv:0912.0201 [astro-ph.IM]}
  \BibitemShut {NoStop}%
\bibitem [{\citenamefont {Abate}\ \emph {et~al.}(2012)\citenamefont {Abate
  et~al.}}]{Abate:2012za}%
  \BibitemOpen
  \bibfield  {author} {\bibinfo {author} {\bibfnamefont {A.}~\bibnamefont
  {Abate}} \bibnamefont {et~al.} (\bibinfo {collaboration} {LSST Dark Energy
  Science}),\ }\bibfield  {title} {\bibinfo {title} {{Large Synoptic Survey
  Telescope: Dark Energy Science Collaboration}},\ }\href@noop {} {\  (\bibinfo
  {year} {2012})},\ \Eprint {https://arxiv.org/abs/1211.0310} {arXiv:1211.0310
  [astro-ph.CO]} \BibitemShut {NoStop}%
\bibitem [{\citenamefont {Laureijs}\ \emph {et~al.}(2011)\citenamefont
  {Laureijs et~al.}}]{Laureijs:2011gra}%
  \BibitemOpen
  \bibfield  {author} {\bibinfo {author} {\bibfnamefont {R.}~\bibnamefont
  {Laureijs}} \bibnamefont {et~al.} (\bibinfo {collaboration} {EUCLID
  Collaboration}),\ }\bibfield  {title} {\bibinfo {title} {{Euclid Definition
  Study Report}},\ }\href@noop {} {\  (\bibinfo {year} {2011})},\ \Eprint
  {https://arxiv.org/abs/1110.3193} {arXiv:1110.3193 [astro-ph.CO]}
  \BibitemShut {NoStop}%
\bibitem [{\citenamefont {Spergel}\ \emph {et~al.}(2013)\citenamefont {Spergel,
  Gehrels, Breckinridge, Donahue, Dressler et~al.}}]{Spergel:2013tha}%
  \BibitemOpen
  \bibfield  {author} {\bibinfo {author} {\bibfnamefont {D.}~\bibnamefont
  {Spergel}}, \bibinfo {author} {\bibfnamefont {N.}~\bibnamefont {Gehrels}},
  \bibinfo {author} {\bibfnamefont {J.}~\bibnamefont {Breckinridge}}, \bibinfo
  {author} {\bibfnamefont {M.}~\bibnamefont {Donahue}}, \bibinfo {author}
  {\bibfnamefont {A.}~\bibnamefont {Dressler}}, \bibnamefont {et~al.},\
  }\bibfield  {title} {\bibinfo {title} {{Wide-Field InfraRed Survey
  Telescope-Astrophysics Focused Telescope Assets WFIRST-AFTA Final Report}},\
  }\href@noop {} {\  (\bibinfo {year} {2013})},\ \Eprint
  {https://arxiv.org/abs/1305.5422} {arXiv:1305.5422 [astro-ph.IM]}
  \BibitemShut {NoStop}%
\bibitem [{\citenamefont {{Dor{\'e}}}\ \emph {et~al.}(2014)\citenamefont
  {{Dor{\'e}}, {Bock}, {Ashby}, {Capak}, {Cooray}, {de Putter}, {Eifler},
  {Flagey}, {Gong}, {Habib} et~al.}}]{2014arXiv1412.4872D}%
  \BibitemOpen
  \bibfield  {author} {\bibinfo {author} {\bibfnamefont {O.}~\bibnamefont
  {{Dor{\'e}}}}, \bibinfo {author} {\bibfnamefont {J.}~\bibnamefont {{Bock}}},
  \bibinfo {author} {\bibfnamefont {M.}~\bibnamefont {{Ashby}}}, \bibinfo
  {author} {\bibfnamefont {P.}~\bibnamefont {{Capak}}}, \bibinfo {author}
  {\bibfnamefont {A.}~\bibnamefont {{Cooray}}}, \bibinfo {author}
  {\bibfnamefont {R.}~\bibnamefont {{de Putter}}}, \bibinfo {author}
  {\bibfnamefont {T.}~\bibnamefont {{Eifler}}}, \bibinfo {author}
  {\bibfnamefont {N.}~\bibnamefont {{Flagey}}}, \bibinfo {author}
  {\bibfnamefont {Y.}~\bibnamefont {{Gong}}}, \bibinfo {author} {\bibfnamefont
  {S.}~\bibnamefont {{Habib}}}, \bibinfo {author} {\bibfnamefont
  {K.}~\bibnamefont {{Heitmann}}}, \bibinfo {author} {\bibfnamefont
  {C.}~\bibnamefont {{Hirata}}}, \bibinfo {author} {\bibfnamefont {W.-S.}\
  \bibnamefont {{Jeong}}}, \bibinfo {author} {\bibfnamefont {R.}~\bibnamefont
  {{Katti}}}, \bibinfo {author} {\bibfnamefont {P.}~\bibnamefont {{Korngut}}},
  \bibinfo {author} {\bibfnamefont {E.}~\bibnamefont {{Krause}}}, \bibnamefont
  {et~al.},\ }\bibfield  {title} {\bibinfo {title} {{Cosmology with the SPHEREX
  All-Sky Spectral Survey}},\ }\href {https://doi.org/10.48550/arXiv.1412.4872}
  {\bibfield  {journal} {\bibinfo  {journal} {arXiv e-prints}\ ,\ \bibinfo
  {eid} {arXiv:1412.4872}} (\bibinfo {year} {2014})},\ \Eprint
  {https://arxiv.org/abs/1412.4872} {arXiv:1412.4872 [astro-ph.CO]}
  \BibitemShut {NoStop}%
\bibitem [{\citenamefont {Chen}\ \emph {et~al.}(2016)\citenamefont {Chen,
  Dvorkin, Huang, Namjoo,\ and\ Verde}}]{Chen_2016}%
  \BibitemOpen
  \bibfield  {author} {\bibinfo {author} {\bibfnamefont {X.}~\bibnamefont
  {Chen}}, \bibinfo {author} {\bibfnamefont {C.}~\bibnamefont {Dvorkin}},
  \bibinfo {author} {\bibfnamefont {Z.}~\bibnamefont {Huang}}, \bibinfo
  {author} {\bibfnamefont {M.~H.}\ \bibnamefont {Namjoo}},\ \bibnamefont {and}\
  \bibinfo {author} {\bibfnamefont {L.}~\bibnamefont {Verde}},\ }\bibfield
  {title} {\bibinfo {title} {The future of primordial features with large-scale
  structure surveys},\ }\href {https://doi.org/10.1088/1475-7516/2016/11/014}
  {\bibfield  {journal} {\bibinfo  {journal} {Journal of Cosmology and
  Astroparticle Physics}\ }\textbf {\bibinfo {volume} {2016}}\bibinfo  {number}
  { (11)},\ \bibinfo {pages} {014}}\BibitemShut {NoStop}%
\bibitem [{\citenamefont {DePorzio}\ \emph {et~al.}(2021)\citenamefont
  {DePorzio, Xu, Mu\~noz,\ and\ Dvorkin}}]{DePorzio:2020wcz}%
  \BibitemOpen
\bibfield  {number} {  }\bibfield  {author} {\bibinfo {author} {\bibfnamefont
  {N.}~\bibnamefont {DePorzio}}, \bibinfo {author} {\bibfnamefont {W.~L.}\
  \bibnamefont {Xu}}, \bibinfo {author} {\bibfnamefont {J.~B.}\ \bibnamefont
  {Mu\~noz}},\ \bibnamefont {and}\ \bibinfo {author} {\bibfnamefont
  {C.}~\bibnamefont {Dvorkin}},\ }\bibfield  {title} {\bibinfo {title}
  {{Finding eV-scale light relics with cosmological observables}},\ }\href
  {https://doi.org/10.1103/PhysRevD.103.023504} {\bibfield  {journal} {\bibinfo
   {journal} {Phys. Rev. D}\ }\textbf {\bibinfo {volume} {103}},\ \bibinfo
  {pages} {023504} (\bibinfo {year} {2021})},\ \Eprint
  {https://arxiv.org/abs/2006.09380} {arXiv:2006.09380 [astro-ph.CO]}
  \BibitemShut {NoStop}%
\bibitem [{\citenamefont {Xu}\ \emph {et~al.}(2022)\citenamefont {Xu, Mu\~noz,\
  and\ Dvorkin}}]{Xu:2021rwg}%
  \BibitemOpen
  \bibfield  {author} {\bibinfo {author} {\bibfnamefont {W.~L.}\ \bibnamefont
  {Xu}}, \bibinfo {author} {\bibfnamefont {J.~B.}\ \bibnamefont {Mu\~noz}},\
  \bibnamefont {and}\ \bibinfo {author} {\bibfnamefont {C.}~\bibnamefont
  {Dvorkin}},\ }\bibfield  {title} {\bibinfo {title} {{Cosmological constraints
  on light but massive relics}},\ }\href
  {https://doi.org/10.1103/PhysRevD.105.095029} {\bibfield  {journal} {\bibinfo
   {journal} {Phys. Rev. D}\ }\textbf {\bibinfo {volume} {105}},\ \bibinfo
  {pages} {095029} (\bibinfo {year} {2022})},\ \Eprint
  {https://arxiv.org/abs/2107.09664} {arXiv:2107.09664 [astro-ph.CO]}
  \BibitemShut {NoStop}%
\bibitem [{\citenamefont {Drlica-Wagner}\ \emph {et~al.}(2019)\citenamefont
  {Drlica-Wagner et~al.}}]{LSSTDarkMatterGroup:2019mwo}%
  \BibitemOpen
  \bibfield  {author} {\bibinfo {author} {\bibfnamefont {A.}~\bibnamefont
  {Drlica-Wagner}} \bibnamefont {et~al.} (\bibinfo {collaboration} {LSST Dark
  Matter Group}),\ }\bibfield  {title} {\bibinfo {title} {{Probing the
  Fundamental Nature of Dark Matter with the Large Synoptic Survey
  Telescope}},\ }\href@noop {} {\  (\bibinfo {year} {2019})},\ \Eprint
  {https://arxiv.org/abs/1902.01055} {arXiv:1902.01055 [astro-ph.CO]}
  \BibitemShut {NoStop}%
\bibitem [{\citenamefont {Lesgourgues}\ and\ \citenamefont
  {Pastor}(2006)\citenamefont {Lesgourgues\ and\ Pastor}}]{LESGOURGUES2006307}%
  \BibitemOpen
  \bibfield  {author} {\bibinfo {author} {\bibfnamefont {J.}~\bibnamefont
  {Lesgourgues}}\ \bibnamefont {and}\ \bibinfo {author} {\bibfnamefont
  {S.}~\bibnamefont {Pastor}},\ }\bibfield  {title} {\bibinfo {title} {Massive
  neutrinos and cosmology},\ }\href
  {https://doi.org/https://doi.org/10.1016/j.physrep.2006.04.001} {\bibfield
  {journal} {\bibinfo  {journal} {Physics Reports}\ }\textbf {\bibinfo {volume}
  {429}},\ \bibinfo {pages} {307} (\bibinfo {year} {2006})}\BibitemShut
  {NoStop}%
\bibitem [{\citenamefont {Dvorkin}\ \emph {et~al.}(2019)\citenamefont {Dvorkin
  et~al.}}]{Dvorkin:2019jgs}%
  \BibitemOpen
  \bibfield  {author} {\bibinfo {author} {\bibfnamefont {C.}~\bibnamefont
  {Dvorkin}} \bibnamefont {et~al.},\ }\bibfield  {title} {\bibinfo {title}
  {{Neutrino Mass from Cosmology: Probing Physics Beyond the Standard Model}},\
  }\href@noop {} {\  (\bibinfo {year} {2019})},\ \Eprint
  {https://arxiv.org/abs/1903.03689} {arXiv:1903.03689 [astro-ph.CO]}
  \BibitemShut {NoStop}%
\bibitem [{\citenamefont {Copeland}\ \emph {et~al.}(2006)\citenamefont
  {Copeland, Sami,\ and\ Tsujikawa}}]{Copeland:2006wr}%
  \BibitemOpen
  \bibfield  {author} {\bibinfo {author} {\bibfnamefont {E.~J.}\ \bibnamefont
  {Copeland}}, \bibinfo {author} {\bibfnamefont {M.}~\bibnamefont {Sami}},\
  \bibnamefont {and}\ \bibinfo {author} {\bibfnamefont {S.}~\bibnamefont
  {Tsujikawa}},\ }\bibfield  {title} {\bibinfo {title} {{Dynamics of dark
  energy}},\ }\href {https://doi.org/10.1142/S021827180600942X} {\bibfield
  {journal} {\bibinfo  {journal} {Int. J. Mod. Phys. D}\ }\textbf {\bibinfo
  {volume} {15}},\ \bibinfo {pages} {1753} (\bibinfo {year} {2006})},\ \Eprint
  {https://arxiv.org/abs/hep-th/0603057} {arXiv:hep-th/0603057} \BibitemShut
  {NoStop}%
\bibitem [{\citenamefont {{Koyama}}(2016)}]{2016RPPh...79d6902K}%
  \BibitemOpen
  \bibfield  {author} {\bibinfo {author} {\bibfnamefont {K.}~\bibnamefont
  {{Koyama}}},\ }\bibfield  {title} {\bibinfo {title} {{Cosmological tests of
  modified gravity}},\ }\href {https://doi.org/10.1088/0034-4885/79/4/046902}
  {\bibfield  {journal} {\bibinfo  {journal} {Reports on Progress in Physics}\
  }\textbf {\bibinfo {volume} {79}},\ \bibinfo {eid} {046902} (\bibinfo {year}
  {2016})},\ \Eprint {https://arxiv.org/abs/1504.04623} {arXiv:1504.04623
  [astro-ph.CO]} \BibitemShut {NoStop}%
\bibitem [{\citenamefont {Ishak}(2019)}]{Ishak:2018his}%
  \BibitemOpen
  \bibfield  {author} {\bibinfo {author} {\bibfnamefont {M.}~\bibnamefont
  {Ishak}},\ }\bibfield  {title} {\bibinfo {title} {{Testing General Relativity
  in Cosmology}},\ }\href {https://doi.org/10.1007/s41114-018-0017-4}
  {\bibfield  {journal} {\bibinfo  {journal} {Living Rev. Rel.}\ }\textbf
  {\bibinfo {volume} {22}},\ \bibinfo {pages} {1} (\bibinfo {year} {2019})},\
  \Eprint {https://arxiv.org/abs/1806.10122} {arXiv:1806.10122 [astro-ph.CO]}
  \BibitemShut {NoStop}%
\bibitem [{\citenamefont
  {Ferreira}(2019)}]{doi:10.1146/annurev-astro-091918-104423}%
  \BibitemOpen
  \bibfield  {author} {\bibinfo {author} {\bibfnamefont {P.~G.}\ \bibnamefont
  {Ferreira}},\ }\bibfield  {title} {\bibinfo {title} {Cosmological tests of
  gravity},\ }\href {https://doi.org/10.1146/annurev-astro-091918-104423}
  {\bibfield  {journal} {\bibinfo  {journal} {Annual Review of Astronomy and
  Astrophysics}\ }\textbf {\bibinfo {volume} {57}},\ \bibinfo {pages} {335}
  (\bibinfo {year} {2019})},\ \Eprint
  {https://arxiv.org/abs/https://doi.org/10.1146/annurev-astro-091918-104423}
  {https://doi.org/10.1146/annurev-astro-091918-104423} \BibitemShut {NoStop}%
\bibitem [{\citenamefont {Alam}\ \emph {et~al.}(2020)\citenamefont {Alam
  et~al.}}]{Alam:2020jdv}%
  \BibitemOpen
  \bibfield  {author} {\bibinfo {author} {\bibfnamefont {S.}~\bibnamefont
  {Alam}} \bibnamefont {et~al.},\ }\bibfield  {title} {\bibinfo {title}
  {{Testing the theory of gravity with DESI: estimators, predictions and
  simulation requirements}},\ }\href@noop {} {\  (\bibinfo {year} {2020})},\
  \Eprint {https://arxiv.org/abs/2011.05771} {arXiv:2011.05771 [astro-ph.CO]}
  \BibitemShut {NoStop}%
\bibitem [{\citenamefont {{Will}}(2006)}]{2006LRR.....9....3W}%
  \BibitemOpen
  \bibfield  {author} {\bibinfo {author} {\bibfnamefont {C.~M.}\ \bibnamefont
  {{Will}}},\ }\bibfield  {title} {\bibinfo {title} {{The Confrontation between
  General Relativity and Experiment}},\ }\href
  {https://doi.org/10.12942/lrr-2006-3} {\bibfield  {journal} {\bibinfo
  {journal} {Living Reviews in Relativity}\ }\textbf {\bibinfo {volume} {9}},\
  \bibinfo {eid} {3} (\bibinfo {year} {2006})},\ \Eprint
  {https://arxiv.org/abs/gr-qc/0510072} {arXiv:gr-qc/0510072 [gr-qc]}
  \BibitemShut {NoStop}%
\bibitem [{\citenamefont {{Psaltis}}(2008)}]{2008LRR....11....9P}%
  \BibitemOpen
  \bibfield  {author} {\bibinfo {author} {\bibfnamefont {D.}~\bibnamefont
  {{Psaltis}}},\ }\bibfield  {title} {\bibinfo {title} {{Probes and Tests of
  Strong-Field Gravity with Observations in the Electromagnetic Spectrum}},\
  }\href {https://doi.org/10.12942/lrr-2008-9} {\bibfield  {journal} {\bibinfo
  {journal} {Living Reviews in Relativity}\ }\textbf {\bibinfo {volume} {11}},\
  \bibinfo {eid} {9} (\bibinfo {year} {2008})},\ \Eprint
  {https://arxiv.org/abs/0806.1531} {arXiv:0806.1531 [astro-ph]} \BibitemShut
  {NoStop}%
\bibitem [{\citenamefont {{Abbott}}\ \emph {et~al.}(2016)\citenamefont
  {{Abbott}, {Abbott}, {Abbott}, {Abernathy}, {Acernese}, {Ackley}, {Adams},
  {Adams}, {Addesso}, {Adhikari} et~al.}}]{2016PhRvL.116f1102A}%
  \BibitemOpen
  \bibfield  {author} {\bibinfo {author} {\bibfnamefont {B.~P.}\ \bibnamefont
  {{Abbott}}}, \bibinfo {author} {\bibfnamefont {R.}~\bibnamefont {{Abbott}}},
  \bibinfo {author} {\bibfnamefont {T.~D.}\ \bibnamefont {{Abbott}}}, \bibinfo
  {author} {\bibfnamefont {M.~R.}\ \bibnamefont {{Abernathy}}}, \bibinfo
  {author} {\bibfnamefont {F.}~\bibnamefont {{Acernese}}}, \bibinfo {author}
  {\bibfnamefont {K.}~\bibnamefont {{Ackley}}}, \bibinfo {author}
  {\bibfnamefont {C.}~\bibnamefont {{Adams}}}, \bibinfo {author} {\bibfnamefont
  {T.}~\bibnamefont {{Adams}}}, \bibinfo {author} {\bibfnamefont
  {P.}~\bibnamefont {{Addesso}}}, \bibinfo {author} {\bibfnamefont {R.~X.}\
  \bibnamefont {{Adhikari}}}, \bibinfo {author} {\bibfnamefont {V.~B.}\
  \bibnamefont {{Adya}}}, \bibinfo {author} {\bibfnamefont {C.}~\bibnamefont
  {{Affeldt}}}, \bibinfo {author} {\bibfnamefont {M.}~\bibnamefont
  {{Agathos}}}, \bibinfo {author} {\bibfnamefont {K.}~\bibnamefont
  {{Agatsuma}}}, \bibinfo {author} {\bibfnamefont {N.}~\bibnamefont
  {{Aggarwal}}}, \bibinfo {author} {\bibfnamefont {O.~D.}\ \bibnamefont
  {{Aguiar}}}, \bibnamefont {et~al.},\ }\bibfield  {title} {\bibinfo {title}
  {{Observation of Gravitational Waves from a Binary Black Hole Merger}},\
  }\href {https://doi.org/10.1103/PhysRevLett.116.061102} {\bibfield  {journal}
  {\bibinfo  {journal} {\prl}\ }\textbf {\bibinfo {volume} {116}},\ \bibinfo
  {eid} {061102} (\bibinfo {year} {2016})},\ \Eprint
  {https://arxiv.org/abs/1602.03837} {arXiv:1602.03837 [gr-qc]} \BibitemShut
  {NoStop}%
\bibitem [{\citenamefont {{Perlmutter}}\ \emph {et~al.}(1999)\citenamefont
  {{Perlmutter}, {Aldering}, {Goldhaber}, {Knop}, {Nugent}, {Castro},
  {Deustua}, {Fabbro}, {Goobar}, {Groom} et~al.}}]{1999ApJ...517..565P}%
  \BibitemOpen
  \bibfield  {author} {\bibinfo {author} {\bibfnamefont {S.}~\bibnamefont
  {{Perlmutter}}}, \bibinfo {author} {\bibfnamefont {G.}~\bibnamefont
  {{Aldering}}}, \bibinfo {author} {\bibfnamefont {G.}~\bibnamefont
  {{Goldhaber}}}, \bibinfo {author} {\bibfnamefont {R.~A.}\ \bibnamefont
  {{Knop}}}, \bibinfo {author} {\bibfnamefont {P.}~\bibnamefont {{Nugent}}},
  \bibinfo {author} {\bibfnamefont {P.~G.}\ \bibnamefont {{Castro}}}, \bibinfo
  {author} {\bibfnamefont {S.}~\bibnamefont {{Deustua}}}, \bibinfo {author}
  {\bibfnamefont {S.}~\bibnamefont {{Fabbro}}}, \bibinfo {author}
  {\bibfnamefont {A.}~\bibnamefont {{Goobar}}}, \bibinfo {author}
  {\bibfnamefont {D.~E.}\ \bibnamefont {{Groom}}}, \bibinfo {author}
  {\bibfnamefont {I.~M.}\ \bibnamefont {{Hook}}}, \bibinfo {author}
  {\bibfnamefont {A.~G.}\ \bibnamefont {{Kim}}}, \bibinfo {author}
  {\bibfnamefont {M.~Y.}\ \bibnamefont {{Kim}}}, \bibinfo {author}
  {\bibfnamefont {J.~C.}\ \bibnamefont {{Lee}}}, \bibinfo {author}
  {\bibfnamefont {N.~J.}\ \bibnamefont {{Nunes}}}, \bibinfo {author}
  {\bibfnamefont {R.}~\bibnamefont {{Pain}}}, \bibnamefont {et~al.},\
  }\bibfield  {title} {\bibinfo {title} {{Measurements of {\ensuremath{\Omega}}
  and {\ensuremath{\Lambda}} from 42 High-Redshift Supernovae}},\ }\href
  {https://doi.org/10.1086/307221} {\bibfield  {journal} {\bibinfo  {journal}
  {\apj}\ }\textbf {\bibinfo {volume} {517}},\ \bibinfo {pages} {565} (\bibinfo
  {year} {1999})},\ \Eprint {https://arxiv.org/abs/astro-ph/9812133}
  {arXiv:astro-ph/9812133 [astro-ph]} \BibitemShut {NoStop}%
\bibitem [{\citenamefont {{Riess}}\ \emph {et~al.}(1998)\citenamefont {{Riess},
  {Filippenko}, {Challis}, {Clocchiatti}, {Diercks}, {Garnavich}, {Gilliland},
  {Hogan}, {Jha}, {Kirshner} et~al.}}]{1998AJ....116.1009R}%
  \BibitemOpen
  \bibfield  {author} {\bibinfo {author} {\bibfnamefont {A.~G.}\ \bibnamefont
  {{Riess}}}, \bibinfo {author} {\bibfnamefont {A.~V.}\ \bibnamefont
  {{Filippenko}}}, \bibinfo {author} {\bibfnamefont {P.}~\bibnamefont
  {{Challis}}}, \bibinfo {author} {\bibfnamefont {A.}~\bibnamefont
  {{Clocchiatti}}}, \bibinfo {author} {\bibfnamefont {A.}~\bibnamefont
  {{Diercks}}}, \bibinfo {author} {\bibfnamefont {P.~M.}\ \bibnamefont
  {{Garnavich}}}, \bibinfo {author} {\bibfnamefont {R.~L.}\ \bibnamefont
  {{Gilliland}}}, \bibinfo {author} {\bibfnamefont {C.~J.}\ \bibnamefont
  {{Hogan}}}, \bibinfo {author} {\bibfnamefont {S.}~\bibnamefont {{Jha}}},
  \bibinfo {author} {\bibfnamefont {R.~P.}\ \bibnamefont {{Kirshner}}},
  \bibinfo {author} {\bibfnamefont {B.}~\bibnamefont {{Leibundgut}}}, \bibinfo
  {author} {\bibfnamefont {M.~M.}\ \bibnamefont {{Phillips}}}, \bibinfo
  {author} {\bibfnamefont {D.}~\bibnamefont {{Reiss}}}, \bibinfo {author}
  {\bibfnamefont {B.~P.}\ \bibnamefont {{Schmidt}}}, \bibinfo {author}
  {\bibfnamefont {R.~A.}\ \bibnamefont {{Schommer}}}, \bibinfo {author}
  {\bibfnamefont {R.~C.}\ \bibnamefont {{Smith}}}, \bibnamefont {et~al.},\
  }\bibfield  {title} {\bibinfo {title} {{Observational Evidence from
  Supernovae for an Accelerating Universe and a Cosmological Constant}},\
  }\href {https://doi.org/10.1086/300499} {\bibfield  {journal} {\bibinfo
  {journal} {\aj}\ }\textbf {\bibinfo {volume} {116}},\ \bibinfo {pages} {1009}
  (\bibinfo {year} {1998})},\ \Eprint {https://arxiv.org/abs/astro-ph/9805201}
  {arXiv:astro-ph/9805201 [astro-ph]} \BibitemShut {NoStop}%
\bibitem [{\citenamefont {{Weinberg}}(1989)}]{1989RvMP...61....1W}%
  \BibitemOpen
  \bibfield  {author} {\bibinfo {author} {\bibfnamefont {S.}~\bibnamefont
  {{Weinberg}}},\ }\bibfield  {title} {\bibinfo {title} {{The cosmological
  constant problem}},\ }\href {https://doi.org/10.1103/RevModPhys.61.1}
  {\bibfield  {journal} {\bibinfo  {journal} {Reviews of Modern Physics}\
  }\textbf {\bibinfo {volume} {61}},\ \bibinfo {pages} {1} (\bibinfo {year}
  {1989})}\BibitemShut {NoStop}%
\bibitem [{\citenamefont {Calderon}\ \emph {et~al.}(2024)\citenamefont
  {Calderon et~al.}}]{DESI:2024aqx}%
  \BibitemOpen
  \bibfield  {author} {\bibinfo {author} {\bibfnamefont {R.}~\bibnamefont
  {Calderon}} \bibnamefont {et~al.} (\bibinfo {collaboration} {DESI}),\
  }\bibfield  {title} {\bibinfo {title} {{DESI 2024: Reconstructing Dark Energy
  using Crossing Statistics with DESI DR1 BAO data}},\ }\href@noop {} {\
  (\bibinfo {year} {2024})},\ \Eprint {https://arxiv.org/abs/2405.04216}
  {arXiv:2405.04216 [astro-ph.CO]} \BibitemShut {NoStop}%
\bibitem [{\citenamefont {Lodha}\ \emph {et~al.}(2024)\citenamefont {Lodha
  et~al.}}]{DESI:2024kob}%
  \BibitemOpen
  \bibfield  {author} {\bibinfo {author} {\bibfnamefont {K.}~\bibnamefont
  {Lodha}} \bibnamefont {et~al.} (\bibinfo {collaboration} {DESI}),\ }\bibfield
   {title} {\bibinfo {title} {{DESI 2024: Constraints on Physics-Focused
  Aspects of Dark Energy using DESI DR1 BAO Data}},\ }\href@noop {} {\
  (\bibinfo {year} {2024})},\ \Eprint {https://arxiv.org/abs/2405.13588}
  {arXiv:2405.13588 [astro-ph.CO]} \BibitemShut {NoStop}%
\bibitem [{\citenamefont {Carron}(2012)}]{PhysRevLett.108.071301}%
  \BibitemOpen
  \bibfield  {author} {\bibinfo {author} {\bibfnamefont {J.}~\bibnamefont
  {Carron}},\ }\bibfield  {title} {\bibinfo {title} {Information escaping the
  correlation hierarchy of the convergence field in the study of cosmological
  parameters},\ }\href {https://doi.org/10.1103/PhysRevLett.108.071301}
  {\bibfield  {journal} {\bibinfo  {journal} {Phys. Rev. Lett.}\ }\textbf
  {\bibinfo {volume} {108}},\ \bibinfo {pages} {071301} (\bibinfo {year}
  {2012})}\BibitemShut {NoStop}%
\bibitem [{\citenamefont {Gil-Marín}\ \emph {et~al.}(2015)\citenamefont
  {Gil-Marín, Noreña, Verde, Percival, Wagner, Manera,\ and\
  Schneider}}]{10.1093/mnras/stv961}%
  \BibitemOpen
  \bibfield  {author} {\bibinfo {author} {\bibfnamefont {H.}~\bibnamefont
  {Gil-Marín}}, \bibinfo {author} {\bibfnamefont {J.}~\bibnamefont {Noreña}},
  \bibinfo {author} {\bibfnamefont {L.}~\bibnamefont {Verde}}, \bibinfo
  {author} {\bibfnamefont {W.~J.}\ \bibnamefont {Percival}}, \bibinfo {author}
  {\bibfnamefont {C.}~\bibnamefont {Wagner}}, \bibinfo {author} {\bibfnamefont
  {M.}~\bibnamefont {Manera}},\ \bibnamefont {and}\ \bibinfo {author}
  {\bibfnamefont {D.~P.}\ \bibnamefont {Schneider}},\ }\bibfield  {title}
  {\bibinfo {title} {{The power spectrum and bispectrum of SDSS DR11 BOSS
  galaxies – I. Bias and gravity}},\ }\href
  {https://doi.org/10.1093/mnras/stv961} {\bibfield  {journal} {\bibinfo
  {journal} {Monthly Notices of the Royal Astronomical Society}\ }\textbf
  {\bibinfo {volume} {451}},\ \bibinfo {pages} {539} (\bibinfo {year}
  {2015})},\ \Eprint
  {https://arxiv.org/abs/https://academic.oup.com/mnras/article-pdf/451/1/539/4177830/stv961.pdf}
  {https://academic.oup.com/mnras/article-pdf/451/1/539/4177830/stv961.pdf}
  \BibitemShut {NoStop}%
\bibitem [{\citenamefont {Gil-Marín}\ \emph {et~al.}(2016)\citenamefont
  {Gil-Marín, Percival, Verde, Brownstein, Chuang, Kitaura,
  Rodríguez-Torres,\ and\ Olmstead}}]{10.1093/mnras/stw2679}%
  \BibitemOpen
  \bibfield  {author} {\bibinfo {author} {\bibfnamefont {H.}~\bibnamefont
  {Gil-Marín}}, \bibinfo {author} {\bibfnamefont {W.~J.}\ \bibnamefont
  {Percival}}, \bibinfo {author} {\bibfnamefont {L.}~\bibnamefont {Verde}},
  \bibinfo {author} {\bibfnamefont {J.~R.}\ \bibnamefont {Brownstein}},
  \bibinfo {author} {\bibfnamefont {C.-H.}\ \bibnamefont {Chuang}}, \bibinfo
  {author} {\bibfnamefont {F.-S.}\ \bibnamefont {Kitaura}}, \bibinfo {author}
  {\bibfnamefont {S.~A.}\ \bibnamefont {Rodríguez-Torres}},\ \bibnamefont
  {and}\ \bibinfo {author} {\bibfnamefont {M.~D.}\ \bibnamefont {Olmstead}},\
  }\bibfield  {title} {\bibinfo {title} {{The clustering of galaxies in the
  SDSS-III Baryon Oscillation Spectroscopic Survey: RSD measurement from the
  power spectrum and bispectrum of the DR12 BOSS galaxies}},\ }\href
  {https://doi.org/10.1093/mnras/stw2679} {\bibfield  {journal} {\bibinfo
  {journal} {Monthly Notices of the Royal Astronomical Society}\ }\textbf
  {\bibinfo {volume} {465}},\ \bibinfo {pages} {1757} (\bibinfo {year}
  {2016})},\ \Eprint
  {https://arxiv.org/abs/https://academic.oup.com/mnras/article-pdf/465/2/1757/8364690/stw2679.pdf}
  {https://academic.oup.com/mnras/article-pdf/465/2/1757/8364690/stw2679.pdf}
  \BibitemShut {NoStop}%
\bibitem [{\citenamefont {Bernardeau}\ \emph {et~al.}(2002)\citenamefont
  {Bernardeau, Colombi, Gaztañaga,\ and\ Scoccimarro}}]{BERNARDEAU20021}%
  \BibitemOpen
  \bibfield  {author} {\bibinfo {author} {\bibfnamefont {F.}~\bibnamefont
  {Bernardeau}}, \bibinfo {author} {\bibfnamefont {S.}~\bibnamefont {Colombi}},
  \bibinfo {author} {\bibfnamefont {E.}~\bibnamefont {Gaztañaga}},\
  \bibnamefont {and}\ \bibinfo {author} {\bibfnamefont {R.}~\bibnamefont
  {Scoccimarro}},\ }\bibfield  {title} {\bibinfo {title} {Large-scale structure
  of the universe and cosmological perturbation theory},\ }\href
  {https://doi.org/https://doi.org/10.1016/S0370-1573(02)00135-7} {\bibfield
  {journal} {\bibinfo  {journal} {Physics Reports}\ }\textbf {\bibinfo {volume}
  {367}},\ \bibinfo {pages} {1} (\bibinfo {year} {2002})}\BibitemShut {NoStop}%
\bibitem [{\citenamefont {Philcox}\ and\ \citenamefont
  {Ivanov}(2022)\citenamefont {Philcox\ and\ Ivanov}}]{PhysRevD.105.043517}%
  \BibitemOpen
  \bibfield  {author} {\bibinfo {author} {\bibfnamefont {O.~H.~E.}\
  \bibnamefont {Philcox}}\ \bibnamefont {and}\ \bibinfo {author} {\bibfnamefont
  {M.~M.}\ \bibnamefont {Ivanov}},\ }\bibfield  {title} {\bibinfo {title} {Boss
  dr12 full-shape cosmology: $\mathrm{\ensuremath{\Lambda}}\mathrm{CDM}$
  constraints from the large-scale galaxy power spectrum and bispectrum
  monopole},\ }\href {https://doi.org/10.1103/PhysRevD.105.043517} {\bibfield
  {journal} {\bibinfo  {journal} {Phys. Rev. D}\ }\textbf {\bibinfo {volume}
  {105}},\ \bibinfo {pages} {043517} (\bibinfo {year} {2022})}\BibitemShut
  {NoStop}%
\bibitem [{\citenamefont {Chen}\ \emph {et~al.}(2021)\citenamefont {Chen, Lee,\
  and\ Dvorkin}}]{Chen:2021vba}%
  \BibitemOpen
  \bibfield  {author} {\bibinfo {author} {\bibfnamefont {S.-F.}\ \bibnamefont
  {Chen}}, \bibinfo {author} {\bibfnamefont {H.}~\bibnamefont {Lee}},\
  \bibnamefont {and}\ \bibinfo {author} {\bibfnamefont {C.}~\bibnamefont
  {Dvorkin}},\ }\bibfield  {title} {\bibinfo {title} {{Precise and accurate
  cosmology with CMB\texttimes{}LSS power spectra and bispectra}},\ }\href
  {https://doi.org/10.1088/1475-7516/2021/05/030} {\bibfield  {journal}
  {\bibinfo  {journal} {JCAP}\ }\textbf {\bibinfo {volume} {05}},\ \bibinfo
  {pages} {030}},\ \Eprint {https://arxiv.org/abs/2103.01229} {arXiv:2103.01229
  [astro-ph.CO]} \BibitemShut {NoStop}%
\bibitem [{\citenamefont {Philcox}\ \emph {et~al.}(2021)\citenamefont {Philcox,
  Hou,\ and\ Slepian}}]{Philcox:2021hbm}%
  \BibitemOpen
  \bibfield  {author} {\bibinfo {author} {\bibfnamefont {O.~H.~E.}\
  \bibnamefont {Philcox}}, \bibinfo {author} {\bibfnamefont {J.}~\bibnamefont
  {Hou}},\ \bibnamefont {and}\ \bibinfo {author} {\bibfnamefont
  {Z.}~\bibnamefont {Slepian}},\ }\bibfield  {title} {\bibinfo {title} {{A
  First Detection of the Connected 4-Point Correlation Function of Galaxies
  Using the BOSS CMASS Sample}},\ }\href@noop {} {\  (\bibinfo {year}
  {2021})},\ \Eprint {https://arxiv.org/abs/2108.01670} {arXiv:2108.01670
  [astro-ph.CO]} \BibitemShut {NoStop}%
\bibitem [{\citenamefont {{Philcox}}\ and\ \citenamefont
  {{Fl{\"o}ss}}(2024)\citenamefont {{Philcox}\ and\
  {Fl{\"o}ss}}}]{2024arXiv240407249P}%
  \BibitemOpen
  \bibfield  {author} {\bibinfo {author} {\bibfnamefont {O.~H.~E.}\
  \bibnamefont {{Philcox}}}\ \bibnamefont {and}\ \bibinfo {author}
  {\bibfnamefont {T.}~\bibnamefont {{Fl{\"o}ss}}},\ }\bibfield  {title}
  {\bibinfo {title} {{PolyBin3D: A Suite of Optimal and Efficient Power
  Spectrum and Bispectrum Estimators for Large-Scale Structure}},\ }\href
  {https://doi.org/10.48550/arXiv.2404.07249} {\bibfield  {journal} {\bibinfo
  {journal} {arXiv e-prints}\ ,\ \bibinfo {eid} {arXiv:2404.07249}} (\bibinfo
  {year} {2024})},\ \Eprint {https://arxiv.org/abs/2404.07249}
  {arXiv:2404.07249 [astro-ph.CO]} \BibitemShut {NoStop}%
\bibitem [{\citenamefont {{Khoury}}(2010)}]{2010arXiv1011.5909K}%
  \BibitemOpen
  \bibfield  {author} {\bibinfo {author} {\bibfnamefont {J.}~\bibnamefont
  {{Khoury}}},\ }\bibfield  {title} {\bibinfo {title} {{Theories of Dark Energy
  with Screening Mechanisms}},\ }\href
  {https://doi.org/10.48550/arXiv.1011.5909} {\bibfield  {journal} {\bibinfo
  {journal} {arXiv e-prints}\ ,\ \bibinfo {eid} {arXiv:1011.5909}} (\bibinfo
  {year} {2010})},\ \Eprint {https://arxiv.org/abs/1011.5909} {arXiv:1011.5909
  [astro-ph.CO]} \BibitemShut {NoStop}%
\bibitem [{\citenamefont {{Khoury}}(2013)}]{2013arXiv1312.2006K}%
  \BibitemOpen
  \bibfield  {author} {\bibinfo {author} {\bibfnamefont {J.}~\bibnamefont
  {{Khoury}}},\ }\bibfield  {title} {\bibinfo {title} {{Les Houches Lectures on
  Physics Beyond the Standard Model of Cosmology}},\ }\href
  {https://doi.org/10.48550/arXiv.1312.2006} {\bibfield  {journal} {\bibinfo
  {journal} {arXiv e-prints}\ ,\ \bibinfo {eid} {arXiv:1312.2006}} (\bibinfo
  {year} {2013})},\ \Eprint {https://arxiv.org/abs/1312.2006} {arXiv:1312.2006
  [astro-ph.CO]} \BibitemShut {NoStop}%
\bibitem [{\citenamefont {{Khoury}}\ and\ \citenamefont
  {{Weltman}}(2004)\citenamefont {{Khoury}\ and\
  {Weltman}}}]{2004PhRvL..93q1104K}%
  \BibitemOpen
  \bibfield  {author} {\bibinfo {author} {\bibfnamefont {J.}~\bibnamefont
  {{Khoury}}}\ \bibnamefont {and}\ \bibinfo {author} {\bibfnamefont
  {A.}~\bibnamefont {{Weltman}}},\ }\bibfield  {title} {\bibinfo {title}
  {{Chameleon Fields: Awaiting Surprises for Tests of Gravity in Space}},\
  }\href {https://doi.org/10.1103/PhysRevLett.93.171104} {\bibfield  {journal}
  {\bibinfo  {journal} {\prl}\ }\textbf {\bibinfo {volume} {93}},\ \bibinfo
  {eid} {171104} (\bibinfo {year} {2004})},\ \Eprint
  {https://arxiv.org/abs/astro-ph/0309300} {arXiv:astro-ph/0309300 [astro-ph]}
  \BibitemShut {NoStop}%
\bibitem [{\citenamefont {{Olive}}\ and\ \citenamefont
  {{Pospelov}}(2008)\citenamefont {{Olive}\ and\
  {Pospelov}}}]{2008PhRvD..77d3524O}%
  \BibitemOpen
  \bibfield  {author} {\bibinfo {author} {\bibfnamefont {K.~A.}\ \bibnamefont
  {{Olive}}}\ \bibnamefont {and}\ \bibinfo {author} {\bibfnamefont
  {M.}~\bibnamefont {{Pospelov}}},\ }\bibfield  {title} {\bibinfo {title}
  {{Environmental dependence of masses and coupling constants}},\ }\href
  {https://doi.org/10.1103/PhysRevD.77.043524} {\bibfield  {journal} {\bibinfo
  {journal} {\prd}\ }\textbf {\bibinfo {volume} {77}},\ \bibinfo {eid} {043524}
  (\bibinfo {year} {2008})},\ \Eprint {https://arxiv.org/abs/0709.3825}
  {arXiv:0709.3825 [hep-ph]} \BibitemShut {NoStop}%
\bibitem [{\citenamefont {{Hinterbichler}}\ and\ \citenamefont
  {{Khoury}}(2010)\citenamefont {{Hinterbichler}\ and\
  {Khoury}}}]{2010PhRvL.104w1301H}%
  \BibitemOpen
  \bibfield  {author} {\bibinfo {author} {\bibfnamefont {K.}~\bibnamefont
  {{Hinterbichler}}}\ \bibnamefont {and}\ \bibinfo {author} {\bibfnamefont
  {J.}~\bibnamefont {{Khoury}}},\ }\bibfield  {title} {\bibinfo {title}
  {{Screening Long-Range Forces through Local Symmetry Restoration}},\ }\href
  {https://doi.org/10.1103/PhysRevLett.104.231301} {\bibfield  {journal}
  {\bibinfo  {journal} {\prl}\ }\textbf {\bibinfo {volume} {104}},\ \bibinfo
  {eid} {231301} (\bibinfo {year} {2010})},\ \Eprint
  {https://arxiv.org/abs/1001.4525} {arXiv:1001.4525 [hep-th]} \BibitemShut
  {NoStop}%
\bibitem [{\citenamefont {{Babichev}}\ \emph {et~al.}(2009)\citenamefont
  {{Babichev}, {Deffayet},\ and\ {Ziour}}}]{2009IJMPD..18.2147B}%
  \BibitemOpen
  \bibfield  {author} {\bibinfo {author} {\bibfnamefont {E.}~\bibnamefont
  {{Babichev}}}, \bibinfo {author} {\bibfnamefont {C.}~\bibnamefont
  {{Deffayet}}},\ \bibnamefont {and}\ \bibinfo {author} {\bibfnamefont
  {R.}~\bibnamefont {{Ziour}}},\ }\bibfield  {title} {\bibinfo {title}
  {{k-MOUFLAGE Gravity}},\ }\href {https://doi.org/10.1142/S0218271809016107}
  {\bibfield  {journal} {\bibinfo  {journal} {International Journal of Modern
  Physics D}\ }\textbf {\bibinfo {volume} {18}},\ \bibinfo {pages} {2147}
  (\bibinfo {year} {2009})},\ \Eprint {https://arxiv.org/abs/0905.2943}
  {arXiv:0905.2943 [hep-th]} \BibitemShut {NoStop}%
\bibitem [{\citenamefont {{Dvali}}\ \emph {et~al.}(2011)\citenamefont {{Dvali},
  {Giudice}, {Gomez},\ and\ {Kehagias}}}]{2011JHEP...08..108D}%
  \BibitemOpen
  \bibfield  {author} {\bibinfo {author} {\bibfnamefont {G.}~\bibnamefont
  {{Dvali}}}, \bibinfo {author} {\bibfnamefont {G.~F.}\ \bibnamefont
  {{Giudice}}}, \bibinfo {author} {\bibfnamefont {C.}~\bibnamefont {{Gomez}}},\
  \bibnamefont {and}\ \bibinfo {author} {\bibfnamefont {A.}~\bibnamefont
  {{Kehagias}}},\ }\bibfield  {title} {\bibinfo {title} {{UV-completion by
  classicalization}},\ }\href {https://doi.org/10.1007/JHEP08(2011)108}
  {\bibfield  {journal} {\bibinfo  {journal} {Journal of High Energy Physics}\
  }\textbf {\bibinfo {volume} {2011}},\ \bibinfo {eid} {108} (\bibinfo {year}
  {2011})},\ \Eprint {https://arxiv.org/abs/1010.1415} {arXiv:1010.1415
  [hep-ph]} \BibitemShut {NoStop}%
\bibitem [{\citenamefont {{Vainshtein}}(1972)}]{1972PhLB...39..393V}%
  \BibitemOpen
  \bibfield  {author} {\bibinfo {author} {\bibfnamefont {A.~I.}\ \bibnamefont
  {{Vainshtein}}},\ }\bibfield  {title} {\bibinfo {title} {{To the problem of
  nonvanishing gravitation mass}},\ }\href
  {https://doi.org/10.1016/0370-2693(72)90147-5} {\bibfield  {journal}
  {\bibinfo  {journal} {Physics Letters B}\ }\textbf {\bibinfo {volume} {39}},\
  \bibinfo {pages} {393} (\bibinfo {year} {1972})}\BibitemShut {NoStop}%
\bibitem [{\citenamefont {White}(2016)}]{White_2016}%
  \BibitemOpen
  \bibfield  {author} {\bibinfo {author} {\bibfnamefont {M.}~\bibnamefont
  {White}},\ }\bibfield  {title} {\bibinfo {title} {A marked correlation
  function for constraining modified gravity models},\ }\href
  {https://doi.org/10.1088/1475-7516/2016/11/057} {\bibfield  {journal}
  {\bibinfo  {journal} {Journal of Cosmology and Astroparticle Physics}\
  }\textbf {\bibinfo {volume} {2016}}\bibinfo  {number} { (11)},\ \bibinfo
  {pages} {057}}\BibitemShut {NoStop}%
\bibitem [{\citenamefont {Valogiannis}\ and\ \citenamefont
  {Bean}(2018)\citenamefont {Valogiannis\ and\ Bean}}]{PhysRevD.97.023535}%
  \BibitemOpen
\bibfield  {number} {  }\bibfield  {author} {\bibinfo {author} {\bibfnamefont
  {G.}~\bibnamefont {Valogiannis}}\ \bibnamefont {and}\ \bibinfo {author}
  {\bibfnamefont {R.}~\bibnamefont {Bean}},\ }\bibfield  {title} {\bibinfo
  {title} {Beyond $\ensuremath{\delta}$: Tailoring marked statistics to reveal
  modified gravity},\ }\href {https://doi.org/10.1103/PhysRevD.97.023535}
  {\bibfield  {journal} {\bibinfo  {journal} {Phys. Rev. D}\ }\textbf {\bibinfo
  {volume} {97}},\ \bibinfo {pages} {023535} (\bibinfo {year}
  {2018})}\BibitemShut {NoStop}%
\bibitem [{\citenamefont {{Hern{\'a}ndez-Aguayo}}\ \emph
  {et~al.}(2018)\citenamefont {{Hern{\'a}ndez-Aguayo}, {Baugh},\ and\
  {Li}}}]{2018MNRAS.479.4824H}%
  \BibitemOpen
  \bibfield  {author} {\bibinfo {author} {\bibfnamefont {C.}~\bibnamefont
  {{Hern{\'a}ndez-Aguayo}}}, \bibinfo {author} {\bibfnamefont {C.~M.}\
  \bibnamefont {{Baugh}}},\ \bibnamefont {and}\ \bibinfo {author}
  {\bibfnamefont {B.}~\bibnamefont {{Li}}},\ }\bibfield  {title} {\bibinfo
  {title} {{Marked clustering statistics in f(R) gravity cosmologies}},\ }\href
  {https://doi.org/10.1093/mnras/sty1822} {\bibfield  {journal} {\bibinfo
  {journal} {MNRAS}\ }\textbf {\bibinfo {volume} {479}},\ \bibinfo {pages}
  {4824} (\bibinfo {year} {2018})},\ \Eprint {https://arxiv.org/abs/1801.08880}
  {arXiv:1801.08880 [astro-ph.CO]} \BibitemShut {NoStop}%
\bibitem [{\citenamefont {{Armijo}}\ \emph {et~al.}(2018)\citenamefont
  {{Armijo}, {Cai}, {Padilla}, {Li},\ and\ {Peacock}}}]{2018MNRAS.478.3627A}%
  \BibitemOpen
  \bibfield  {author} {\bibinfo {author} {\bibfnamefont {J.}~\bibnamefont
  {{Armijo}}}, \bibinfo {author} {\bibfnamefont {Y.-C.}\ \bibnamefont {{Cai}}},
  \bibinfo {author} {\bibfnamefont {N.}~\bibnamefont {{Padilla}}}, \bibinfo
  {author} {\bibfnamefont {B.}~\bibnamefont {{Li}}},\ \bibnamefont {and}\
  \bibinfo {author} {\bibfnamefont {J.~A.}\ \bibnamefont {{Peacock}}},\
  }\bibfield  {title} {\bibinfo {title} {{Testing modified gravity using a
  marked correlation function}},\ }\href
  {https://doi.org/10.1093/mnras/sty1335} {\bibfield  {journal} {\bibinfo
  {journal} {\mnras}\ }\textbf {\bibinfo {volume} {478}},\ \bibinfo {pages}
  {3627} (\bibinfo {year} {2018})},\ \Eprint {https://arxiv.org/abs/1801.08975}
  {arXiv:1801.08975 [astro-ph.CO]} \BibitemShut {NoStop}%
\bibitem [{\citenamefont {{Armijo}}\ \emph {et~al.}(2024)\citenamefont
  {{Armijo}, {Baugh}, {Norberg},\ and\ {Padilla}}}]{2024MNRAS.528.6631A}%
  \BibitemOpen
  \bibfield  {author} {\bibinfo {author} {\bibfnamefont {J.}~\bibnamefont
  {{Armijo}}}, \bibinfo {author} {\bibfnamefont {C.~M.}\ \bibnamefont
  {{Baugh}}}, \bibinfo {author} {\bibfnamefont {P.}~\bibnamefont {{Norberg}}},\
  \bibnamefont {and}\ \bibinfo {author} {\bibfnamefont {N.~D.}\ \bibnamefont
  {{Padilla}}},\ }\bibfield  {title} {\bibinfo {title} {{A new test of gravity
  - II. Application of marked correlation functions to luminous red galaxy
  samples}},\ }\href {https://doi.org/10.1093/mnras/stae449} {\bibfield
  {journal} {\bibinfo  {journal} {\mnras}\ }\textbf {\bibinfo {volume} {528}},\
  \bibinfo {pages} {6631} (\bibinfo {year} {2024})},\ \Eprint
  {https://arxiv.org/abs/2309.09636} {arXiv:2309.09636 [astro-ph.CO]}
  \BibitemShut {NoStop}%
\bibitem [{\citenamefont {Pisani}\ \emph {et~al.}(2019)\citenamefont {Pisani
  et~al.}}]{Pisani:2019cvo}%
  \BibitemOpen
  \bibfield  {author} {\bibinfo {author} {\bibfnamefont {A.}~\bibnamefont
  {Pisani}} \bibnamefont {et~al.},\ }\bibfield  {title} {\bibinfo {title}
  {{Cosmic voids: a novel probe to shed light on our Universe}},\ }\href@noop
  {} {\  (\bibinfo {year} {2019})},\ \Eprint {https://arxiv.org/abs/1903.05161}
  {arXiv:1903.05161 [astro-ph.CO]} \BibitemShut {NoStop}%
\bibitem [{\citenamefont {Massara}\ \emph {et~al.}(2015)\citenamefont {Massara,
  Villaescusa-Navarro, Viel,\ and\ Sutter}}]{Massara_2015}%
  \BibitemOpen
  \bibfield  {author} {\bibinfo {author} {\bibfnamefont {E.}~\bibnamefont
  {Massara}}, \bibinfo {author} {\bibfnamefont {F.}~\bibnamefont
  {Villaescusa-Navarro}}, \bibinfo {author} {\bibfnamefont {M.}~\bibnamefont
  {Viel}},\ \bibnamefont {and}\ \bibinfo {author} {\bibfnamefont
  {P.}~\bibnamefont {Sutter}},\ }\bibfield  {title} {\bibinfo {title} {Voids in
  massive neutrino cosmologies},\ }\href
  {https://doi.org/10.1088/1475-7516/2015/11/018} {\bibfield  {journal}
  {\bibinfo  {journal} {Journal of Cosmology and Astroparticle Physics}\
  }\textbf {\bibinfo {volume} {2015}}\bibinfo  {number} { (11)},\ \bibinfo
  {pages} {018}}\BibitemShut {NoStop}%
\bibitem [{\citenamefont {Kreisch}\ \emph {et~al.}(2019)\citenamefont {Kreisch,
  Pisani, Carbone, Liu, Hawken, Massara, Spergel,\ and\
  Wandelt}}]{10.1093/mnras/stz1944}%
  \BibitemOpen
\bibfield  {number} {  }\bibfield  {author} {\bibinfo {author} {\bibfnamefont
  {C.~D.}\ \bibnamefont {Kreisch}}, \bibinfo {author} {\bibfnamefont
  {A.}~\bibnamefont {Pisani}}, \bibinfo {author} {\bibfnamefont
  {C.}~\bibnamefont {Carbone}}, \bibinfo {author} {\bibfnamefont
  {J.}~\bibnamefont {Liu}}, \bibinfo {author} {\bibfnamefont {A.~J.}\
  \bibnamefont {Hawken}}, \bibinfo {author} {\bibfnamefont {E.}~\bibnamefont
  {Massara}}, \bibinfo {author} {\bibfnamefont {D.~N.}\ \bibnamefont
  {Spergel}},\ \bibnamefont {and}\ \bibinfo {author} {\bibfnamefont {B.~D.}\
  \bibnamefont {Wandelt}},\ }\bibfield  {title} {\bibinfo {title} {{Massive
  neutrinos leave fingerprints on cosmic voids}},\ }\href
  {https://doi.org/10.1093/mnras/stz1944} {\bibfield  {journal} {\bibinfo
  {journal} {Monthly Notices of the Royal Astronomical Society}\ }\textbf
  {\bibinfo {volume} {488}},\ \bibinfo {pages} {4413} (\bibinfo {year}
  {2019})},\ \Eprint
  {https://arxiv.org/abs/https://academic.oup.com/mnras/article-pdf/488/3/4413/29113022/stz1944.pdf}
  {https://academic.oup.com/mnras/article-pdf/488/3/4413/29113022/stz1944.pdf}
  \BibitemShut {NoStop}%
\bibitem [{\citenamefont {Cai}\ \emph {et~al.}(2015)\citenamefont {Cai,
  Padilla,\ and\ Li}}]{10.1093/mnras/stv777}%
  \BibitemOpen
  \bibfield  {author} {\bibinfo {author} {\bibfnamefont {Y.-C.}\ \bibnamefont
  {Cai}}, \bibinfo {author} {\bibfnamefont {N.}~\bibnamefont {Padilla}},\
  \bibnamefont {and}\ \bibinfo {author} {\bibfnamefont {B.}~\bibnamefont
  {Li}},\ }\bibfield  {title} {\bibinfo {title} {{Testing gravity using cosmic
  voids}},\ }\href {https://doi.org/10.1093/mnras/stv777} {\bibfield  {journal}
  {\bibinfo  {journal} {Monthly Notices of the Royal Astronomical Society}\
  }\textbf {\bibinfo {volume} {451}},\ \bibinfo {pages} {1036} (\bibinfo {year}
  {2015})},\ \Eprint
  {https://arxiv.org/abs/https://academic.oup.com/mnras/article-pdf/451/1/1036/4166730/stv777.pdf}
  {https://academic.oup.com/mnras/article-pdf/451/1/1036/4166730/stv777.pdf}
  \BibitemShut {NoStop}%
\bibitem [{\citenamefont {Hamaus}\ \emph {et~al.}(2015)\citenamefont {Hamaus,
  Sutter, Lavaux,\ and\ Wandelt}}]{Hamaus_2015}%
  \BibitemOpen
  \bibfield  {author} {\bibinfo {author} {\bibfnamefont {N.}~\bibnamefont
  {Hamaus}}, \bibinfo {author} {\bibfnamefont {P.}~\bibnamefont {Sutter}},
  \bibinfo {author} {\bibfnamefont {G.}~\bibnamefont {Lavaux}},\ \bibnamefont
  {and}\ \bibinfo {author} {\bibfnamefont {B.~D.}\ \bibnamefont {Wandelt}},\
  }\bibfield  {title} {\bibinfo {title} {Probing cosmology and gravity with
  redshift-space distortions around voids},\ }\href
  {https://doi.org/10.1088/1475-7516/2015/11/036} {\bibfield  {journal}
  {\bibinfo  {journal} {Journal of Cosmology and Astroparticle Physics}\
  }\textbf {\bibinfo {volume} {2015}}\bibinfo  {number} { (11)},\ \bibinfo
  {pages} {036}}\BibitemShut {NoStop}%
\bibitem [{\citenamefont {Kreisch}\ \emph {et~al.}(2021)\citenamefont {Kreisch,
  Pisani, Villaescusa-Navarro, Spergel, Wandelt, Hamaus,\ and\
  Bayer}}]{Kreisch:2021xzq}%
  \BibitemOpen
\bibfield  {number} {  }\bibfield  {author} {\bibinfo {author} {\bibfnamefont
  {C.~D.}\ \bibnamefont {Kreisch}}, \bibinfo {author} {\bibfnamefont
  {A.}~\bibnamefont {Pisani}}, \bibinfo {author} {\bibfnamefont
  {F.}~\bibnamefont {Villaescusa-Navarro}}, \bibinfo {author} {\bibfnamefont
  {D.~N.}\ \bibnamefont {Spergel}}, \bibinfo {author} {\bibfnamefont {B.~D.}\
  \bibnamefont {Wandelt}}, \bibinfo {author} {\bibfnamefont {N.}~\bibnamefont
  {Hamaus}},\ \bibnamefont {and}\ \bibinfo {author} {\bibfnamefont {A.~E.}\
  \bibnamefont {Bayer}},\ }\bibfield  {title} {\bibinfo {title} {{The GIGANTES
  dataset: precision cosmology from voids in the machine learning era}},\
  }\href@noop {} {\  (\bibinfo {year} {2021})},\ \Eprint
  {https://arxiv.org/abs/2107.02304} {arXiv:2107.02304 [astro-ph.CO]}
  \BibitemShut {NoStop}%
\bibitem [{\citenamefont {Bonnaire}\ \emph {et~al.}(2021)\citenamefont
  {Bonnaire, Aghanim, Kuruvilla,\ and\ Decelle}}]{Bonnaire:2021sie}%
  \BibitemOpen
  \bibfield  {author} {\bibinfo {author} {\bibfnamefont {T.}~\bibnamefont
  {Bonnaire}}, \bibinfo {author} {\bibfnamefont {N.}~\bibnamefont {Aghanim}},
  \bibinfo {author} {\bibfnamefont {J.}~\bibnamefont {Kuruvilla}},\
  \bibnamefont {and}\ \bibinfo {author} {\bibfnamefont {A.}~\bibnamefont
  {Decelle}},\ }\bibfield  {title} {\bibinfo {title} {{Cosmology with cosmic
  web environments I. Real-space power spectra}},\ }\href@noop {} {\  (\bibinfo
  {year} {2021})},\ \Eprint {https://arxiv.org/abs/2112.03926}
  {arXiv:2112.03926 [astro-ph.CO]} \BibitemShut {NoStop}%
\bibitem [{\citenamefont {{Radinovi{\'c}}}\ \emph {et~al.}(2023)\citenamefont
  {{Radinovi{\'c}}, {Nadathur}, {Winther}, {Percival}, {Woodfinden}, {Massara},
  {Paillas}, {Contarini}, {Hamaus}, {Kovacs} et~al.}}]{2023arXiv230205302R}%
  \BibitemOpen
  \bibfield  {author} {\bibinfo {author} {\bibfnamefont {S.}~\bibnamefont
  {{Radinovi{\'c}}}}, \bibinfo {author} {\bibfnamefont {S.}~\bibnamefont
  {{Nadathur}}}, \bibinfo {author} {\bibfnamefont {H.~A.}\ \bibnamefont
  {{Winther}}}, \bibinfo {author} {\bibfnamefont {W.~J.}\ \bibnamefont
  {{Percival}}}, \bibinfo {author} {\bibfnamefont {A.}~\bibnamefont
  {{Woodfinden}}}, \bibinfo {author} {\bibfnamefont {E.}~\bibnamefont
  {{Massara}}}, \bibinfo {author} {\bibfnamefont {E.}~\bibnamefont
  {{Paillas}}}, \bibinfo {author} {\bibfnamefont {S.}~\bibnamefont
  {{Contarini}}}, \bibinfo {author} {\bibfnamefont {N.}~\bibnamefont
  {{Hamaus}}}, \bibinfo {author} {\bibfnamefont {A.}~\bibnamefont {{Kovacs}}},
  \bibinfo {author} {\bibfnamefont {A.}~\bibnamefont {{Pisani}}}, \bibinfo
  {author} {\bibfnamefont {G.}~\bibnamefont {{Verza}}}, \bibinfo {author}
  {\bibfnamefont {M.}~\bibnamefont {{Aubert}}}, \bibinfo {author}
  {\bibfnamefont {A.}~\bibnamefont {{Amara}}}, \bibinfo {author} {\bibfnamefont
  {N.}~\bibnamefont {{Auricchio}}}, \bibinfo {author} {\bibfnamefont
  {M.}~\bibnamefont {{Baldi}}}, \bibnamefont {et~al.},\ }\bibfield  {title}
  {\bibinfo {title} {{Euclid: Cosmology forecasts from the void-galaxy
  cross-correlation function with reconstruction}},\ }\href
  {https://doi.org/10.48550/arXiv.2302.05302} {\bibfield  {journal} {\bibinfo
  {journal} {arXiv e-prints}\ ,\ \bibinfo {eid} {arXiv:2302.05302}} (\bibinfo
  {year} {2023})},\ \Eprint {https://arxiv.org/abs/2302.05302}
  {arXiv:2302.05302 [astro-ph.CO]} \BibitemShut {NoStop}%
\bibitem [{\citenamefont {Neyrinck}\ \emph {et~al.}(2009)\citenamefont
  {Neyrinck, Szapudi,\ and\ Szalay}}]{Neyrinck_2009}%
  \BibitemOpen
  \bibfield  {author} {\bibinfo {author} {\bibfnamefont {M.~C.}\ \bibnamefont
  {Neyrinck}}, \bibinfo {author} {\bibfnamefont {I.}~\bibnamefont {Szapudi}},\
  \bibnamefont {and}\ \bibinfo {author} {\bibfnamefont {A.~S.}\ \bibnamefont
  {Szalay}},\ }\bibfield  {title} {\bibinfo {title} {{REJUVENATING} {THE}
  {MATTER} {POWER} {SPECTRUM}: {RESTORING} {INFORMATION} {WITH} a {LOGARITHMIC}
  {DENSITY} {MAPPING}},\ }\href {https://doi.org/10.1088/0004-637x/698/2/l90}
  {\bibfield  {journal} {\bibinfo  {journal} {The Astrophysical Journal}\
  }\textbf {\bibinfo {volume} {698}},\ \bibinfo {pages} {L90} (\bibinfo {year}
  {2009})}\BibitemShut {NoStop}%
\bibitem [{\citenamefont {Simpson}\ \emph {et~al.}(2011)\citenamefont {Simpson,
  James, Heavens,\ and\ Heymans}}]{PhysRevLett.107.271301}%
  \BibitemOpen
  \bibfield  {author} {\bibinfo {author} {\bibfnamefont {F.}~\bibnamefont
  {Simpson}}, \bibinfo {author} {\bibfnamefont {J.~B.}\ \bibnamefont {James}},
  \bibinfo {author} {\bibfnamefont {A.~F.}\ \bibnamefont {Heavens}},\
  \bibnamefont {and}\ \bibinfo {author} {\bibfnamefont {C.}~\bibnamefont
  {Heymans}},\ }\bibfield  {title} {\bibinfo {title} {Clipping the cosmos: The
  bias and bispectrum of large scale structure},\ }\href
  {https://doi.org/10.1103/PhysRevLett.107.271301} {\bibfield  {journal}
  {\bibinfo  {journal} {Phys. Rev. Lett.}\ }\textbf {\bibinfo {volume} {107}},\
  \bibinfo {pages} {271301} (\bibinfo {year} {2011})}\BibitemShut {NoStop}%
\bibitem [{\citenamefont {Massara}\ \emph {et~al.}(2021)\citenamefont {Massara,
  Villaescusa-Navarro, Ho, Dalal,\ and\ Spergel}}]{PhysRevLett.126.011301}%
  \BibitemOpen
  \bibfield  {author} {\bibinfo {author} {\bibfnamefont {E.}~\bibnamefont
  {Massara}}, \bibinfo {author} {\bibfnamefont {F.}~\bibnamefont
  {Villaescusa-Navarro}}, \bibinfo {author} {\bibfnamefont {S.}~\bibnamefont
  {Ho}}, \bibinfo {author} {\bibfnamefont {N.}~\bibnamefont {Dalal}},\
  \bibnamefont {and}\ \bibinfo {author} {\bibfnamefont {D.~N.}\ \bibnamefont
  {Spergel}},\ }\bibfield  {title} {\bibinfo {title} {Using the marked power
  spectrum to detect the signature of neutrinos in large-scale structure},\
  }\href {https://doi.org/10.1103/PhysRevLett.126.011301} {\bibfield  {journal}
  {\bibinfo  {journal} {Phys. Rev. Lett.}\ }\textbf {\bibinfo {volume} {126}},\
  \bibinfo {pages} {011301} (\bibinfo {year} {2021})}\BibitemShut {NoStop}%
\bibitem [{\citenamefont {{Massara}}\ \emph {et~al.}(2022)\citenamefont
  {{Massara}, {Villaescusa-Navarro}, {Hahn}, {Abidi}, {Eickenberg}, {Ho},
  {Lemos}, {Moradinezhad Dizgah},\ and\ {R{\'e}galdo-Saint
  Blancard}}}]{2022arXiv220601709M}%
  \BibitemOpen
  \bibfield  {author} {\bibinfo {author} {\bibfnamefont {E.}~\bibnamefont
  {{Massara}}}, \bibinfo {author} {\bibfnamefont {F.}~\bibnamefont
  {{Villaescusa-Navarro}}}, \bibinfo {author} {\bibfnamefont {C.}~\bibnamefont
  {{Hahn}}}, \bibinfo {author} {\bibfnamefont {M.~M.}\ \bibnamefont {{Abidi}}},
  \bibinfo {author} {\bibfnamefont {M.}~\bibnamefont {{Eickenberg}}}, \bibinfo
  {author} {\bibfnamefont {S.}~\bibnamefont {{Ho}}}, \bibinfo {author}
  {\bibfnamefont {P.}~\bibnamefont {{Lemos}}}, \bibinfo {author} {\bibfnamefont
  {A.}~\bibnamefont {{Moradinezhad Dizgah}}},\ \bibnamefont {and}\ \bibinfo
  {author} {\bibfnamefont {B.}~\bibnamefont {{R{\'e}galdo-Saint Blancard}}},\
  }\bibfield  {title} {\bibinfo {title} {{Cosmological Information in the
  Marked Power Spectrum of the Galaxy Field}},\ }\href
  {https://doi.org/10.48550/arXiv.2206.01709} {\bibfield  {journal} {\bibinfo
  {journal} {arXiv e-prints}\ ,\ \bibinfo {eid} {arXiv:2206.01709}} (\bibinfo
  {year} {2022})},\ \Eprint {https://arxiv.org/abs/2206.01709}
  {arXiv:2206.01709 [astro-ph.CO]} \BibitemShut {NoStop}%
\bibitem [{\citenamefont {{Massara}}\ \emph {et~al.}(2024)\citenamefont
  {{Massara}, {Hahn}, {Eickenberg}, {Ho}, {Hou}, {Lemos}, {Modi}, {Moradinezhad
  Dizgah}, {Parker},\ and\ {R{\'e}galdo-Saint
  Blancard}}}]{2024arXiv240404228M}%
  \BibitemOpen
  \bibfield  {author} {\bibinfo {author} {\bibfnamefont {E.}~\bibnamefont
  {{Massara}}}, \bibinfo {author} {\bibfnamefont {C.}~\bibnamefont {{Hahn}}},
  \bibinfo {author} {\bibfnamefont {M.}~\bibnamefont {{Eickenberg}}}, \bibinfo
  {author} {\bibfnamefont {S.}~\bibnamefont {{Ho}}}, \bibinfo {author}
  {\bibfnamefont {J.}~\bibnamefont {{Hou}}}, \bibinfo {author} {\bibfnamefont
  {P.}~\bibnamefont {{Lemos}}}, \bibinfo {author} {\bibfnamefont
  {C.}~\bibnamefont {{Modi}}}, \bibinfo {author} {\bibfnamefont
  {A.}~\bibnamefont {{Moradinezhad Dizgah}}}, \bibinfo {author} {\bibfnamefont
  {L.}~\bibnamefont {{Parker}}},\ \bibnamefont {and}\ \bibinfo {author}
  {\bibfnamefont {B.}~\bibnamefont {{R{\'e}galdo-Saint Blancard}}},\ }\bibfield
   {title} {\bibinfo {title} {"simbig: Cosmological constraints using
  simulation-based inference of galaxy clustering with marked power spectra"},\
  }\href {https://doi.org/10.48550/arXiv.2404.04228} {\bibfield  {journal}
  {\bibinfo  {journal} {arXiv e-prints}\ ,\ \bibinfo {eid} {arXiv:2404.04228}}
  (\bibinfo {year} {2024})},\ \Eprint {https://arxiv.org/abs/2404.04228}
  {arXiv:2404.04228 [astro-ph.CO]} \BibitemShut {NoStop}%
\bibitem [{\citenamefont {Schmalzing}\ \emph {et~al.}(1999)\citenamefont
  {Schmalzing, Gottlöber, Klypin,\ and\
  Kravtsov}}]{10.1046/j.1365-8711.1999.02912.x}%
  \BibitemOpen
  \bibfield  {author} {\bibinfo {author} {\bibfnamefont {J.}~\bibnamefont
  {Schmalzing}}, \bibinfo {author} {\bibfnamefont {S.}~\bibnamefont
  {Gottlöber}}, \bibinfo {author} {\bibfnamefont {A.~A.}\ \bibnamefont
  {Klypin}},\ \bibnamefont {and}\ \bibinfo {author} {\bibfnamefont {A.~V.}\
  \bibnamefont {Kravtsov}},\ }\bibfield  {title} {\bibinfo {title}
  {{Quantifying the evolution of higher order clustering}},\ }\href
  {https://doi.org/10.1046/j.1365-8711.1999.02912.x} {\bibfield  {journal}
  {\bibinfo  {journal} {Monthly Notices of the Royal Astronomical Society}\
  }\textbf {\bibinfo {volume} {309}},\ \bibinfo {pages} {1007} (\bibinfo {year}
  {1999})},\ \Eprint
  {https://arxiv.org/abs/https://academic.oup.com/mnras/article-pdf/309/4/1007/3785859/309-4-1007.pdf}
  {https://academic.oup.com/mnras/article-pdf/309/4/1007/3785859/309-4-1007.pdf}
  \BibitemShut {NoStop}%
\bibitem [{\citenamefont {Pratten}\ and\ \citenamefont
  {Munshi}(2012)\citenamefont {Pratten\ and\
  Munshi}}]{10.1111/j.1365-2966.2012.21103.x}%
  \BibitemOpen
  \bibfield  {author} {\bibinfo {author} {\bibfnamefont {G.}~\bibnamefont
  {Pratten}}\ \bibnamefont {and}\ \bibinfo {author} {\bibfnamefont
  {D.}~\bibnamefont {Munshi}},\ }\bibfield  {title} {\bibinfo {title}
  {{Non-Gaussianity in large-scale structure and Minkowski functionals}},\
  }\href {https://doi.org/10.1111/j.1365-2966.2012.21103.x} {\bibfield
  {journal} {\bibinfo  {journal} {Monthly Notices of the Royal Astronomical
  Society}\ }\textbf {\bibinfo {volume} {423}},\ \bibinfo {pages} {3209}
  (\bibinfo {year} {2012})},\ \Eprint
  {https://arxiv.org/abs/https://academic.oup.com/mnras/article-pdf/423/4/3209/4885960/mnras0423-3209.pdf}
  {https://academic.oup.com/mnras/article-pdf/423/4/3209/4885960/mnras0423-3209.pdf}
  \BibitemShut {NoStop}%
\bibitem [{\citenamefont {Codis}\ \emph {et~al.}(2013)\citenamefont {Codis,
  Pichon, Pogosyan, Bernardeau,\ and\ Matsubara}}]{10.1093/mnras/stt1316}%
  \BibitemOpen
  \bibfield  {author} {\bibinfo {author} {\bibfnamefont {S.}~\bibnamefont
  {Codis}}, \bibinfo {author} {\bibfnamefont {C.}~\bibnamefont {Pichon}},
  \bibinfo {author} {\bibfnamefont {D.}~\bibnamefont {Pogosyan}}, \bibinfo
  {author} {\bibfnamefont {F.}~\bibnamefont {Bernardeau}},\ \bibnamefont {and}\
  \bibinfo {author} {\bibfnamefont {T.}~\bibnamefont {Matsubara}},\ }\bibfield
  {title} {\bibinfo {title} {{Non-Gaussian Minkowski functionals and extrema
  counts in redshift space}},\ }\href {https://doi.org/10.1093/mnras/stt1316}
  {\bibfield  {journal} {\bibinfo  {journal} {Monthly Notices of the Royal
  Astronomical Society}\ }\textbf {\bibinfo {volume} {435}},\ \bibinfo {pages}
  {531} (\bibinfo {year} {2013})},\ \Eprint
  {https://arxiv.org/abs/https://academic.oup.com/mnras/article-pdf/435/1/531/3888130/stt1316.pdf}
  {https://academic.oup.com/mnras/article-pdf/435/1/531/3888130/stt1316.pdf}
  \BibitemShut {NoStop}%
\bibitem [{\citenamefont {Fang}\ \emph {et~al.}(2017)\citenamefont {Fang, Li,\
  and\ Zhao}}]{PhysRevLett.118.181301}%
  \BibitemOpen
  \bibfield  {author} {\bibinfo {author} {\bibfnamefont {W.}~\bibnamefont
  {Fang}}, \bibinfo {author} {\bibfnamefont {B.}~\bibnamefont {Li}},\
  \bibnamefont {and}\ \bibinfo {author} {\bibfnamefont {G.-B.}\ \bibnamefont
  {Zhao}},\ }\bibfield  {title} {\bibinfo {title} {New probe of departures from
  general relativity using minkowski functionals},\ }\href
  {https://doi.org/10.1103/PhysRevLett.118.181301} {\bibfield  {journal}
  {\bibinfo  {journal} {Phys. Rev. Lett.}\ }\textbf {\bibinfo {volume} {118}},\
  \bibinfo {pages} {181301} (\bibinfo {year} {2017})}\BibitemShut {NoStop}%
\bibitem [{\citenamefont {{Liu}}\ \emph {et~al.}(2023)\citenamefont {{Liu},
  {Jiang},\ and\ {Fang}}}]{2023arXiv230208162L}%
  \BibitemOpen
  \bibfield  {author} {\bibinfo {author} {\bibfnamefont {W.}~\bibnamefont
  {{Liu}}}, \bibinfo {author} {\bibfnamefont {A.}~\bibnamefont {{Jiang}}},\
  \bibnamefont {and}\ \bibinfo {author} {\bibfnamefont {W.}~\bibnamefont
  {{Fang}}},\ }\bibfield  {title} {\bibinfo {title} {{Probing massive neutrinos
  with the Minkowski functionals of the galaxy distribution}},\ }\href
  {https://doi.org/10.48550/arXiv.2302.08162} {\bibfield  {journal} {\bibinfo
  {journal} {arXiv e-prints}\ ,\ \bibinfo {eid} {arXiv:2302.08162}} (\bibinfo
  {year} {2023})},\ \Eprint {https://arxiv.org/abs/2302.08162}
  {arXiv:2302.08162 [astro-ph.CO]} \BibitemShut {NoStop}%
\bibitem [{\citenamefont {{Yip}}\ \emph {et~al.}(2024)\citenamefont {{Yip},
  {Biagetti}, {Cole}, {Viswanathan},\ and\ {Shiu}}}]{2024arXiv240313985Y}%
  \BibitemOpen
  \bibfield  {author} {\bibinfo {author} {\bibfnamefont {J.~H.~T.}\
  \bibnamefont {{Yip}}}, \bibinfo {author} {\bibfnamefont {M.}~\bibnamefont
  {{Biagetti}}}, \bibinfo {author} {\bibfnamefont {A.}~\bibnamefont {{Cole}}},
  \bibinfo {author} {\bibfnamefont {K.}~\bibnamefont {{Viswanathan}}},\
  \bibnamefont {and}\ \bibinfo {author} {\bibfnamefont {G.}~\bibnamefont
  {{Shiu}}},\ }\bibfield  {title} {\bibinfo {title} {{Cosmology with Persistent
  Homology: a Fisher Forecast}},\ }\href
  {https://doi.org/10.48550/arXiv.2403.13985} {\bibfield  {journal} {\bibinfo
  {journal} {arXiv e-prints}\ ,\ \bibinfo {eid} {arXiv:2403.13985}} (\bibinfo
  {year} {2024})},\ \Eprint {https://arxiv.org/abs/2403.13985}
  {arXiv:2403.13985 [astro-ph.CO]} \BibitemShut {NoStop}%
\bibitem [{\citenamefont {Paillas}\ \emph {et~al.}(2021)\citenamefont {Paillas,
  Cai, Padilla,\ and\ Sánchez}}]{10.1093/mnras/stab1654}%
  \BibitemOpen
  \bibfield  {author} {\bibinfo {author} {\bibfnamefont {E.}~\bibnamefont
  {Paillas}}, \bibinfo {author} {\bibfnamefont {Y.-C.}\ \bibnamefont {Cai}},
  \bibinfo {author} {\bibfnamefont {N.}~\bibnamefont {Padilla}},\ \bibnamefont
  {and}\ \bibinfo {author} {\bibfnamefont {A.~G.}\ \bibnamefont {Sánchez}},\
  }\bibfield  {title} {\bibinfo {title} {{Redshift-space distortions with split
  densities}},\ }\href {https://doi.org/10.1093/mnras/stab1654} {\bibfield
  {journal} {\bibinfo  {journal} {Monthly Notices of the Royal Astronomical
  Society}\ }\textbf {\bibinfo {volume} {505}},\ \bibinfo {pages} {5731}
  (\bibinfo {year} {2021})},\ \Eprint
  {https://arxiv.org/abs/https://academic.oup.com/mnras/article-pdf/505/4/5731/38864518/stab1654.pdf}
  {https://academic.oup.com/mnras/article-pdf/505/4/5731/38864518/stab1654.pdf}
  \BibitemShut {NoStop}%
\bibitem [{\citenamefont {{Paillas}}\ \emph {et~al.}(2022)\citenamefont
  {{Paillas}, {Cuesta-Lazaro}, {Zarrouk}, {Cai}, {Percival}, {Nadathur},
  {Pinon}, {de Mattia},\ and\ {Beutler}}}]{2022arXiv220904310P}%
  \BibitemOpen
  \bibfield  {author} {\bibinfo {author} {\bibfnamefont {E.}~\bibnamefont
  {{Paillas}}}, \bibinfo {author} {\bibfnamefont {C.}~\bibnamefont
  {{Cuesta-Lazaro}}}, \bibinfo {author} {\bibfnamefont {P.}~\bibnamefont
  {{Zarrouk}}}, \bibinfo {author} {\bibfnamefont {Y.-C.}\ \bibnamefont
  {{Cai}}}, \bibinfo {author} {\bibfnamefont {W.~J.}\ \bibnamefont
  {{Percival}}}, \bibinfo {author} {\bibfnamefont {S.}~\bibnamefont
  {{Nadathur}}}, \bibinfo {author} {\bibfnamefont {M.}~\bibnamefont {{Pinon}}},
  \bibinfo {author} {\bibfnamefont {A.}~\bibnamefont {{de Mattia}}},\
  \bibnamefont {and}\ \bibinfo {author} {\bibfnamefont {F.}~\bibnamefont
  {{Beutler}}},\ }\bibfield  {title} {\bibinfo {title} {{Constraining $\nu
  \Lambda$CDM with density-split clustering}},\ }\href
  {https://doi.org/10.48550/arXiv.2209.04310} {\bibfield  {journal} {\bibinfo
  {journal} {arXiv e-prints}\ ,\ \bibinfo {eid} {arXiv:2209.04310}} (\bibinfo
  {year} {2022})},\ \Eprint {https://arxiv.org/abs/2209.04310}
  {arXiv:2209.04310 [astro-ph.CO]} \BibitemShut {NoStop}%
\bibitem [{\citenamefont {Bayer}\ \emph {et~al.}(2021)\citenamefont {Bayer,
  Villaescusa-Navarro, Massara, Liu, Spergel, Verde, Wandelt, Viel,\ and\
  Ho}}]{Bayer:2021iyb}%
  \BibitemOpen
  \bibfield  {author} {\bibinfo {author} {\bibfnamefont {A.~E.}\ \bibnamefont
  {Bayer}}, \bibinfo {author} {\bibfnamefont {F.}~\bibnamefont
  {Villaescusa-Navarro}}, \bibinfo {author} {\bibfnamefont {E.}~\bibnamefont
  {Massara}}, \bibinfo {author} {\bibfnamefont {J.}~\bibnamefont {Liu}},
  \bibinfo {author} {\bibfnamefont {D.~N.}\ \bibnamefont {Spergel}}, \bibinfo
  {author} {\bibfnamefont {L.}~\bibnamefont {Verde}}, \bibinfo {author}
  {\bibfnamefont {B.}~\bibnamefont {Wandelt}}, \bibinfo {author} {\bibfnamefont
  {M.}~\bibnamefont {Viel}},\ \bibnamefont {and}\ \bibinfo {author}
  {\bibfnamefont {S.}~\bibnamefont {Ho}},\ }\bibfield  {title} {\bibinfo
  {title} {{Detecting neutrino mass by combining matter clustering, halos, and
  voids}},\ }\href@noop {} {\  (\bibinfo {year} {2021})},\ \Eprint
  {https://arxiv.org/abs/2102.05049} {arXiv:2102.05049 [astro-ph.CO]}
  \BibitemShut {NoStop}%
\bibitem [{\citenamefont {{Paillas}}\ \emph {et~al.}(2023)\citenamefont
  {{Paillas}, {Cuesta-Lazaro}, {Percival}, {Nadathur}, {Cai}, {Yuan},
  {Beutler}, {de Mattia}, {Eisenstein}, {Forero-Sanchez}
  et~al.}}]{Paillas2023:2309.16541}%
  \BibitemOpen
  \bibfield  {author} {\bibinfo {author} {\bibfnamefont {E.}~\bibnamefont
  {{Paillas}}}, \bibinfo {author} {\bibfnamefont {C.}~\bibnamefont
  {{Cuesta-Lazaro}}}, \bibinfo {author} {\bibfnamefont {W.~J.}\ \bibnamefont
  {{Percival}}}, \bibinfo {author} {\bibfnamefont {S.}~\bibnamefont
  {{Nadathur}}}, \bibinfo {author} {\bibfnamefont {Y.-C.}\ \bibnamefont
  {{Cai}}}, \bibinfo {author} {\bibfnamefont {S.}~\bibnamefont {{Yuan}}},
  \bibinfo {author} {\bibfnamefont {F.}~\bibnamefont {{Beutler}}}, \bibinfo
  {author} {\bibfnamefont {A.}~\bibnamefont {{de Mattia}}}, \bibinfo {author}
  {\bibfnamefont {D.}~\bibnamefont {{Eisenstein}}}, \bibinfo {author}
  {\bibfnamefont {D.}~\bibnamefont {{Forero-Sanchez}}}, \bibinfo {author}
  {\bibfnamefont {N.}~\bibnamefont {{Padilla}}}, \bibinfo {author}
  {\bibfnamefont {M.}~\bibnamefont {{Pinon}}}, \bibinfo {author} {\bibfnamefont
  {V.}~\bibnamefont {{Ruhlmann-Kleider}}}, \bibinfo {author} {\bibfnamefont
  {A.~G.}\ \bibnamefont {{S{\'a}nchez}}}, \bibinfo {author} {\bibfnamefont
  {G.}~\bibnamefont {{Valogiannis}}},\ \bibnamefont {and}\ \bibinfo {author}
  {\bibfnamefont {P.}~\bibnamefont {{Zarrouk}}},\ }\bibfield  {title} {\bibinfo
  {title} {{Cosmological constraints from density-split clustering in the BOSS
  CMASS galaxy sample}},\ }\href@noop {} {\bibfield  {journal} {\bibinfo
  {journal} {arXiv e-prints}\ ,\ \bibinfo {eid} {arXiv:2309.16541}} (\bibinfo
  {year} {2023})},\ \Eprint {https://arxiv.org/abs/2309.16541}
  {arXiv:2309.16541 [astro-ph.CO]} \BibitemShut {NoStop}%
\bibitem [{\citenamefont {{Cuesta-Lazaro}}\ \emph {et~al.}(2023)\citenamefont
  {{Cuesta-Lazaro}, {Paillas}, {Yuan}, {Cai}, {Nadathur}, {Percival},
  {Beutler}, {de Mattia}, {Eisenstein}, {Forero-Sanchez}
  et~al.}}]{Cuesta-Lazaro2023:2309.16539}%
  \BibitemOpen
  \bibfield  {author} {\bibinfo {author} {\bibfnamefont {C.}~\bibnamefont
  {{Cuesta-Lazaro}}}, \bibinfo {author} {\bibfnamefont {E.}~\bibnamefont
  {{Paillas}}}, \bibinfo {author} {\bibfnamefont {S.}~\bibnamefont {{Yuan}}},
  \bibinfo {author} {\bibfnamefont {Y.-C.}\ \bibnamefont {{Cai}}}, \bibinfo
  {author} {\bibfnamefont {S.}~\bibnamefont {{Nadathur}}}, \bibinfo {author}
  {\bibfnamefont {W.~J.}\ \bibnamefont {{Percival}}}, \bibinfo {author}
  {\bibfnamefont {F.}~\bibnamefont {{Beutler}}}, \bibinfo {author}
  {\bibfnamefont {A.}~\bibnamefont {{de Mattia}}}, \bibinfo {author}
  {\bibfnamefont {D.}~\bibnamefont {{Eisenstein}}}, \bibinfo {author}
  {\bibfnamefont {D.}~\bibnamefont {{Forero-Sanchez}}}, \bibinfo {author}
  {\bibfnamefont {N.}~\bibnamefont {{Padilla}}}, \bibinfo {author}
  {\bibfnamefont {M.}~\bibnamefont {{Pinon}}}, \bibinfo {author} {\bibfnamefont
  {V.}~\bibnamefont {{Ruhlmann-Kleider}}}, \bibinfo {author} {\bibfnamefont
  {A.~G.}\ \bibnamefont {{S{\'a}nchez}}}, \bibinfo {author} {\bibfnamefont
  {G.}~\bibnamefont {{Valogiannis}}},\ \bibnamefont {and}\ \bibinfo {author}
  {\bibfnamefont {P.}~\bibnamefont {{Zarrouk}}},\ }\bibfield  {title} {\bibinfo
  {title} {{SUNBIRD: A simulation-based model for full-shape density-split
  clustering}},\ }\href@noop {} {\bibfield  {journal} {\bibinfo  {journal}
  {arXiv e-prints}\ ,\ \bibinfo {eid} {arXiv:2309.16539}} (\bibinfo {year}
  {2023})},\ \Eprint {https://arxiv.org/abs/2309.16539} {arXiv:2309.16539
  [astro-ph.CO]} \BibitemShut {NoStop}%
\bibitem [{\citenamefont {Schmittfull}\ \emph {et~al.}(2015)\citenamefont
  {Schmittfull, Baldauf,\ and\ Seljak}}]{PhysRevD.91.043530}%
  \BibitemOpen
  \bibfield  {author} {\bibinfo {author} {\bibfnamefont {M.}~\bibnamefont
  {Schmittfull}}, \bibinfo {author} {\bibfnamefont {T.}~\bibnamefont
  {Baldauf}},\ \bibnamefont {and}\ \bibinfo {author} {\bibfnamefont {U.~c.~v.}\
  \bibnamefont {Seljak}},\ }\bibfield  {title} {\bibinfo {title} {Near optimal
  bispectrum estimators for large-scale structure},\ }\href
  {https://doi.org/10.1103/PhysRevD.91.043530} {\bibfield  {journal} {\bibinfo
  {journal} {Phys. Rev. D}\ }\textbf {\bibinfo {volume} {91}},\ \bibinfo
  {pages} {043530} (\bibinfo {year} {2015})}\BibitemShut {NoStop}%
\bibitem [{\citenamefont {Peel}\ \emph {et~al.}(2018)\citenamefont {Peel,
  Pettorino, Giocoli, Starck,\ and\ Baldi}}]{Peel:2018aly}%
  \BibitemOpen
  \bibfield  {author} {\bibinfo {author} {\bibfnamefont {A.}~\bibnamefont
  {Peel}}, \bibinfo {author} {\bibfnamefont {V.}~\bibnamefont {Pettorino}},
  \bibinfo {author} {\bibfnamefont {C.}~\bibnamefont {Giocoli}}, \bibinfo
  {author} {\bibfnamefont {J.-L.}\ \bibnamefont {Starck}},\ \bibnamefont {and}\
  \bibinfo {author} {\bibfnamefont {M.}~\bibnamefont {Baldi}},\ }\bibfield
  {title} {\bibinfo {title} {{Breaking degeneracies in modified gravity with
  higher (than 2nd) order weak-lensing statistics}},\ }\href
  {https://doi.org/10.1051/0004-6361/201833481} {\bibfield  {journal} {\bibinfo
   {journal} {Astron. Astrophys.}\ }\textbf {\bibinfo {volume} {619}},\
  \bibinfo {pages} {A38} (\bibinfo {year} {2018})},\ \Eprint
  {https://arxiv.org/abs/1805.05146} {arXiv:1805.05146 [astro-ph.CO]}
  \BibitemShut {NoStop}%
\bibitem [{\citenamefont {Dizgah}\ \emph {et~al.}(2020)\citenamefont {Dizgah,
  Lee, Schmittfull,\ and\ Dvorkin}}]{Dizgah_2020}%
  \BibitemOpen
  \bibfield  {author} {\bibinfo {author} {\bibfnamefont {A.~M.}\ \bibnamefont
  {Dizgah}}, \bibinfo {author} {\bibfnamefont {H.}~\bibnamefont {Lee}},
  \bibinfo {author} {\bibfnamefont {M.}~\bibnamefont {Schmittfull}},\
  \bibnamefont {and}\ \bibinfo {author} {\bibfnamefont {C.}~\bibnamefont
  {Dvorkin}},\ }\bibfield  {title} {\bibinfo {title} {Capturing non-gaussianity
  of the large-scale structure with weighted skew-spectra},\ }\href
  {https://doi.org/10.1088/1475-7516/2020/04/011} {\bibfield  {journal}
  {\bibinfo  {journal} {Journal of Cosmology and Astroparticle Physics}\
  }\textbf {\bibinfo {volume} {2020}}\bibinfo  {number} { (04)},\ \bibinfo
  {pages} {011}}\BibitemShut {NoStop}%
\bibitem [{\citenamefont {Chakraborty}\ \emph {et~al.}(2022)\citenamefont
  {Chakraborty, Chen,\ and\ Dvorkin}}]{Chakraborty:2022aok}%
  \BibitemOpen
\bibfield  {number} {  }\bibfield  {author} {\bibinfo {author} {\bibfnamefont
  {P.}~\bibnamefont {Chakraborty}}, \bibinfo {author} {\bibfnamefont {S.-F.}\
  \bibnamefont {Chen}},\ \bibnamefont {and}\ \bibinfo {author} {\bibfnamefont
  {C.}~\bibnamefont {Dvorkin}},\ }\bibfield  {title} {\bibinfo {title}
  {{Skewing the CMB$\times$LSS: a Fast Method for Bispectrum Analysis}},\
  }\href@noop {} {\  (\bibinfo {year} {2022})},\ \Eprint
  {https://arxiv.org/abs/2202.11724} {arXiv:2202.11724 [astro-ph.CO]}
  \BibitemShut {NoStop}%
\bibitem [{\citenamefont {{Hou}}\ \emph {et~al.}(2023)\citenamefont {{Hou},
  {Moradinezhad Dizgah}, {Hahn},\ and\ {Massara}}}]{2023JCAP...03..045H}%
  \BibitemOpen
  \bibfield  {author} {\bibinfo {author} {\bibfnamefont {J.}~\bibnamefont
  {{Hou}}}, \bibinfo {author} {\bibfnamefont {A.}~\bibnamefont {{Moradinezhad
  Dizgah}}}, \bibinfo {author} {\bibfnamefont {C.}~\bibnamefont {{Hahn}}},\
  \bibnamefont {and}\ \bibinfo {author} {\bibfnamefont {E.}~\bibnamefont
  {{Massara}}},\ }\bibfield  {title} {\bibinfo {title} {{Cosmological
  information in skew spectra of biased tracers in redshift space}},\ }\href
  {https://doi.org/10.1088/1475-7516/2023/03/045} {\bibfield  {journal}
  {\bibinfo  {journal} {Journal of Cosmology and Astroparticle Physics}\
  }\textbf {\bibinfo {volume} {2023}}\bibfield  {number} {\bibinfo  {number} {
  (3)},\ \bibinfo {eid} {045}},\ }\Eprint {https://arxiv.org/abs/2210.12743}
  {arXiv:2210.12743 [astro-ph.CO]} \BibitemShut {NoStop}%
\bibitem [{\citenamefont {{Chen}}\ \emph {et~al.}(2024)\citenamefont {{Chen},
  {Chakraborty},\ and\ {Dvorkin}}}]{2024arXiv240113036C}%
  \BibitemOpen
  \bibfield  {author} {\bibinfo {author} {\bibfnamefont {S.-F.}\ \bibnamefont
  {{Chen}}}, \bibinfo {author} {\bibfnamefont {P.}~\bibnamefont
  {{Chakraborty}}},\ \bibnamefont {and}\ \bibinfo {author} {\bibfnamefont
  {C.}~\bibnamefont {{Dvorkin}}},\ }\bibfield  {title} {\bibinfo {title}
  {{Analysis of BOSS Galaxy Data with Weighted Skew-Spectra}},\ }\href
  {https://doi.org/10.48550/arXiv.2401.13036} {\bibfield  {journal} {\bibinfo
  {journal} {arXiv e-prints}\ ,\ \bibinfo {eid} {arXiv:2401.13036}} (\bibinfo
  {year} {2024})},\ \Eprint {https://arxiv.org/abs/2401.13036}
  {arXiv:2401.13036 [astro-ph.CO]} \BibitemShut {NoStop}%
\bibitem [{\citenamefont {{Hou}}\ \emph {et~al.}(2024)\citenamefont {{Hou},
  {Moradinezhad Dizgah}, {Hahn}, {Eickenberg}, {Ho}, {Lemos}, {Massara},
  {Modi}, {Parker},\ and\ {R{\'e}galdo-Saint Blancard}}}]{2024arXiv240115074H}%
  \BibitemOpen
  \bibfield  {author} {\bibinfo {author} {\bibfnamefont {J.}~\bibnamefont
  {{Hou}}}, \bibinfo {author} {\bibfnamefont {A.}~\bibnamefont {{Moradinezhad
  Dizgah}}}, \bibinfo {author} {\bibfnamefont {C.}~\bibnamefont {{Hahn}}},
  \bibinfo {author} {\bibfnamefont {M.}~\bibnamefont {{Eickenberg}}}, \bibinfo
  {author} {\bibfnamefont {S.}~\bibnamefont {{Ho}}}, \bibinfo {author}
  {\bibfnamefont {P.}~\bibnamefont {{Lemos}}}, \bibinfo {author} {\bibfnamefont
  {E.}~\bibnamefont {{Massara}}}, \bibinfo {author} {\bibfnamefont
  {C.}~\bibnamefont {{Modi}}}, \bibinfo {author} {\bibfnamefont
  {L.}~\bibnamefont {{Parker}}},\ \bibnamefont {and}\ \bibinfo {author}
  {\bibfnamefont {B.}~\bibnamefont {{R{\'e}galdo-Saint Blancard}}},\ }\bibfield
   {title} {\bibinfo {title} {{${\rm S{\scriptsize IM}BIG}$: Cosmological
  Constraints from the Redshift-Space Galaxy Skew Spectra}},\ }\href
  {https://doi.org/10.48550/arXiv.2401.15074} {\bibfield  {journal} {\bibinfo
  {journal} {arXiv e-prints}\ ,\ \bibinfo {eid} {arXiv:2401.15074}} (\bibinfo
  {year} {2024})},\ \Eprint {https://arxiv.org/abs/2401.15074}
  {arXiv:2401.15074 [astro-ph.CO]} \BibitemShut {NoStop}%
\bibitem [{\citenamefont {Banerjee}\ and\ \citenamefont
  {Abel}(2020)\citenamefont {Banerjee\ and\ Abel}}]{Banerjee:2020umh}%
  \BibitemOpen
  \bibfield  {author} {\bibinfo {author} {\bibfnamefont {A.}~\bibnamefont
  {Banerjee}}\ \bibnamefont {and}\ \bibinfo {author} {\bibfnamefont
  {T.}~\bibnamefont {Abel}},\ }\bibfield  {title} {\bibinfo {title} {{Nearest
  neighbour distributions: New statistical measures for cosmological
  clustering}},\ }\href {https://doi.org/10.1093/mnras/staa3604} {\bibfield
  {journal} {\bibinfo  {journal} {Mon. Not. Roy. Astron. Soc.}\ }\textbf
  {\bibinfo {volume} {500}},\ \bibinfo {pages} {5479} (\bibinfo {year}
  {2020})},\ \Eprint {https://arxiv.org/abs/2007.13342} {arXiv:2007.13342
  [astro-ph.CO]} \BibitemShut {NoStop}%
\bibitem [{\citenamefont {Banerjee}\ and\ \citenamefont
  {Abel}(2021)\citenamefont {Banerjee\ and\ Abel}}]{10.1093/mnras/stab961}%
  \BibitemOpen
  \bibfield  {author} {\bibinfo {author} {\bibfnamefont {A.}~\bibnamefont
  {Banerjee}}\ \bibnamefont {and}\ \bibinfo {author} {\bibfnamefont
  {T.}~\bibnamefont {Abel}},\ }\bibfield  {title} {\bibinfo {title}
  {{Cosmological cross-correlations and nearest neighbour distributions}},\
  }\href {https://doi.org/10.1093/mnras/stab961} {\bibfield  {journal}
  {\bibinfo  {journal} {Monthly Notices of the Royal Astronomical Society}\
  }\textbf {\bibinfo {volume} {504}},\ \bibinfo {pages} {2911} (\bibinfo {year}
  {2021})},\ \Eprint
  {https://arxiv.org/abs/https://academic.oup.com/mnras/article-pdf/504/2/2911/37787124/stab961.pdf}
  {https://academic.oup.com/mnras/article-pdf/504/2/2911/37787124/stab961.pdf}
  \BibitemShut {NoStop}%
\bibitem [{\citenamefont {Uhlemann}\ \emph {et~al.}(2020)\citenamefont
  {Uhlemann, Friedrich, Villaescusa-Navarro, Banerjee,\ and\
  Codis}}]{Uhlemann:2019gni}%
  \BibitemOpen
  \bibfield  {author} {\bibinfo {author} {\bibfnamefont {C.}~\bibnamefont
  {Uhlemann}}, \bibinfo {author} {\bibfnamefont {O.}~\bibnamefont {Friedrich}},
  \bibinfo {author} {\bibfnamefont {F.}~\bibnamefont {Villaescusa-Navarro}},
  \bibinfo {author} {\bibfnamefont {A.}~\bibnamefont {Banerjee}},\ \bibnamefont
  {and}\ \bibinfo {author} {\bibfnamefont {S.}~\bibnamefont {Codis}},\
  }\bibfield  {title} {\bibinfo {title} {{Fisher for complements: Extracting
  cosmology and neutrino mass from the counts-in-cells PDF}},\ }\href
  {https://doi.org/10.1093/mnras/staa1155} {\bibfield  {journal} {\bibinfo
  {journal} {Mon. Not. Roy. Astron. Soc.}\ }\textbf {\bibinfo {volume} {495}},\
  \bibinfo {pages} {4006} (\bibinfo {year} {2020})},\ \Eprint
  {https://arxiv.org/abs/1911.11158} {arXiv:1911.11158 [astro-ph.CO]}
  \BibitemShut {NoStop}%
\bibitem [{\citenamefont {{Jamieson}}\ and\ \citenamefont
  {{Loverde}}(2020)\citenamefont {{Jamieson}\ and\
  {Loverde}}}]{2020PhRvD.102l3546J}%
  \BibitemOpen
  \bibfield  {author} {\bibinfo {author} {\bibfnamefont {D.}~\bibnamefont
  {{Jamieson}}}\ \bibnamefont {and}\ \bibinfo {author} {\bibfnamefont
  {M.}~\bibnamefont {{Loverde}}},\ }\bibfield  {title} {\bibinfo {title}
  {{Position-dependent matter density probability distribution function}},\
  }\href {https://doi.org/10.1103/PhysRevD.102.123546} {\bibfield  {journal}
  {\bibinfo  {journal} {Phys. Rev. D}\ }\textbf {\bibinfo {volume} {102}},\
  \bibinfo {eid} {123546} (\bibinfo {year} {2020})},\ \Eprint
  {https://arxiv.org/abs/2010.07235} {arXiv:2010.07235 [astro-ph.CO]}
  \BibitemShut {NoStop}%
\bibitem [{\citenamefont {Naidoo}\ \emph {et~al.}(2021)\citenamefont {Naidoo,
  Massara,\ and\ Lahav}}]{Naidoo:2021dxz}%
  \BibitemOpen
  \bibfield  {author} {\bibinfo {author} {\bibfnamefont {K.}~\bibnamefont
  {Naidoo}}, \bibinfo {author} {\bibfnamefont {E.}~\bibnamefont {Massara}},\
  \bibnamefont {and}\ \bibinfo {author} {\bibfnamefont {O.}~\bibnamefont
  {Lahav}},\ }\bibfield  {title} {\bibinfo {title} {{Cosmology and neutrino
  mass with the Minimum Spanning Tree}},\ }\href@noop {} {\  (\bibinfo {year}
  {2021})},\ \Eprint {https://arxiv.org/abs/2111.12088} {arXiv:2111.12088
  [astro-ph.CO]} \BibitemShut {NoStop}%
\bibitem [{\citenamefont {{Euclid Collaboration}}\ \emph
  {et~al.}(2023)\citenamefont {{Euclid Collaboration}, {Ajani}, {Baldi},
  {Barthelemy}, {Boyle}, {Burger}, {Cardone}, {Cheng}, {Codis}, {Giocoli}
  et~al.}}]{2023arXiv230112890E}%
  \BibitemOpen
  \bibfield  {author} {\bibinfo {author} {\bibnamefont {{Euclid
  Collaboration}}}, \bibinfo {author} {\bibfnamefont {V.}~\bibnamefont
  {{Ajani}}}, \bibinfo {author} {\bibfnamefont {M.}~\bibnamefont {{Baldi}}},
  \bibinfo {author} {\bibfnamefont {A.}~\bibnamefont {{Barthelemy}}}, \bibinfo
  {author} {\bibfnamefont {A.}~\bibnamefont {{Boyle}}}, \bibinfo {author}
  {\bibfnamefont {P.}~\bibnamefont {{Burger}}}, \bibinfo {author}
  {\bibfnamefont {V.~F.}\ \bibnamefont {{Cardone}}}, \bibinfo {author}
  {\bibfnamefont {S.}~\bibnamefont {{Cheng}}}, \bibinfo {author} {\bibfnamefont
  {S.}~\bibnamefont {{Codis}}}, \bibinfo {author} {\bibfnamefont
  {C.}~\bibnamefont {{Giocoli}}}, \bibinfo {author} {\bibfnamefont
  {J.}~\bibnamefont {{Harnois-D{\'e}raps}}}, \bibinfo {author} {\bibfnamefont
  {S.}~\bibnamefont {{Heydenreich}}}, \bibinfo {author} {\bibfnamefont
  {V.}~\bibnamefont {{Kansal}}}, \bibinfo {author} {\bibfnamefont
  {M.}~\bibnamefont {{Kilbinger}}}, \bibinfo {author} {\bibfnamefont
  {L.}~\bibnamefont {{Linke}}}, \bibinfo {author} {\bibfnamefont
  {C.}~\bibnamefont {{Llinares}}}, \bibnamefont {et~al.},\ }\bibfield  {title}
  {\bibinfo {title} {{Euclid Preparation XXIX: Forecasts for 10 different
  higher-order weak lensing statistics}},\ }\href
  {https://doi.org/10.48550/arXiv.2301.12890} {\bibfield  {journal} {\bibinfo
  {journal} {arXiv e-prints}\ ,\ \bibinfo {eid} {arXiv:2301.12890}} (\bibinfo
  {year} {2023})},\ \Eprint {https://arxiv.org/abs/2301.12890}
  {arXiv:2301.12890 [astro-ph.CO]} \BibitemShut {NoStop}%
\bibitem [{\citenamefont {Gupta}\ \emph {et~al.}(2018)\citenamefont {Gupta,
  Matilla, Hsu,\ and\ Haiman}}]{PhysRevD.97.103515}%
  \BibitemOpen
  \bibfield  {author} {\bibinfo {author} {\bibfnamefont {A.}~\bibnamefont
  {Gupta}}, \bibinfo {author} {\bibfnamefont {J.~M.~Z.}\ \bibnamefont
  {Matilla}}, \bibinfo {author} {\bibfnamefont {D.}~\bibnamefont {Hsu}},\
  \bibnamefont {and}\ \bibinfo {author} {\bibfnamefont {Z.}~\bibnamefont
  {Haiman}},\ }\bibfield  {title} {\bibinfo {title} {Non-gaussian information
  from weak lensing data via deep learning},\ }\href
  {https://doi.org/10.1103/PhysRevD.97.103515} {\bibfield  {journal} {\bibinfo
  {journal} {Phys. Rev. D}\ }\textbf {\bibinfo {volume} {97}},\ \bibinfo
  {pages} {103515} (\bibinfo {year} {2018})}\BibitemShut {NoStop}%
\bibitem [{\citenamefont {Peel}\ \emph {et~al.}(2019)\citenamefont {Peel,
  Lalande, Starck, Pettorino, Merten, Giocoli, Meneghetti,\ and\
  Baldi}}]{Peel:2018aei}%
  \BibitemOpen
  \bibfield  {author} {\bibinfo {author} {\bibfnamefont {A.}~\bibnamefont
  {Peel}}, \bibinfo {author} {\bibfnamefont {F.}~\bibnamefont {Lalande}},
  \bibinfo {author} {\bibfnamefont {J.-L.}\ \bibnamefont {Starck}}, \bibinfo
  {author} {\bibfnamefont {V.}~\bibnamefont {Pettorino}}, \bibinfo {author}
  {\bibfnamefont {J.}~\bibnamefont {Merten}}, \bibinfo {author} {\bibfnamefont
  {C.}~\bibnamefont {Giocoli}}, \bibinfo {author} {\bibfnamefont
  {M.}~\bibnamefont {Meneghetti}},\ \bibnamefont {and}\ \bibinfo {author}
  {\bibfnamefont {M.}~\bibnamefont {Baldi}},\ }\bibfield  {title} {\bibinfo
  {title} {{Distinguishing standard and modified gravity cosmologies with
  machine learning}},\ }\href {https://doi.org/10.1103/PhysRevD.100.023508}
  {\bibfield  {journal} {\bibinfo  {journal} {Phys. Rev. D}\ }\textbf {\bibinfo
  {volume} {100}},\ \bibinfo {pages} {023508} (\bibinfo {year} {2019})},\
  \Eprint {https://arxiv.org/abs/1810.11030} {arXiv:1810.11030 [astro-ph.CO]}
  \BibitemShut {NoStop}%
\bibitem [{\citenamefont {Merten}\ \emph {et~al.}(2019)\citenamefont {Merten,
  Giocoli, Baldi, Meneghetti, Peel, Lalande, Starck,\ and\
  Pettorino}}]{Merten:2018bgr}%
  \BibitemOpen
  \bibfield  {author} {\bibinfo {author} {\bibfnamefont {J.}~\bibnamefont
  {Merten}}, \bibinfo {author} {\bibfnamefont {C.}~\bibnamefont {Giocoli}},
  \bibinfo {author} {\bibfnamefont {M.}~\bibnamefont {Baldi}}, \bibinfo
  {author} {\bibfnamefont {M.}~\bibnamefont {Meneghetti}}, \bibinfo {author}
  {\bibfnamefont {A.}~\bibnamefont {Peel}}, \bibinfo {author} {\bibfnamefont
  {F.}~\bibnamefont {Lalande}}, \bibinfo {author} {\bibfnamefont {J.-L.}\
  \bibnamefont {Starck}},\ \bibnamefont {and}\ \bibinfo {author} {\bibfnamefont
  {V.}~\bibnamefont {Pettorino}},\ }\bibfield  {title} {\bibinfo {title} {{On
  the dissection of degenerate cosmologies with machine learning}},\ }\href
  {https://doi.org/10.1093/mnras/stz972} {\bibfield  {journal} {\bibinfo
  {journal} {Mon. Not. Roy. Astron. Soc.}\ }\textbf {\bibinfo {volume} {487}},\
  \bibinfo {pages} {104} (\bibinfo {year} {2019})},\ \Eprint
  {https://arxiv.org/abs/1810.11027} {arXiv:1810.11027 [astro-ph.CO]}
  \BibitemShut {NoStop}%
\bibitem [{\citenamefont {Villaescusa-Navarro}\ \emph
  {et~al.}(2021)\citenamefont {Villaescusa-Navarro, Angl{\'{e}}s-Alc{\'{a}}zar,
  Genel, Spergel, Somerville, Dave, Pillepich, Hernquist, Nelson, Torrey
  et~al.}}]{Villaescusa_Navarro_2021}%
  \BibitemOpen
  \bibfield  {author} {\bibinfo {author} {\bibfnamefont {F.}~\bibnamefont
  {Villaescusa-Navarro}}, \bibinfo {author} {\bibfnamefont {D.}~\bibnamefont
  {Angl{\'{e}}s-Alc{\'{a}}zar}}, \bibinfo {author} {\bibfnamefont
  {S.}~\bibnamefont {Genel}}, \bibinfo {author} {\bibfnamefont {D.~N.}\
  \bibnamefont {Spergel}}, \bibinfo {author} {\bibfnamefont {R.~S.}\
  \bibnamefont {Somerville}}, \bibinfo {author} {\bibfnamefont
  {R.}~\bibnamefont {Dave}}, \bibinfo {author} {\bibfnamefont {A.}~\bibnamefont
  {Pillepich}}, \bibinfo {author} {\bibfnamefont {L.}~\bibnamefont
  {Hernquist}}, \bibinfo {author} {\bibfnamefont {D.}~\bibnamefont {Nelson}},
  \bibinfo {author} {\bibfnamefont {P.}~\bibnamefont {Torrey}}, \bibinfo
  {author} {\bibfnamefont {D.}~\bibnamefont {Narayanan}}, \bibinfo {author}
  {\bibfnamefont {Y.}~\bibnamefont {Li}}, \bibinfo {author} {\bibfnamefont
  {O.}~\bibnamefont {Philcox}}, \bibinfo {author} {\bibfnamefont {V.~L.}\
  \bibnamefont {Torre}}, \bibinfo {author} {\bibfnamefont {A.~M.}\ \bibnamefont
  {Delgado}}, \bibinfo {author} {\bibfnamefont {S.}~\bibnamefont {Ho}},
  \bibnamefont {et~al.},\ }\bibfield  {title} {\bibinfo {title} {The {CAMELS}
  project: Cosmology and astrophysics with machine-learning simulations},\
  }\href {https://doi.org/10.3847/1538-4357/abf7ba} {\bibfield  {journal}
  {\bibinfo  {journal} {The Astrophysical Journal}\ }\textbf {\bibinfo {volume}
  {915}},\ \bibinfo {pages} {71} (\bibinfo {year} {2021})}\BibitemShut
  {NoStop}%
\bibitem [{\citenamefont {Perez}\ \emph {et~al.}(2022)\citenamefont {Perez,
  Genel, Villaescusa-Navarro, Somerville, Gabrielpillai, Angl\'es-Alc\'azar,
  Wandelt,\ and\ Yung}}]{Perez:2022nlv}%
  \BibitemOpen
  \bibfield  {author} {\bibinfo {author} {\bibfnamefont {L.~A.}\ \bibnamefont
  {Perez}}, \bibinfo {author} {\bibfnamefont {S.}~\bibnamefont {Genel}},
  \bibinfo {author} {\bibfnamefont {F.}~\bibnamefont {Villaescusa-Navarro}},
  \bibinfo {author} {\bibfnamefont {R.~S.}\ \bibnamefont {Somerville}},
  \bibinfo {author} {\bibfnamefont {A.}~\bibnamefont {Gabrielpillai}}, \bibinfo
  {author} {\bibfnamefont {D.}~\bibnamefont {Angl\'es-Alc\'azar}}, \bibinfo
  {author} {\bibfnamefont {B.~D.}\ \bibnamefont {Wandelt}},\ \bibnamefont
  {and}\ \bibinfo {author} {\bibfnamefont {L.~Y.~A.}\ \bibnamefont {Yung}},\
  }\bibfield  {title} {\bibinfo {title} {{Constraining cosmology with machine
  learning and galaxy clustering: the CAMELS-SAM suite}},\ }\href@noop {} {\
  (\bibinfo {year} {2022})},\ \Eprint {https://arxiv.org/abs/2204.02408}
  {arXiv:2204.02408 [astro-ph.GA]} \BibitemShut {NoStop}%
\bibitem [{\citenamefont {{de Santi}}\ \emph {et~al.}(2023)\citenamefont {{de
  Santi}, {Shao}, {Villaescusa-Navarro}, {Abramo}, {Teyssier},
  {Villanueva-Domingo}, {Ni}, {Angles-Alcazar}, {Genel}, {Hernandez-Martinez}
  et~al.}}]{2023ApJ...952...69D}%
  \BibitemOpen
  \bibfield  {author} {\bibinfo {author} {\bibfnamefont {N.~S.~M.}\
  \bibnamefont {{de Santi}}}, \bibinfo {author} {\bibfnamefont
  {H.}~\bibnamefont {{Shao}}}, \bibinfo {author} {\bibfnamefont
  {F.}~\bibnamefont {{Villaescusa-Navarro}}}, \bibinfo {author} {\bibfnamefont
  {L.~R.}\ \bibnamefont {{Abramo}}}, \bibinfo {author} {\bibfnamefont
  {R.}~\bibnamefont {{Teyssier}}}, \bibinfo {author} {\bibfnamefont
  {P.}~\bibnamefont {{Villanueva-Domingo}}}, \bibinfo {author} {\bibfnamefont
  {Y.}~\bibnamefont {{Ni}}}, \bibinfo {author} {\bibfnamefont {D.}~\bibnamefont
  {{Angles-Alcazar}}}, \bibinfo {author} {\bibfnamefont {S.}~\bibnamefont
  {{Genel}}}, \bibinfo {author} {\bibfnamefont {E.}~\bibnamefont
  {{Hernandez-Martinez}}}, \bibinfo {author} {\bibfnamefont {U.~P.}\
  \bibnamefont {{Steinwandel}}}, \bibinfo {author} {\bibfnamefont {C.~C.}\
  \bibnamefont {{Lovell}}}, \bibinfo {author} {\bibfnamefont {K.}~\bibnamefont
  {{Dolag}}}, \bibinfo {author} {\bibfnamefont {T.}~\bibnamefont {{Castro}}},\
  \bibnamefont {and}\ \bibinfo {author} {\bibfnamefont {M.}~\bibnamefont
  {{Vogelsberger}}},\ }\bibfield  {title} {\bibinfo {title} {{Robust
  Field-level Likelihood-free Inference with Galaxies}},\ }\href
  {https://doi.org/10.3847/1538-4357/acd1e2} {\bibfield  {journal} {\bibinfo
  {journal} {The Astrophysical Journal}\ }\textbf {\bibinfo {volume} {952}},\
  \bibinfo {eid} {69} (\bibinfo {year} {2023})},\ \Eprint
  {https://arxiv.org/abs/2302.14101} {arXiv:2302.14101 [astro-ph.CO]}
  \BibitemShut {NoStop}%
\bibitem [{\citenamefont {Dai}\ and\ \citenamefont {Seljak}(2023)\citenamefont
  {Dai\ and\ Seljak}}]{Dai:2023lcb}%
  \BibitemOpen
  \bibfield  {author} {\bibinfo {author} {\bibfnamefont {B.}~\bibnamefont
  {Dai}}\ \bibnamefont {and}\ \bibinfo {author} {\bibfnamefont
  {U.}~\bibnamefont {Seljak}},\ }\bibfield  {title} {\bibinfo {title}
  {{Multiscale Flow for Robust and Optimal Cosmological Analysis}},\
  }\href@noop {} {\  (\bibinfo {year} {2023})},\ \Eprint
  {https://arxiv.org/abs/2306.04689} {arXiv:2306.04689 [astro-ph.CO]}
  \BibitemShut {NoStop}%
\bibitem [{\citenamefont {Sharma}\ \emph {et~al.}(2024)\citenamefont {Sharma,
  Dai,\ and\ Seljak}}]{Sharma:2024pth}%
  \BibitemOpen
  \bibfield  {author} {\bibinfo {author} {\bibfnamefont {D.}~\bibnamefont
  {Sharma}}, \bibinfo {author} {\bibfnamefont {B.}~\bibnamefont {Dai}},\
  \bibnamefont {and}\ \bibinfo {author} {\bibfnamefont {U.}~\bibnamefont
  {Seljak}},\ }\bibfield  {title} {\bibinfo {title} {{A comparative study of
  cosmological constraints from weak lensing using Convolutional Neural
  Networks}},\ }\href@noop {} {\  (\bibinfo {year} {2024})},\ \Eprint
  {https://arxiv.org/abs/2403.03490} {arXiv:2403.03490 [astro-ph.CO]}
  \BibitemShut {NoStop}%
\bibitem [{\citenamefont {Mallat}(2012)}]{https://doi.org/10.1002/cpa.21413}%
  \BibitemOpen
  \bibfield  {author} {\bibinfo {author} {\bibfnamefont {S.}~\bibnamefont
  {Mallat}},\ }\bibfield  {title} {\bibinfo {title} {Group invariant
  scattering},\ }\href {https://doi.org/https://doi.org/10.1002/cpa.21413}
  {\bibfield  {journal} {\bibinfo  {journal} {Communications on Pure and
  Applied Mathematics}\ }\textbf {\bibinfo {volume} {65}},\ \bibinfo {pages}
  {1331} (\bibinfo {year} {2012})},\ \Eprint
  {https://arxiv.org/abs/https://onlinelibrary.wiley.com/doi/pdf/10.1002/cpa.21413}
  {https://onlinelibrary.wiley.com/doi/pdf/10.1002/cpa.21413} \BibitemShut
  {NoStop}%
\bibitem [{\citenamefont {Bruna}\ and\ \citenamefont
  {Mallat}(2013)\citenamefont {Bruna\ and\ Mallat}}]{6522407}%
  \BibitemOpen
  \bibfield  {author} {\bibinfo {author} {\bibfnamefont {J.}~\bibnamefont
  {Bruna}}\ \bibnamefont {and}\ \bibinfo {author} {\bibfnamefont
  {S.}~\bibnamefont {Mallat}},\ }\bibfield  {title} {\bibinfo {title}
  {Invariant scattering convolution networks},\ }\href
  {https://doi.org/10.1109/TPAMI.2012.230} {\bibfield  {journal} {\bibinfo
  {journal} {IEEE Transactions on Pattern Analysis and Machine Intelligence}\
  }\textbf {\bibinfo {volume} {35}},\ \bibinfo {pages} {1872} (\bibinfo {year}
  {2013})}\BibitemShut {NoStop}%
\bibitem [{\citenamefont {Sifre}\ and\ \citenamefont
  {Mallat}(2013)\citenamefont {Sifre\ and\ Mallat}}]{Sifre_2013_CVPR}%
  \BibitemOpen
  \bibfield  {author} {\bibinfo {author} {\bibfnamefont {L.}~\bibnamefont
  {Sifre}}\ \bibnamefont {and}\ \bibinfo {author} {\bibfnamefont
  {S.}~\bibnamefont {Mallat}},\ }\bibfield  {title} {\bibinfo {title}
  {Rotation, scaling and deformation invariant scattering for texture
  discrimination},\ }in\ \href@noop {} {\emph {\bibinfo {booktitle}
  {Proceedings of the IEEE Conference on Computer Vision and Pattern
  Recognition (CVPR)}}}\ (\bibinfo {year} {2013})\BibitemShut {NoStop}%
\bibitem [{\citenamefont {Bruna}\ \emph {et~al.}(2015)\citenamefont {Bruna,
  Mallat, Bacry,\ and\ Muzy}}]{10.1214/14-AOS1276}%
  \BibitemOpen
  \bibfield  {author} {\bibinfo {author} {\bibfnamefont {J.}~\bibnamefont
  {Bruna}}, \bibinfo {author} {\bibfnamefont {S.}~\bibnamefont {Mallat}},
  \bibinfo {author} {\bibfnamefont {E.}~\bibnamefont {Bacry}},\ \bibnamefont
  {and}\ \bibinfo {author} {\bibfnamefont {J.-F.}\ \bibnamefont {Muzy}},\
  }\bibfield  {title} {\bibinfo {title} {{Intermittent process analysis with
  scattering moments}},\ }\href {https://doi.org/10.1214/14-AOS1276} {\bibfield
   {journal} {\bibinfo  {journal} {The Annals of Statistics}\ }\textbf
  {\bibinfo {volume} {43}},\ \bibinfo {pages} {323 } (\bibinfo {year}
  {2015})}\BibitemShut {NoStop}%
\bibitem [{\citenamefont {Andén}\ and\ \citenamefont
  {Mallat}(2014)\citenamefont {Andén\ and\ Mallat}}]{6822556}%
  \BibitemOpen
  \bibfield  {author} {\bibinfo {author} {\bibfnamefont {J.}~\bibnamefont
  {Andén}}\ \bibnamefont {and}\ \bibinfo {author} {\bibfnamefont
  {S.}~\bibnamefont {Mallat}},\ }\bibfield  {title} {\bibinfo {title} {Deep
  scattering spectrum},\ }\href {https://doi.org/10.1109/TSP.2014.2326991}
  {\bibfield  {journal} {\bibinfo  {journal} {IEEE Transactions on Signal
  Processing}\ }\textbf {\bibinfo {volume} {62}},\ \bibinfo {pages} {4114}
  (\bibinfo {year} {2014})}\BibitemShut {NoStop}%
\bibitem [{\citenamefont {Cheng}\ and\ \citenamefont
  {Menard}(2021)\citenamefont {Cheng\ and\ Menard}}]{cheng:2021xdw}%
  \BibitemOpen
  \bibfield  {author} {\bibinfo {author} {\bibfnamefont {S.}~\bibnamefont
  {Cheng}}\ \bibnamefont {and}\ \bibinfo {author} {\bibfnamefont
  {B.}~\bibnamefont {Menard}},\ }\bibfield  {title} {\bibinfo {title} {{How to
  quantify fields or textures? A guide to the scattering transform}},\
  }\href@noop {} {\  (\bibinfo {year} {2021})},\ \Eprint
  {https://arxiv.org/abs/2112.01288} {arXiv:2112.01288 [astro-ph.IM]}
  \BibitemShut {NoStop}%
\bibitem [{\citenamefont {{Allys, E.}}\ \emph {et~al.}(2019)\citenamefont
  {{Allys, E.}, {Levrier, F.}, {Zhang, S.}, {Colling, C.}, {Regaldo-Saint
  Blancard, B.}, {Boulanger, F.}, {Hennebelle, P.},\ and\ {Mallat,
  S.}}}]{refId0}%
  \BibitemOpen
  \bibfield  {author} {\bibinfo {author} {\bibnamefont {{Allys, E.}}}, \bibinfo
  {author} {\bibnamefont {{Levrier, F.}}}, \bibinfo {author} {\bibnamefont
  {{Zhang, S.}}}, \bibinfo {author} {\bibnamefont {{Colling, C.}}}, \bibinfo
  {author} {\bibnamefont {{Regaldo-Saint Blancard, B.}}}, \bibinfo {author}
  {\bibnamefont {{Boulanger, F.}}}, \bibinfo {author} {\bibnamefont
  {{Hennebelle, P.}}},\ \bibnamefont {and}\ \bibinfo {author} {\bibnamefont
  {{Mallat, S.}}},\ }\bibfield  {title} {\bibinfo {title} {The rwst, a
  comprehensive statistical description of the non-gaussian structures in the
  ism},\ }\href {https://doi.org/10.1051/0004-6361/201834975} {\bibfield
  {journal} {\bibinfo  {journal} {A\&A}\ }\textbf {\bibinfo {volume} {629}},\
  \bibinfo {pages} {A115} (\bibinfo {year} {2019})}\BibitemShut {NoStop}%
\bibitem [{\citenamefont {{Saydjari}}\ \emph {et~al.}(2021)\citenamefont
  {{Saydjari}, {Portillo}, {Slepian}, {Kahraman}, {Burkhart},\ and\
  {Finkbeiner}}}]{Saydjari_2021}%
  \BibitemOpen
  \bibfield  {author} {\bibinfo {author} {\bibfnamefont {A.~K.}\ \bibnamefont
  {{Saydjari}}}, \bibinfo {author} {\bibfnamefont {S.~K.~N.}\ \bibnamefont
  {{Portillo}}}, \bibinfo {author} {\bibfnamefont {Z.}~\bibnamefont
  {{Slepian}}}, \bibinfo {author} {\bibfnamefont {S.}~\bibnamefont
  {{Kahraman}}}, \bibinfo {author} {\bibfnamefont {B.}~\bibnamefont
  {{Burkhart}}},\ \bibnamefont {and}\ \bibinfo {author} {\bibfnamefont {D.~P.}\
  \bibnamefont {{Finkbeiner}}},\ }\bibfield  {title} {\bibinfo {title}
  {{Classification of Magnetohydrodynamic Simulations Using Wavelet Scattering
  Transforms}},\ }\href {https://doi.org/10.3847/1538-4357/abe46d} {\bibfield
  {journal} {\bibinfo  {journal} {\apj}\ }\textbf {\bibinfo {volume} {910}},\
  \bibinfo {eid} {122} (\bibinfo {year} {2021})},\ \Eprint
  {https://arxiv.org/abs/2010.11963} {arXiv:2010.11963 [astro-ph.GA]}
  \BibitemShut {NoStop}%
\bibitem [{\citenamefont {{Regaldo-Saint Blancard, B.}}\ \emph
  {et~al.}(2020)\citenamefont {{Regaldo-Saint Blancard, B.}, {Levrier, F.},
  {Allys, E.}, {Bellomi, E.},\ and\ {Boulanger, F.}}}]{refdust}%
  \BibitemOpen
  \bibfield  {author} {\bibinfo {author} {\bibnamefont {{Regaldo-Saint
  Blancard, B.}}}, \bibinfo {author} {\bibnamefont {{Levrier, F.}}}, \bibinfo
  {author} {\bibnamefont {{Allys, E.}}}, \bibinfo {author} {\bibnamefont
  {{Bellomi, E.}}},\ \bibnamefont {and}\ \bibinfo {author} {\bibnamefont
  {{Boulanger, F.}}},\ }\bibfield  {title} {\bibinfo {title} {Statistical
  description of dust polarized emission from the diffuse interstellar medium -
  a rwst approach},\ }\href {https://doi.org/10.1051/0004-6361/202038044}
  {\bibfield  {journal} {\bibinfo  {journal} {A\&A}\ }\textbf {\bibinfo
  {volume} {642}},\ \bibinfo {pages} {A217} (\bibinfo {year}
  {2020})}\BibitemShut {NoStop}%
\bibitem [{\citenamefont {Cheng}\ \emph {et~al.}(2020)\citenamefont {Cheng,
  Ting, Ménard,\ and\ Bruna}}]{10.1093/mnras/staa3165}%
  \BibitemOpen
  \bibfield  {author} {\bibinfo {author} {\bibfnamefont {S.}~\bibnamefont
  {Cheng}}, \bibinfo {author} {\bibfnamefont {Y.-S.}\ \bibnamefont {Ting}},
  \bibinfo {author} {\bibfnamefont {B.}~\bibnamefont {Ménard}},\ \bibnamefont
  {and}\ \bibinfo {author} {\bibfnamefont {J.}~\bibnamefont {Bruna}},\
  }\bibfield  {title} {\bibinfo {title} {{A new approach to observational
  cosmology using the scattering transform}},\ }\href
  {https://doi.org/10.1093/mnras/staa3165} {\bibfield  {journal} {\bibinfo
  {journal} {Monthly Notices of the Royal Astronomical Society}\ }\textbf
  {\bibinfo {volume} {499}},\ \bibinfo {pages} {5902} (\bibinfo {year}
  {2020})},\ \Eprint
  {https://arxiv.org/abs/https://academic.oup.com/mnras/article-pdf/499/4/5902/34157889/staa3165.pdf}
  {https://academic.oup.com/mnras/article-pdf/499/4/5902/34157889/staa3165.pdf}
  \BibitemShut {NoStop}%
\bibitem [{\citenamefont {Cheng}\ and\ \citenamefont
  {Ménard}(2021)\citenamefont {Cheng\ and\ Ménard}}]{10.1093/mnras/stab2102}%
  \BibitemOpen
  \bibfield  {author} {\bibinfo {author} {\bibfnamefont {S.}~\bibnamefont
  {Cheng}}\ \bibnamefont {and}\ \bibinfo {author} {\bibfnamefont
  {B.}~\bibnamefont {Ménard}},\ }\bibfield  {title} {\bibinfo {title} {{Weak
  lensing scattering transform: dark energy and neutrino mass sensitivity}},\
  }\href {https://doi.org/10.1093/mnras/stab2102} {\bibfield  {journal}
  {\bibinfo  {journal} {Monthly Notices of the Royal Astronomical Society}\
  }\textbf {\bibinfo {volume} {507}},\ \bibinfo {pages} {1012} (\bibinfo {year}
  {2021})},\ \Eprint
  {https://arxiv.org/abs/https://academic.oup.com/mnras/article-pdf/507/1/1012/39811822/stab2102.pdf}
  {https://academic.oup.com/mnras/article-pdf/507/1/1012/39811822/stab2102.pdf}
  \BibitemShut {NoStop}%
\bibitem [{\citenamefont {Valogiannis}\ and\ \citenamefont
  {Dvorkin}(2022{\natexlab{a}})\citenamefont {Valogiannis\ and\
  Dvorkin}}]{PhysRevD.105.103534}%
  \BibitemOpen
  \bibfield  {author} {\bibinfo {author} {\bibfnamefont {G.}~\bibnamefont
  {Valogiannis}}\ \bibnamefont {and}\ \bibinfo {author} {\bibfnamefont
  {C.}~\bibnamefont {Dvorkin}},\ }\bibfield  {title} {\bibinfo {title} {Towards
  an optimal estimation of cosmological parameters with the wavelet scattering
  transform},\ }\href {https://doi.org/10.1103/PhysRevD.105.103534} {\bibfield
  {journal} {\bibinfo  {journal} {Phys. Rev. D}\ }\textbf {\bibinfo {volume}
  {105}},\ \bibinfo {pages} {103534} (\bibinfo {year}
  {2022}{\natexlab{a}})}\BibitemShut {NoStop}%
\bibitem [{\citenamefont {Valogiannis}\ and\ \citenamefont
  {Dvorkin}(2022{\natexlab{b}})\citenamefont {Valogiannis\ and\
  Dvorkin}}]{PhysRevD.106.103509}%
  \BibitemOpen
  \bibfield  {author} {\bibinfo {author} {\bibfnamefont {G.}~\bibnamefont
  {Valogiannis}}\ \bibnamefont {and}\ \bibinfo {author} {\bibfnamefont
  {C.}~\bibnamefont {Dvorkin}},\ }\bibfield  {title} {\bibinfo {title} {Going
  beyond the galaxy power spectrum: An analysis of boss data with wavelet
  scattering transforms},\ }\href {https://doi.org/10.1103/PhysRevD.106.103509}
  {\bibfield  {journal} {\bibinfo  {journal} {Phys. Rev. D}\ }\textbf {\bibinfo
  {volume} {106}},\ \bibinfo {pages} {103509} (\bibinfo {year}
  {2022}{\natexlab{b}})}\BibitemShut {NoStop}%
\bibitem [{\citenamefont {Valogiannis}\ \emph {et~al.}(2024)\citenamefont
  {Valogiannis, Yuan,\ and\ Dvorkin}}]{PhysRevD.109.103503}%
  \BibitemOpen
  \bibfield  {author} {\bibinfo {author} {\bibfnamefont {G.}~\bibnamefont
  {Valogiannis}}, \bibinfo {author} {\bibfnamefont {S.}~\bibnamefont {Yuan}},\
  \bibnamefont {and}\ \bibinfo {author} {\bibfnamefont {C.}~\bibnamefont
  {Dvorkin}},\ }\bibfield  {title} {\bibinfo {title} {Precise cosmological
  constraints from boss galaxy clustering with a simulation-based emulator of
  the wavelet scattering transform},\ }\href
  {https://doi.org/10.1103/PhysRevD.109.103503} {\bibfield  {journal} {\bibinfo
   {journal} {Phys. Rev. D}\ }\textbf {\bibinfo {volume} {109}},\ \bibinfo
  {pages} {103503} (\bibinfo {year} {2024})}\BibitemShut {NoStop}%
\bibitem [{\citenamefont {Allys}\ \emph {et~al.}(2020)\citenamefont {Allys,
  Marchand, Cardoso, Villaescusa-Navarro, Ho,\ and\
  Mallat}}]{PhysRevD.102.103506}%
  \BibitemOpen
  \bibfield  {author} {\bibinfo {author} {\bibfnamefont {E.}~\bibnamefont
  {Allys}}, \bibinfo {author} {\bibfnamefont {T.}~\bibnamefont {Marchand}},
  \bibinfo {author} {\bibfnamefont {J.-F.}\ \bibnamefont {Cardoso}}, \bibinfo
  {author} {\bibfnamefont {F.}~\bibnamefont {Villaescusa-Navarro}}, \bibinfo
  {author} {\bibfnamefont {S.}~\bibnamefont {Ho}},\ \bibnamefont {and}\
  \bibinfo {author} {\bibfnamefont {S.}~\bibnamefont {Mallat}},\ }\bibfield
  {title} {\bibinfo {title} {New interpretable statistics for large-scale
  structure analysis and generation},\ }\href
  {https://doi.org/10.1103/PhysRevD.102.103506} {\bibfield  {journal} {\bibinfo
   {journal} {Phys. Rev. D}\ }\textbf {\bibinfo {volume} {102}},\ \bibinfo
  {pages} {103506} (\bibinfo {year} {2020})}\BibitemShut {NoStop}%
\bibitem [{\citenamefont {Greig}\ \emph {et~al.}(2022)\citenamefont {Greig,
  Ting,\ and\ Kaurov}}]{10.1093/mnras/stac977}%
  \BibitemOpen
  \bibfield  {author} {\bibinfo {author} {\bibfnamefont {B.}~\bibnamefont
  {Greig}}, \bibinfo {author} {\bibfnamefont {Y.-S.}\ \bibnamefont {Ting}},\
  \bibnamefont {and}\ \bibinfo {author} {\bibfnamefont {A.~A.}\ \bibnamefont
  {Kaurov}},\ }\bibfield  {title} {\bibinfo {title} {{Exploring the cosmic
  21-cm signal from the epoch of reionization using the wavelet scattering
  transform}},\ }\href {https://doi.org/10.1093/mnras/stac977} {\bibfield
  {journal} {\bibinfo  {journal} {Monthly Notices of the Royal Astronomical
  Society}\ }\textbf {\bibinfo {volume} {513}},\ \bibinfo {pages} {1719}
  (\bibinfo {year} {2022})},\ \Eprint
  {https://arxiv.org/abs/https://academic.oup.com/mnras/article-pdf/513/2/1719/45181861/stac977.pdf}
  {https://academic.oup.com/mnras/article-pdf/513/2/1719/45181861/stac977.pdf}
  \BibitemShut {NoStop}%
\bibitem [{\citenamefont {{Eickenberg}}\ \emph {et~al.}(2022)\citenamefont
  {{Eickenberg}, {Allys}, {Moradinezhad Dizgah}, {Lemos}, {Massara}, {Abidi},
  {Hahn}, {Hassan}, {Regaldo-Saint Blancard}, {Ho}
  et~al.}}]{2022arXiv220407646E}%
  \BibitemOpen
  \bibfield  {author} {\bibinfo {author} {\bibfnamefont {M.}~\bibnamefont
  {{Eickenberg}}}, \bibinfo {author} {\bibfnamefont {E.}~\bibnamefont
  {{Allys}}}, \bibinfo {author} {\bibfnamefont {A.}~\bibnamefont {{Moradinezhad
  Dizgah}}}, \bibinfo {author} {\bibfnamefont {P.}~\bibnamefont {{Lemos}}},
  \bibinfo {author} {\bibfnamefont {E.}~\bibnamefont {{Massara}}}, \bibinfo
  {author} {\bibfnamefont {M.}~\bibnamefont {{Abidi}}}, \bibinfo {author}
  {\bibfnamefont {C.}~\bibnamefont {{Hahn}}}, \bibinfo {author} {\bibfnamefont
  {S.}~\bibnamefont {{Hassan}}}, \bibinfo {author} {\bibfnamefont
  {B.}~\bibnamefont {{Regaldo-Saint Blancard}}}, \bibinfo {author}
  {\bibfnamefont {S.}~\bibnamefont {{Ho}}}, \bibinfo {author} {\bibfnamefont
  {S.}~\bibnamefont {{Mallat}}}, \bibinfo {author} {\bibfnamefont
  {J.}~\bibnamefont {{Anden}}},\ \bibnamefont {and}\ \bibinfo {author}
  {\bibfnamefont {F.}~\bibnamefont {{Villaescusa-Navarro}}},\ }\bibfield
  {title} {\bibinfo {title} {{Wavelet Moments for Cosmological Parameter
  Estimation}},\ }\href@noop {} {\bibfield  {journal} {\bibinfo  {journal}
  {arXiv e-prints}\ ,\ \bibinfo {eid} {arXiv:2204.07646}} (\bibinfo {year}
  {2022})},\ \Eprint {https://arxiv.org/abs/2204.07646} {arXiv:2204.07646
  [astro-ph.CO]} \BibitemShut {NoStop}%
\bibitem [{\citenamefont {{Regaldo-Saint Blancard}}\ \emph
  {et~al.}(2023)\citenamefont {{Regaldo-Saint Blancard}, {Hahn}, {Ho}, {Hou},
  {Lemos}, {Massara}, {Modi}, {Moradinezhad Dizgah}, {Parker}, {Yao}
  et~al.}}]{2023arXiv231015250R}%
  \BibitemOpen
  \bibfield  {author} {\bibinfo {author} {\bibfnamefont {B.}~\bibnamefont
  {{Regaldo-Saint Blancard}}}, \bibinfo {author} {\bibfnamefont
  {C.}~\bibnamefont {{Hahn}}}, \bibinfo {author} {\bibfnamefont
  {S.}~\bibnamefont {{Ho}}}, \bibinfo {author} {\bibfnamefont {J.}~\bibnamefont
  {{Hou}}}, \bibinfo {author} {\bibfnamefont {P.}~\bibnamefont {{Lemos}}},
  \bibinfo {author} {\bibfnamefont {E.}~\bibnamefont {{Massara}}}, \bibinfo
  {author} {\bibfnamefont {C.}~\bibnamefont {{Modi}}}, \bibinfo {author}
  {\bibfnamefont {A.}~\bibnamefont {{Moradinezhad Dizgah}}}, \bibinfo {author}
  {\bibfnamefont {L.}~\bibnamefont {{Parker}}}, \bibinfo {author}
  {\bibfnamefont {Y.}~\bibnamefont {{Yao}}},\ \bibnamefont {and}\ \bibinfo
  {author} {\bibfnamefont {M.}~\bibnamefont {{Eickenberg}}},\ }\bibfield
  {title} {\bibinfo {title} {{${\rm S{\scriptsize IM}BIG}$: Galaxy Clustering
  Analysis with the Wavelet Scattering Transform}},\ }\href
  {https://doi.org/10.48550/arXiv.2310.15250} {\bibfield  {journal} {\bibinfo
  {journal} {arXiv e-prints}\ ,\ \bibinfo {eid} {arXiv:2310.15250}} (\bibinfo
  {year} {2023})},\ \Eprint {https://arxiv.org/abs/2310.15250}
  {arXiv:2310.15250 [astro-ph.CO]} \BibitemShut {NoStop}%
\bibitem [{\citenamefont {Chung}(2022)}]{10.1093/mnras/stac2662}%
  \BibitemOpen
  \bibfield  {author} {\bibinfo {author} {\bibfnamefont {D.~T.}\ \bibnamefont
  {Chung}},\ }\bibfield  {title} {\bibinfo {title} {{Exploration of 3D wavelet
  scattering transform coefficients for line-intensity mapping measurements}},\
  }\href {https://doi.org/10.1093/mnras/stac2662} {\bibfield  {journal}
  {\bibinfo  {journal} {Monthly Notices of the Royal Astronomical Society}\
  }\textbf {\bibinfo {volume} {517}},\ \bibinfo {pages} {1625} (\bibinfo {year}
  {2022})},\ \Eprint
  {https://arxiv.org/abs/https://academic.oup.com/mnras/article-pdf/517/2/1625/46454615/stac2662.pdf}
  {https://academic.oup.com/mnras/article-pdf/517/2/1625/46454615/stac2662.pdf}
  \BibitemShut {NoStop}%
\bibitem [{\citenamefont {Gatti}\ \emph {et~al.}(2024)\citenamefont {Gatti
  et~al.}}]{DES:2023qwe}%
  \BibitemOpen
  \bibfield  {author} {\bibinfo {author} {\bibfnamefont {M.}~\bibnamefont
  {Gatti}} \bibnamefont {et~al.} (\bibinfo {collaboration} {DES}),\ }\bibfield
  {title} {\bibinfo {title} {{Dark Energy Survey Year 3 results:
  Simulation-based cosmological inference with wavelet harmonics, scattering
  transforms, and moments of weak lensing mass maps. Validation on
  simulations}},\ }\href {https://doi.org/10.1103/PhysRevD.109.063534}
  {\bibfield  {journal} {\bibinfo  {journal} {Phys. Rev. D}\ }\textbf {\bibinfo
  {volume} {109}},\ \bibinfo {pages} {063534} (\bibinfo {year} {2024})},\
  \Eprint {https://arxiv.org/abs/2310.17557} {arXiv:2310.17557 [astro-ph.CO]}
  \BibitemShut {NoStop}%
\bibitem [{\citenamefont {Peron}\ \emph {et~al.}(2024)\citenamefont {Peron,
  Jung, Liguori,\ and\ Pietroni}}]{Peron:2024xaw}%
  \BibitemOpen
  \bibfield  {author} {\bibinfo {author} {\bibfnamefont {M.}~\bibnamefont
  {Peron}}, \bibinfo {author} {\bibfnamefont {G.}~\bibnamefont {Jung}},
  \bibinfo {author} {\bibfnamefont {M.}~\bibnamefont {Liguori}},\ \bibnamefont
  {and}\ \bibinfo {author} {\bibfnamefont {M.}~\bibnamefont {Pietroni}},\
  }\bibfield  {title} {\bibinfo {title} {{Constraining Primordial
  Non-Gaussianity from Large Scale Structure with the Wavelet Scattering
  Transform}},\ }\href@noop {} {\  (\bibinfo {year} {2024})},\ \Eprint
  {https://arxiv.org/abs/2403.17657} {arXiv:2403.17657 [astro-ph.CO]}
  \BibitemShut {NoStop}%
\bibitem [{\citenamefont {{Cheng}}\ \emph {et~al.}(2024)\citenamefont {{Cheng},
  {Marques}, {Grand{\'o}n}, {Thiele}, {Shirasaki}, {M{\'e}nard},\ and\
  {Liu}}}]{2024arXiv240416085C}%
  \BibitemOpen
  \bibfield  {author} {\bibinfo {author} {\bibfnamefont {S.}~\bibnamefont
  {{Cheng}}}, \bibinfo {author} {\bibfnamefont {G.~A.}\ \bibnamefont
  {{Marques}}}, \bibinfo {author} {\bibfnamefont {D.}~\bibnamefont
  {{Grand{\'o}n}}}, \bibinfo {author} {\bibfnamefont {L.}~\bibnamefont
  {{Thiele}}}, \bibinfo {author} {\bibfnamefont {M.}~\bibnamefont
  {{Shirasaki}}}, \bibinfo {author} {\bibfnamefont {B.}~\bibnamefont
  {{M{\'e}nard}}},\ \bibnamefont {and}\ \bibinfo {author} {\bibfnamefont
  {J.}~\bibnamefont {{Liu}}},\ }\bibfield  {title} {\bibinfo {title}
  {{Cosmological constraints from weak lensing scattering transform using HSC
  Y1 data}},\ }\href@noop {} {\bibfield  {journal} {\bibinfo  {journal} {arXiv
  e-prints}\ ,\ \bibinfo {eid} {arXiv:2404.16085}} (\bibinfo {year} {2024})},\
  \Eprint {https://arxiv.org/abs/2404.16085} {arXiv:2404.16085 [astro-ph.CO]}
  \BibitemShut {NoStop}%
\bibitem [{\citenamefont {Eickenberg}\ \emph {et~al.}(2017)\citenamefont
  {Eickenberg, Exarchakis, Hirn,\ and\ Mallat}}]{10.5555/3295222.3295400}%
  \BibitemOpen
  \bibfield  {author} {\bibinfo {author} {\bibfnamefont {M.}~\bibnamefont
  {Eickenberg}}, \bibinfo {author} {\bibfnamefont {G.}~\bibnamefont
  {Exarchakis}}, \bibinfo {author} {\bibfnamefont {M.}~\bibnamefont {Hirn}},\
  \bibnamefont {and}\ \bibinfo {author} {\bibfnamefont {S.}~\bibnamefont
  {Mallat}},\ }\bibfield  {title} {\bibinfo {title} {Solid harmonic wavelet
  scattering: Predicting quantum molecular energy from invariant descriptors of
  3d electronic densities},\ }in\ \href@noop {} {\emph {\bibinfo {booktitle}
  {Proceedings of the 31st International Conference on Neural Information
  Processing Systems}}},\ \bibinfo {series and number} {NIPS'17}\ (\bibinfo
  {publisher} {Curran Associates Inc.},\ \bibinfo {address} {Red Hook, NY,
  USA},\ \bibinfo {year} {2017})\ p.\ \bibinfo {pages}
  {6543–6552}\BibitemShut {NoStop}%
\bibitem [{\citenamefont {Eickenberg}\ \emph {et~al.}(2018)\citenamefont
  {Eickenberg, Exarchakis, Hirn, Mallat,\ and\ Thiry}}]{doi:10.1063/1.5023798}%
  \BibitemOpen
  \bibfield  {author} {\bibinfo {author} {\bibfnamefont {M.}~\bibnamefont
  {Eickenberg}}, \bibinfo {author} {\bibfnamefont {G.}~\bibnamefont
  {Exarchakis}}, \bibinfo {author} {\bibfnamefont {M.}~\bibnamefont {Hirn}},
  \bibinfo {author} {\bibfnamefont {S.}~\bibnamefont {Mallat}},\ \bibnamefont
  {and}\ \bibinfo {author} {\bibfnamefont {L.}~\bibnamefont {Thiry}},\
  }\bibfield  {title} {\bibinfo {title} {Solid harmonic wavelet scattering for
  predictions of molecule properties},\ }\href
  {https://doi.org/10.1063/1.5023798} {\bibfield  {journal} {\bibinfo
  {journal} {The Journal of Chemical Physics}\ }\textbf {\bibinfo {volume}
  {148}},\ \bibinfo {pages} {241732} (\bibinfo {year} {2018})},\ \Eprint
  {https://arxiv.org/abs/https://doi.org/10.1063/1.5023798}
  {https://doi.org/10.1063/1.5023798} \BibitemShut {NoStop}%
\bibitem [{\citenamefont {Villaescusa-Navarro}\ \emph
  {et~al.}(2020)\citenamefont {Villaescusa-Navarro, Hahn, Massara, Banerjee,
  Delgado, Ramanah, Charnock, Giusarma, Li, Allys
  et~al.}}]{Villaescusa_Navarro_2020}%
  \BibitemOpen
  \bibfield  {author} {\bibinfo {author} {\bibfnamefont {F.}~\bibnamefont
  {Villaescusa-Navarro}}, \bibinfo {author} {\bibfnamefont {C.}~\bibnamefont
  {Hahn}}, \bibinfo {author} {\bibfnamefont {E.}~\bibnamefont {Massara}},
  \bibinfo {author} {\bibfnamefont {A.}~\bibnamefont {Banerjee}}, \bibinfo
  {author} {\bibfnamefont {A.~M.}\ \bibnamefont {Delgado}}, \bibinfo {author}
  {\bibfnamefont {D.~K.}\ \bibnamefont {Ramanah}}, \bibinfo {author}
  {\bibfnamefont {T.}~\bibnamefont {Charnock}}, \bibinfo {author}
  {\bibfnamefont {E.}~\bibnamefont {Giusarma}}, \bibinfo {author}
  {\bibfnamefont {Y.}~\bibnamefont {Li}}, \bibinfo {author} {\bibfnamefont
  {E.}~\bibnamefont {Allys}}, \bibinfo {author} {\bibfnamefont
  {A.}~\bibnamefont {Brochard}}, \bibinfo {author} {\bibfnamefont
  {C.}~\bibnamefont {Uhlemann}}, \bibinfo {author} {\bibfnamefont {C.-T.}\
  \bibnamefont {Chiang}}, \bibinfo {author} {\bibfnamefont {S.}~\bibnamefont
  {He}}, \bibinfo {author} {\bibfnamefont {A.}~\bibnamefont {Pisani}}, \bibinfo
  {author} {\bibfnamefont {A.}~\bibnamefont {Obuljen}}, \bibnamefont {et~al.},\
  }\bibfield  {title} {\bibinfo {title} {The quijote simulations},\ }\href
  {https://doi.org/10.3847/1538-4365/ab9d82} {\bibfield  {journal} {\bibinfo
  {journal} {The Astrophysical Journal Supplement Series}\ }\textbf {\bibinfo
  {volume} {250}},\ \bibinfo {pages} {2} (\bibinfo {year} {2020})}\BibitemShut
  {NoStop}%
\bibitem [{\citenamefont {{Maksimova}}\ \emph {et~al.}(2021)\citenamefont
  {{Maksimova}, {Garrison}, {Eisenstein}, {Hadzhiyska}, {Bose},\ and\
  {Satterthwaite}}}]{2021Maksimova}%
  \BibitemOpen
  \bibfield  {author} {\bibinfo {author} {\bibfnamefont {N.~A.}\ \bibnamefont
  {{Maksimova}}}, \bibinfo {author} {\bibfnamefont {L.~H.}\ \bibnamefont
  {{Garrison}}}, \bibinfo {author} {\bibfnamefont {D.~J.}\ \bibnamefont
  {{Eisenstein}}}, \bibinfo {author} {\bibfnamefont {B.}~\bibnamefont
  {{Hadzhiyska}}}, \bibinfo {author} {\bibfnamefont {S.}~\bibnamefont
  {{Bose}}},\ \bibnamefont {and}\ \bibinfo {author} {\bibfnamefont {T.~P.}\
  \bibnamefont {{Satterthwaite}}},\ }\bibfield  {title} {\bibinfo {title}
  {{ABACUSSUMMIT: A Massive Set of High-Accuracy, High-Resolution N-Body
  Simulations}},\ }\bibfield  {journal} {\bibinfo  {journal} {Monthly Notices
  of the Royal Astronomical Society}\ }\href
  {https://doi.org/10.1093/mnras/stab2484} {10.1093/mnras/stab2484} (\bibinfo
  {year} {2021}),\ \Eprint {https://arxiv.org/abs/2110.11398} {arXiv:2110.11398
  [astro-ph.CO]} \BibitemShut {NoStop}%
\bibitem [{\citenamefont {{Hu}}\ and\ \citenamefont
  {{Sawicki}}(2007)\citenamefont {{Hu}\ and\ {Sawicki}}}]{2007PhRvD..76f4004H}%
  \BibitemOpen
  \bibfield  {author} {\bibinfo {author} {\bibfnamefont {W.}~\bibnamefont
  {{Hu}}}\ \bibnamefont {and}\ \bibinfo {author} {\bibfnamefont
  {I.}~\bibnamefont {{Sawicki}}},\ }\bibfield  {title} {\bibinfo {title}
  {{Models of f(R) cosmic acceleration that evade solar system tests}},\ }\href
  {https://doi.org/10.1103/PhysRevD.76.064004} {\bibfield  {journal} {\bibinfo
  {journal} {Phys. Rev. D}\ }\textbf {\bibinfo {volume} {76}},\ \bibinfo {eid}
  {064004} (\bibinfo {year} {2007})},\ \Eprint
  {https://arxiv.org/abs/0705.1158} {arXiv:0705.1158 [astro-ph]} \BibitemShut
  {NoStop}%
\bibitem [{\citenamefont {Baldi}\ and\ \citenamefont {Villaescusa-Navarro~et
  al.}()\citenamefont {Baldi\ and\ Villaescusa-Navarro~et al.}}]{QujiMG}%
  \BibitemOpen
  \bibfield  {author} {\bibinfo {author} {\bibfnamefont {M.}~\bibnamefont
  {Baldi}}\ \bibnamefont {and}\ \bibinfo {author} {\bibfnamefont
  {F.}~\bibnamefont {Villaescusa-Navarro~et al.}},\ }\bibfield  {title}
  {\bibinfo {title} {{2024, in prep.}},\ }\href@noop {} {\ }\BibitemShut
  {NoStop}%
\bibitem [{\citenamefont {{Andreux}}\ \emph {et~al.}(2018)\citenamefont
  {{Andreux}, {Angles}, {Exarchakis}, {Leonarduzzi}, {Rochette}, {Thiry},
  {Zarka}, {Mallat}, {And{\'e}n}, {Belilovsky} et~al.}}]{2018arXiv181211214A}%
  \BibitemOpen
  \bibfield  {author} {\bibinfo {author} {\bibfnamefont {M.}~\bibnamefont
  {{Andreux}}}, \bibinfo {author} {\bibfnamefont {T.}~\bibnamefont {{Angles}}},
  \bibinfo {author} {\bibfnamefont {G.}~\bibnamefont {{Exarchakis}}}, \bibinfo
  {author} {\bibfnamefont {R.}~\bibnamefont {{Leonarduzzi}}}, \bibinfo {author}
  {\bibfnamefont {G.}~\bibnamefont {{Rochette}}}, \bibinfo {author}
  {\bibfnamefont {L.}~\bibnamefont {{Thiry}}}, \bibinfo {author} {\bibfnamefont
  {J.}~\bibnamefont {{Zarka}}}, \bibinfo {author} {\bibfnamefont
  {S.}~\bibnamefont {{Mallat}}}, \bibinfo {author} {\bibfnamefont
  {J.}~\bibnamefont {{And{\'e}n}}}, \bibinfo {author} {\bibfnamefont
  {E.}~\bibnamefont {{Belilovsky}}}, \bibinfo {author} {\bibfnamefont
  {J.}~\bibnamefont {{Bruna}}}, \bibinfo {author} {\bibfnamefont
  {V.}~\bibnamefont {{Lostanlen}}}, \bibinfo {author} {\bibfnamefont {M.~J.}\
  \bibnamefont {{Hirn}}}, \bibinfo {author} {\bibfnamefont {E.}~\bibnamefont
  {{Oyallon}}}, \bibinfo {author} {\bibfnamefont {S.}~\bibnamefont {{Zhang}}},
  \bibinfo {author} {\bibfnamefont {C.}~\bibnamefont {{Cella}}}, \bibnamefont
  {et~al.},\ }\bibfield  {title} {\bibinfo {title} {{Kymatio: Scattering
  Transforms in Python}},\ }\href@noop {} {\bibfield  {journal} {\bibinfo
  {journal} {arXiv e-prints}\ ,\ \bibinfo {eid} {arXiv:1812.11214}} (\bibinfo
  {year} {2018})},\ \Eprint {https://arxiv.org/abs/1812.11214}
  {arXiv:1812.11214 [cs.LG]} \BibitemShut {NoStop}%
\bibitem [{\citenamefont {De~Felice}\ and\ \citenamefont
  {Tsujikawa}(2010)\citenamefont {De~Felice\ and\
  Tsujikawa}}]{DeFelice:2010aj}%
  \BibitemOpen
  \bibfield  {author} {\bibinfo {author} {\bibfnamefont {A.}~\bibnamefont
  {De~Felice}}\ \bibnamefont {and}\ \bibinfo {author} {\bibfnamefont
  {S.}~\bibnamefont {Tsujikawa}},\ }\bibfield  {title} {\bibinfo {title} {{f(R)
  theories}},\ }\href {https://doi.org/10.12942/lrr-2010-3} {\bibfield
  {journal} {\bibinfo  {journal} {Living Rev. Rel.}\ }\textbf {\bibinfo
  {volume} {13}},\ \bibinfo {pages} {3} (\bibinfo {year} {2010})},\ \Eprint
  {https://arxiv.org/abs/1002.4928} {arXiv:1002.4928 [gr-qc]} \BibitemShut
  {NoStop}%
\bibitem [{\citenamefont {Carroll}\ \emph {et~al.}(2004)\citenamefont {Carroll,
  Duvvuri, Trodden,\ and\ Turner}}]{Carroll:2003wy}%
  \BibitemOpen
  \bibfield  {author} {\bibinfo {author} {\bibfnamefont {S.~M.}\ \bibnamefont
  {Carroll}}, \bibinfo {author} {\bibfnamefont {V.}~\bibnamefont {Duvvuri}},
  \bibinfo {author} {\bibfnamefont {M.}~\bibnamefont {Trodden}},\ \bibnamefont
  {and}\ \bibinfo {author} {\bibfnamefont {M.~S.}\ \bibnamefont {Turner}},\
  }\bibfield  {title} {\bibinfo {title} {{Is cosmic speed - up due to new
  gravitational physics?}},\ }\href
  {https://doi.org/10.1103/PhysRevD.70.043528} {\bibfield  {journal} {\bibinfo
  {journal} {Phys. Rev.}\ }\textbf {\bibinfo {volume} {D70}},\ \bibinfo {pages}
  {043528} (\bibinfo {year} {2004})},\ \Eprint
  {https://arxiv.org/abs/astro-ph/0306438} {arXiv:astro-ph/0306438 [astro-ph]}
  \BibitemShut {NoStop}%
\bibitem [{\citenamefont {Springel}(2005)}]{Springel:2005mi}%
  \BibitemOpen
  \bibfield  {author} {\bibinfo {author} {\bibfnamefont {V.}~\bibnamefont
  {Springel}},\ }\bibfield  {title} {\bibinfo {title} {{The Cosmological
  simulation code GADGET-2}},\ }\href
  {https://doi.org/10.1111/j.1365-2966.2005.09655.x} {\bibfield  {journal}
  {\bibinfo  {journal} {Mon. Not. Roy. Astron. Soc.}\ }\textbf {\bibinfo
  {volume} {364}},\ \bibinfo {pages} {1105} (\bibinfo {year} {2005})},\ \Eprint
  {https://arxiv.org/abs/astro-ph/0505010} {arXiv:astro-ph/0505010}
  \BibitemShut {NoStop}%
\bibitem [{\citenamefont {Valogiannis}\ and\ \citenamefont
  {Bean}(2019)\citenamefont {Valogiannis\ and\ Bean}}]{Valogiannis:2019xed}%
  \BibitemOpen
  \bibfield  {author} {\bibinfo {author} {\bibfnamefont {G.}~\bibnamefont
  {Valogiannis}}\ \bibnamefont {and}\ \bibinfo {author} {\bibfnamefont
  {R.}~\bibnamefont {Bean}},\ }\bibfield  {title} {\bibinfo {title}
  {{Convolution Lagrangian perturbation theory for biased tracers beyond
  general relativity}},\ }\href {https://doi.org/10.1103/PhysRevD.99.063526}
  {\bibfield  {journal} {\bibinfo  {journal} {Phys. Rev. D}\ }\textbf {\bibinfo
  {volume} {99}},\ \bibinfo {pages} {063526} (\bibinfo {year} {2019})},\
  \Eprint {https://arxiv.org/abs/1901.03763} {arXiv:1901.03763 [astro-ph.CO]}
  \BibitemShut {NoStop}%
\bibitem [{\citenamefont {Valogiannis}\ \emph {et~al.}(2020)\citenamefont
  {Valogiannis, Bean,\ and\ Aviles}}]{Valogiannis:2019nfz}%
  \BibitemOpen
  \bibfield  {author} {\bibinfo {author} {\bibfnamefont {G.}~\bibnamefont
  {Valogiannis}}, \bibinfo {author} {\bibfnamefont {R.}~\bibnamefont {Bean}},\
  \bibnamefont {and}\ \bibinfo {author} {\bibfnamefont {A.}~\bibnamefont
  {Aviles}},\ }\bibfield  {title} {\bibinfo {title} {{An accurate perturbative
  approach to redshift space clustering of biased tracers in modified
  gravity}},\ }\href {https://doi.org/10.1088/1475-7516/2020/01/055} {\bibfield
   {journal} {\bibinfo  {journal} {JCAP}\ }\textbf {\bibinfo {volume} {01}},\
  \bibinfo {pages} {055}},\ \Eprint {https://arxiv.org/abs/1909.05261}
  {arXiv:1909.05261 [astro-ph.CO]} \BibitemShut {NoStop}%
\bibitem [{\citenamefont {Aviles}\ \emph {et~al.}(2021)\citenamefont {Aviles,
  Valogiannis, Rodriguez-Meza, Cervantes-Cota, Li,\ and\
  Bean}}]{Aviles:2020wme}%
  \BibitemOpen
  \bibfield  {author} {\bibinfo {author} {\bibfnamefont {A.}~\bibnamefont
  {Aviles}}, \bibinfo {author} {\bibfnamefont {G.}~\bibnamefont {Valogiannis}},
  \bibinfo {author} {\bibfnamefont {M.~A.}\ \bibnamefont {Rodriguez-Meza}},
  \bibinfo {author} {\bibfnamefont {J.~L.}\ \bibnamefont {Cervantes-Cota}},
  \bibinfo {author} {\bibfnamefont {B.}~\bibnamefont {Li}},\ \bibnamefont
  {and}\ \bibinfo {author} {\bibfnamefont {R.}~\bibnamefont {Bean}},\
  }\bibfield  {title} {\bibinfo {title} {{Redshift space power spectrum beyond
  Einstein-de Sitter kernels}},\ }\href
  {https://doi.org/10.1088/1475-7516/2021/04/039} {\bibfield  {journal}
  {\bibinfo  {journal} {JCAP}\ }\textbf {\bibinfo {volume} {04}},\ \bibinfo
  {pages} {039}},\ \Eprint {https://arxiv.org/abs/2012.05077} {arXiv:2012.05077
  [astro-ph.CO]} \BibitemShut {NoStop}%
\bibitem [{\citenamefont {Liu}\ \emph {et~al.}(2021)\citenamefont {Liu,
  Valogiannis, Battaglia,\ and\ Bean}}]{Liu:2021weo}%
  \BibitemOpen
  \bibfield  {author} {\bibinfo {author} {\bibfnamefont {R.}~\bibnamefont
  {Liu}}, \bibinfo {author} {\bibfnamefont {G.}~\bibnamefont {Valogiannis}},
  \bibinfo {author} {\bibfnamefont {N.}~\bibnamefont {Battaglia}},\
  \bibnamefont {and}\ \bibinfo {author} {\bibfnamefont {R.}~\bibnamefont
  {Bean}},\ }\bibfield  {title} {\bibinfo {title} {{Constraints on f(R) and
  normal-branch Dvali-Gabadadze-Porrati modified gravity model parameters with
  cluster abundances and galaxy clustering}},\ }\href
  {https://doi.org/10.1103/PhysRevD.104.103519} {\bibfield  {journal} {\bibinfo
   {journal} {Phys. Rev. D}\ }\textbf {\bibinfo {volume} {104}},\ \bibinfo
  {pages} {103519} (\bibinfo {year} {2021})},\ \Eprint
  {https://arxiv.org/abs/2101.08728} {arXiv:2101.08728 [astro-ph.CO]}
  \BibitemShut {NoStop}%
\bibitem [{\citenamefont {Rodriguez-Meza}\ \emph {et~al.}(2024)\citenamefont
  {Rodriguez-Meza, Aviles, Noriega, Ruan, Li, Vargas-Maga\~na,\ and\
  Cervantes-Cota}}]{Rodriguez-Meza:2023rga}%
  \BibitemOpen
  \bibfield  {author} {\bibinfo {author} {\bibfnamefont {M.~A.}\ \bibnamefont
  {Rodriguez-Meza}}, \bibinfo {author} {\bibfnamefont {A.}~\bibnamefont
  {Aviles}}, \bibinfo {author} {\bibfnamefont {H.~E.}\ \bibnamefont {Noriega}},
  \bibinfo {author} {\bibfnamefont {C.-Z.}\ \bibnamefont {Ruan}}, \bibinfo
  {author} {\bibfnamefont {B.}~\bibnamefont {Li}}, \bibinfo {author}
  {\bibfnamefont {M.}~\bibnamefont {Vargas-Maga\~na}},\ \bibnamefont {and}\
  \bibinfo {author} {\bibfnamefont {J.~L.}\ \bibnamefont {Cervantes-Cota}},\
  }\bibfield  {title} {\bibinfo {title} {{fkPT: constraining scale-dependent
  modified gravity with the full-shape galaxy power spectrum}},\ }\href
  {https://doi.org/10.1088/1475-7516/2024/03/049} {\bibfield  {journal}
  {\bibinfo  {journal} {JCAP}\ }\textbf {\bibinfo {volume} {03}},\ \bibinfo
  {pages} {049}},\ \Eprint {https://arxiv.org/abs/2312.10510} {arXiv:2312.10510
  [astro-ph.CO]} \BibitemShut {NoStop}%
\bibitem [{\citenamefont {{Ramachandra}}\ \emph {et~al.}(2021)\citenamefont
  {{Ramachandra}, {Valogiannis}, {Ishak}, {Heitmann},\ and\ {LSST Dark Energy
  Science Collaboration}}}]{2021PhRvD.103l3525R}%
  \BibitemOpen
  \bibfield  {author} {\bibinfo {author} {\bibfnamefont {N.}~\bibnamefont
  {{Ramachandra}}}, \bibinfo {author} {\bibfnamefont {G.}~\bibnamefont
  {{Valogiannis}}}, \bibinfo {author} {\bibfnamefont {M.}~\bibnamefont
  {{Ishak}}}, \bibinfo {author} {\bibfnamefont {K.}~\bibnamefont
  {{Heitmann}}},\ \bibnamefont {and}\ \bibinfo {author} {\bibnamefont {{LSST
  Dark Energy Science Collaboration}}},\ }\bibfield  {title} {\bibinfo {title}
  {{Matter power spectrum emulator for f (R ) modified gravity cosmologies}},\
  }\href {https://doi.org/10.1103/PhysRevD.103.123525} {\bibfield  {journal}
  {\bibinfo  {journal} {\prd}\ }\textbf {\bibinfo {volume} {103}},\ \bibinfo
  {eid} {123525} (\bibinfo {year} {2021})},\ \Eprint
  {https://arxiv.org/abs/2010.00596} {arXiv:2010.00596 [astro-ph.CO]}
  \BibitemShut {NoStop}%
\bibitem [{\citenamefont {Bai}\ and\ \citenamefont {Xia}(2024)\citenamefont
  {Bai\ and\ Xia}}]{Bai:2024cgt}%
  \BibitemOpen
  \bibfield  {author} {\bibinfo {author} {\bibfnamefont {J.}~\bibnamefont
  {Bai}}\ \bibnamefont {and}\ \bibinfo {author} {\bibfnamefont
  {J.}~\bibnamefont {Xia}},\ }\bibfield  {title} {\bibinfo {title} {{FREmu:
  Power Spectrum Emulator for $f(R)$ Gravity}},\ }\href@noop {} {\  (\bibinfo
  {year} {2024})},\ \Eprint {https://arxiv.org/abs/2405.05840}
  {arXiv:2405.05840 [astro-ph.CO]} \BibitemShut {NoStop}%
\bibitem [{\citenamefont {Valogiannis}\ and\ \citenamefont
  {Bean}(2017)\citenamefont {Valogiannis\ and\ Bean}}]{Valogiannis:2016ane}%
  \BibitemOpen
  \bibfield  {author} {\bibinfo {author} {\bibfnamefont {G.}~\bibnamefont
  {Valogiannis}}\ \bibnamefont {and}\ \bibinfo {author} {\bibfnamefont
  {R.}~\bibnamefont {Bean}},\ }\bibfield  {title} {\bibinfo {title} {{Efficient
  simulations of large scale structure in modified gravity cosmologies with
  comoving Lagrangian acceleration}},\ }\href
  {https://doi.org/10.1103/PhysRevD.95.103515} {\bibfield  {journal} {\bibinfo
  {journal} {Phys. Rev. D}\ }\textbf {\bibinfo {volume} {95}},\ \bibinfo
  {pages} {103515} (\bibinfo {year} {2017})},\ \Eprint
  {https://arxiv.org/abs/1612.06469} {arXiv:1612.06469 [astro-ph.CO]}
  \BibitemShut {NoStop}%
\bibitem [{\citenamefont {Wright}\ \emph {et~al.}(2023)\citenamefont {Wright,
  Sen~Gupta, Baker, Valogiannis,\ and\ Fiorini}}]{Wright:2022krq}%
  \BibitemOpen
  \bibfield  {author} {\bibinfo {author} {\bibfnamefont {B.~S.}\ \bibnamefont
  {Wright}}, \bibinfo {author} {\bibfnamefont {A.}~\bibnamefont {Sen~Gupta}},
  \bibinfo {author} {\bibfnamefont {T.}~\bibnamefont {Baker}}, \bibinfo
  {author} {\bibfnamefont {G.}~\bibnamefont {Valogiannis}},\ \bibnamefont
  {and}\ \bibinfo {author} {\bibfnamefont {B.}~\bibnamefont {Fiorini}}
  (\bibinfo {collaboration} {LSST Dark Energy Science}),\ }\bibfield  {title}
  {\bibinfo {title} {{Hi-COLA: fast, approximate simulations of structure
  formation in Horndeski gravity}},\ }\href
  {https://doi.org/10.1088/1475-7516/2023/03/040} {\bibfield  {journal}
  {\bibinfo  {journal} {JCAP}\ }\textbf {\bibinfo {volume} {03}},\ \bibinfo
  {pages} {040}},\ \Eprint {https://arxiv.org/abs/2209.01666} {arXiv:2209.01666
  [astro-ph.CO]} \BibitemShut {NoStop}%
\bibitem [{\citenamefont {Gupta}\ \emph {et~al.}(2024)\citenamefont {Gupta,
  Fiorini,\ and\ Baker}}]{Gupta:2024seu}%
  \BibitemOpen
  \bibfield  {author} {\bibinfo {author} {\bibfnamefont {A.~S.}\ \bibnamefont
  {Gupta}}, \bibinfo {author} {\bibfnamefont {B.}~\bibnamefont {Fiorini}},\
  \bibnamefont {and}\ \bibinfo {author} {\bibfnamefont {T.}~\bibnamefont
  {Baker}},\ }\bibfield  {title} {\bibinfo {title} {{K-mouflage at high k:
  extending the reach of $\texttt{Hi-COLA}$}},\ }\href@noop {} {\  (\bibinfo
  {year} {2024})},\ \Eprint {https://arxiv.org/abs/2407.00855}
  {arXiv:2407.00855 [astro-ph.CO]} \BibitemShut {NoStop}%
\bibitem [{\citenamefont {Puchwein}\ \emph {et~al.}(2013)\citenamefont
  {Puchwein, Baldi,\ and\ Springel}}]{Puchwein:2013lza}%
  \BibitemOpen
  \bibfield  {author} {\bibinfo {author} {\bibfnamefont {E.}~\bibnamefont
  {Puchwein}}, \bibinfo {author} {\bibfnamefont {M.}~\bibnamefont {Baldi}},\
  \bibnamefont {and}\ \bibinfo {author} {\bibfnamefont {V.}~\bibnamefont
  {Springel}},\ }\bibfield  {title} {\bibinfo {title} {{Modified
  Gravity-GADGET: A new code for cosmological hydrodynamical simulations of
  modified gravity models}},\ }\href {https://doi.org/10.1093/mnras/stt1575}
  {\bibfield  {journal} {\bibinfo  {journal} {Mon. Not. Roy. Astron. Soc.}\
  }\textbf {\bibinfo {volume} {436}},\ \bibinfo {pages} {348} (\bibinfo {year}
  {2013})},\ \Eprint {https://arxiv.org/abs/1305.2418} {arXiv:1305.2418
  [astro-ph.CO]} \BibitemShut {NoStop}%
\bibitem [{\citenamefont {{Carron, J.}}(2013)}]{refId0Car}%
  \BibitemOpen
  \bibfield  {author} {\bibinfo {author} {\bibnamefont {{Carron, J.}}},\
  }\bibfield  {title} {\bibinfo {title} {On the assumption of gaussianity for
  cosmological two-point statistics and parameter dependent covariance
  matrices},\ }\href {https://doi.org/10.1051/0004-6361/201220538} {\bibfield
  {journal} {\bibinfo  {journal} {A\&A}\ }\textbf {\bibinfo {volume} {551}},\
  \bibinfo {pages} {A88} (\bibinfo {year} {2013})}\BibitemShut {NoStop}%
\bibitem [{\citenamefont {Tegmark}\ \emph {et~al.}(1997)\citenamefont {Tegmark,
  Taylor,\ and\ Heavens}}]{Tegmark_1997}%
  \BibitemOpen
  \bibfield  {author} {\bibinfo {author} {\bibfnamefont {M.}~\bibnamefont
  {Tegmark}}, \bibinfo {author} {\bibfnamefont {A.~N.}\ \bibnamefont
  {Taylor}},\ \bibnamefont {and}\ \bibinfo {author} {\bibfnamefont {A.~F.}\
  \bibnamefont {Heavens}},\ }\bibfield  {title} {\bibinfo {title}
  {Karhunen-loeve eigenvalue problems in cosmology: How should we tackle large
  data sets?},\ }\href {https://doi.org/10.1086/303939} {\bibfield  {journal}
  {\bibinfo  {journal} {The Astrophysical Journal}\ }\textbf {\bibinfo {volume}
  {480}},\ \bibinfo {pages} {22} (\bibinfo {year} {1997})}\BibitemShut
  {NoStop}%
\bibitem [{\citenamefont {{Hartlap, J.}}\ \emph {et~al.}(2007)\citenamefont
  {{Hartlap, J.}, {Simon, P.},\ and\ {Schneider, P.}}}]{refId22}%
  \BibitemOpen
  \bibfield  {author} {\bibinfo {author} {\bibnamefont {{Hartlap, J.}}},
  \bibinfo {author} {\bibnamefont {{Simon, P.}}},\ \bibnamefont {and}\ \bibinfo
  {author} {\bibnamefont {{Schneider, P.}}},\ }\bibfield  {title} {\bibinfo
  {title} {Why your model parameter confidences might be too optimistic.
  unbiased estimation of the inverse covariance matrix},\ }\href
  {https://doi.org/10.1051/0004-6361:20066170} {\bibfield  {journal} {\bibinfo
  {journal} {A\&A}\ }\textbf {\bibinfo {volume} {464}},\ \bibinfo {pages} {399}
  (\bibinfo {year} {2007})}\BibitemShut {NoStop}%
\bibitem [{\citenamefont {Bayer}\ \emph {et~al.}(2022)\citenamefont {Bayer,
  Banerjee,\ and\ Seljak}}]{Bayer:2021kwg}%
  \BibitemOpen
  \bibfield  {author} {\bibinfo {author} {\bibfnamefont {A.~E.}\ \bibnamefont
  {Bayer}}, \bibinfo {author} {\bibfnamefont {A.}~\bibnamefont {Banerjee}},\
  \bibnamefont {and}\ \bibinfo {author} {\bibfnamefont {U.}~\bibnamefont
  {Seljak}},\ }\bibfield  {title} {\bibinfo {title} {{Beware of fake
  \ensuremath{\nu}\textquoteright{}s: The effect of massive neutrinos on the
  nonlinear evolution of cosmic structure}},\ }\href
  {https://doi.org/10.1103/PhysRevD.105.123510} {\bibfield  {journal} {\bibinfo
   {journal} {Phys. Rev. D}\ }\textbf {\bibinfo {volume} {105}},\ \bibinfo
  {pages} {123510} (\bibinfo {year} {2022})},\ \Eprint
  {https://arxiv.org/abs/2108.04215} {arXiv:2108.04215 [astro-ph.CO]}
  \BibitemShut {NoStop}%
\bibitem [{\citenamefont {Lombriser}(2014)}]{Lombriser:2014dua}%
  \BibitemOpen
  \bibfield  {author} {\bibinfo {author} {\bibfnamefont {L.}~\bibnamefont
  {Lombriser}},\ }\bibfield  {title} {\bibinfo {title} {{Constraining chameleon
  models with cosmology}},\ }\href {https://doi.org/10.1002/andp.201400058}
  {\bibfield  {journal} {\bibinfo  {journal} {Annalen Phys.}\ }\textbf
  {\bibinfo {volume} {526}},\ \bibinfo {pages} {259} (\bibinfo {year}
  {2014})},\ \Eprint {https://arxiv.org/abs/1403.4268} {arXiv:1403.4268
  [astro-ph.CO]} \BibitemShut {NoStop}%
\bibitem [{\citenamefont {{Villaescusa-Navarro}}(2018)}]{Pylians}%
  \BibitemOpen
  \bibfield  {author} {\bibinfo {author} {\bibfnamefont {F.}~\bibnamefont
  {{Villaescusa-Navarro}}},\ }\href@noop {} {\bibinfo {title} {{Pylians: Python
  libraries for the analysis of numerical simulations}}},\ \bibinfo
  {howpublished} {Astrophysics Source Code Library, record ascl:1811.008}
  (\bibinfo {year} {2018}),\ \Eprint {https://arxiv.org/abs/1811.008}
  {ascl:1811.008} \BibitemShut {NoStop}%
\bibitem [{\citenamefont {Desmond}\ and\ \citenamefont
  {Ferreira}(2020)\citenamefont {Desmond\ and\ Ferreira}}]{Desmond:2020gzn}%
  \BibitemOpen
  \bibfield  {author} {\bibinfo {author} {\bibfnamefont {H.}~\bibnamefont
  {Desmond}}\ \bibnamefont {and}\ \bibinfo {author} {\bibfnamefont {P.~G.}\
  \bibnamefont {Ferreira}},\ }\bibfield  {title} {\bibinfo {title} {{Galaxy
  morphology rules out astrophysically relevant Hu-Sawicki $f(R)$ gravity}},\
  }\href {https://doi.org/10.1103/PhysRevD.102.104060} {\bibfield  {journal}
  {\bibinfo  {journal} {Phys. Rev. D}\ }\textbf {\bibinfo {volume} {102}},\
  \bibinfo {pages} {104060} (\bibinfo {year} {2020})},\ \Eprint
  {https://arxiv.org/abs/2009.08743} {arXiv:2009.08743 [astro-ph.CO]}
  \BibitemShut {NoStop}%
\bibitem [{\citenamefont {Burrage}\ \emph {et~al.}(2024)\citenamefont {Burrage,
  March,\ and\ Naik}}]{Burrage:2023eol}%
  \BibitemOpen
  \bibfield  {author} {\bibinfo {author} {\bibfnamefont {C.}~\bibnamefont
  {Burrage}}, \bibinfo {author} {\bibfnamefont {B.}~\bibnamefont {March}},\
  \bibnamefont {and}\ \bibinfo {author} {\bibfnamefont {A.~P.}\ \bibnamefont
  {Naik}},\ }\bibfield  {title} {\bibinfo {title} {{Accurate computation of the
  screening of scalar fifth forces in galaxies}},\ }\href
  {https://doi.org/10.1088/1475-7516/2024/04/004} {\bibfield  {journal}
  {\bibinfo  {journal} {JCAP}\ }\textbf {\bibinfo {volume} {04}},\ \bibinfo
  {pages} {004}},\ \Eprint {https://arxiv.org/abs/2310.19955} {arXiv:2310.19955
  [astro-ph.CO]} \BibitemShut {NoStop}%
\bibitem [{\citenamefont {Nguyen}\ \emph {et~al.}(2024)\citenamefont {Nguyen,
  Schmidt, Tucci, Reinecke,\ and\ Kosti\'c}}]{Nguyen:2024yth}%
  \BibitemOpen
  \bibfield  {author} {\bibinfo {author} {\bibfnamefont {N.-M.}\ \bibnamefont
  {Nguyen}}, \bibinfo {author} {\bibfnamefont {F.}~\bibnamefont {Schmidt}},
  \bibinfo {author} {\bibfnamefont {B.}~\bibnamefont {Tucci}}, \bibinfo
  {author} {\bibfnamefont {M.}~\bibnamefont {Reinecke}},\ \bibnamefont {and}\
  \bibinfo {author} {\bibfnamefont {A.}~\bibnamefont {Kosti\'c}},\ }\bibfield
  {title} {\bibinfo {title} {{How much information can be extracted from galaxy
  clustering at the field level?}},\ }\href@noop {} {\  (\bibinfo {year}
  {2024})},\ \Eprint {https://arxiv.org/abs/2403.03220} {arXiv:2403.03220
  [astro-ph.CO]} \BibitemShut {NoStop}%
\bibitem [{\citenamefont {{Saydjari}}\ and\ \citenamefont
  {{Finkbeiner}}(2021)\citenamefont {{Saydjari}\ and\
  {Finkbeiner}}}]{2021arXiv210411244S}%
  \BibitemOpen
  \bibfield  {author} {\bibinfo {author} {\bibfnamefont {A.~K.}\ \bibnamefont
  {{Saydjari}}}\ \bibnamefont {and}\ \bibinfo {author} {\bibfnamefont {D.~P.}\
  \bibnamefont {{Finkbeiner}}},\ }\bibfield  {title} {\bibinfo {title}
  {{Equivariant Wavelets: Fast Rotation and Translation Invariant Wavelet
  Scattering Transforms}},\ }\href@noop {} {\bibfield  {journal} {\bibinfo
  {journal} {arXiv e-prints}\ ,\ \bibinfo {eid} {arXiv:2104.11244}} (\bibinfo
  {year} {2021})},\ \Eprint {https://arxiv.org/abs/2104.11244}
  {arXiv:2104.11244 [cs.CV]} \BibitemShut {NoStop}%
\bibitem [{\citenamefont {Horndeski}(1974)}]{Horndeski:1974wa}%
  \BibitemOpen
  \bibfield  {author} {\bibinfo {author} {\bibfnamefont {G.~W.}\ \bibnamefont
  {Horndeski}},\ }\bibfield  {title} {\bibinfo {title} {{Second-order
  scalar-tensor field equations in a four-dimensional space}},\ }\href
  {https://doi.org/10.1007/BF01807638} {\bibfield  {journal} {\bibinfo
  {journal} {Int. J. Theor. Phys.}\ }\textbf {\bibinfo {volume} {10}},\
  \bibinfo {pages} {363} (\bibinfo {year} {1974})}\BibitemShut {NoStop}%
\bibitem [{\citenamefont {{Friedrich}}\ \emph {et~al.}(2021)\citenamefont
  {{Friedrich}, {Andrade-Oliveira}, {Camacho}, {Alves}, {Rosenfeld}, {Sanchez},
  {Fang}, {Eifler}, {Krause}, {Chang} et~al.}}]{2021MNRAS.508.3125F}%
  \BibitemOpen
  \bibfield  {author} {\bibinfo {author} {\bibfnamefont {O.}~\bibnamefont
  {{Friedrich}}}, \bibinfo {author} {\bibfnamefont {F.}~\bibnamefont
  {{Andrade-Oliveira}}}, \bibinfo {author} {\bibfnamefont {H.}~\bibnamefont
  {{Camacho}}}, \bibinfo {author} {\bibfnamefont {O.}~\bibnamefont {{Alves}}},
  \bibinfo {author} {\bibfnamefont {R.}~\bibnamefont {{Rosenfeld}}}, \bibinfo
  {author} {\bibfnamefont {J.}~\bibnamefont {{Sanchez}}}, \bibinfo {author}
  {\bibfnamefont {X.}~\bibnamefont {{Fang}}}, \bibinfo {author} {\bibfnamefont
  {T.~F.}\ \bibnamefont {{Eifler}}}, \bibinfo {author} {\bibfnamefont
  {E.}~\bibnamefont {{Krause}}}, \bibinfo {author} {\bibfnamefont
  {C.}~\bibnamefont {{Chang}}}, \bibinfo {author} {\bibfnamefont
  {Y.}~\bibnamefont {{Omori}}}, \bibinfo {author} {\bibfnamefont
  {A.}~\bibnamefont {{Amon}}}, \bibinfo {author} {\bibfnamefont
  {E.}~\bibnamefont {{Baxter}}}, \bibinfo {author} {\bibfnamefont
  {J.}~\bibnamefont {{Elvin-Poole}}}, \bibinfo {author} {\bibfnamefont
  {D.}~\bibnamefont {{Huterer}}}, \bibinfo {author} {\bibfnamefont
  {A.}~\bibnamefont {{Porredon}}}, \bibnamefont {et~al.},\ }\bibfield  {title}
  {\bibinfo {title} {{Dark Energy Survey year 3 results: covariance modelling
  and its impact on parameter estimation and quality of fit}},\ }\href
  {https://doi.org/10.1093/mnras/stab2384} {\bibfield  {journal} {\bibinfo
  {journal} {"Mon. Not. Roy. Astron. Soc."}\ }\textbf {\bibinfo {volume}
  {508}},\ \bibinfo {pages} {3125} (\bibinfo {year} {2021})},\ \Eprint
  {https://arxiv.org/abs/2012.08568} {arXiv:2012.08568 [astro-ph.CO]}
  \BibitemShut {NoStop}%
\bibitem [{\citenamefont {{Yuan}}\ \emph {et~al.}(2023)\citenamefont {{Yuan},
  {Hadzhiyska},\ and\ {Abel}}}]{2023Yuan}%
  \BibitemOpen
  \bibfield  {author} {\bibinfo {author} {\bibfnamefont {S.}~\bibnamefont
  {{Yuan}}}, \bibinfo {author} {\bibfnamefont {B.}~\bibnamefont
  {{Hadzhiyska}}},\ \bibnamefont {and}\ \bibinfo {author} {\bibfnamefont
  {T.}~\bibnamefont {{Abel}}},\ }\bibfield  {title} {\bibinfo {title} {{Full
  forward model of galaxy clustering statistics with ABACUSSUMMIT light
  cones}},\ }\href {https://doi.org/10.1093/mnras/stad550} {\bibfield
  {journal} {\bibinfo  {journal} {Monthly Notices of the Royal Astronomical
  Society}\ }\textbf {\bibinfo {volume} {520}},\ \bibinfo {pages} {6283}
  (\bibinfo {year} {2023})},\ \Eprint {https://arxiv.org/abs/2211.02068}
  {arXiv:2211.02068 [astro-ph.CO]} \BibitemShut {NoStop}%
\bibitem [{\citenamefont {Park}\ \emph {et~al.}(2022)\citenamefont {Park,
  Allys, Villaescusa-Navarro,\ and\ Finkbeiner}}]{Park:2022hzj}%
  \BibitemOpen
  \bibfield  {author} {\bibinfo {author} {\bibfnamefont {C.~F.}\ \bibnamefont
  {Park}}, \bibinfo {author} {\bibfnamefont {E.}~\bibnamefont {Allys}},
  \bibinfo {author} {\bibfnamefont {F.}~\bibnamefont {Villaescusa-Navarro}},\
  \bibnamefont {and}\ \bibinfo {author} {\bibfnamefont {D.~P.}\ \bibnamefont
  {Finkbeiner}},\ }\bibfield  {title} {\bibinfo {title} {{Quantification of
  high dimensional non-Gaussianities and its implication to Fisher analysis in
  cosmology}},\ }\href@noop {} {\  (\bibinfo {year} {2022})},\ \Eprint
  {https://arxiv.org/abs/2204.05435} {arXiv:2204.05435 [astro-ph.CO]}
  \BibitemShut {NoStop}%
\bibitem [{\citenamefont {Coulton}\ \emph {et~al.}(2023)\citenamefont {Coulton,
  Villaescusa-Navarro, Jamieson, Baldi, Jung, Karagiannis, Liguori, Verde,\
  and\ Wandelt}}]{Coulton:2022rir}%
  \BibitemOpen
  \bibfield  {author} {\bibinfo {author} {\bibfnamefont {W.~R.}\ \bibnamefont
  {Coulton}}, \bibinfo {author} {\bibfnamefont {F.}~\bibnamefont
  {Villaescusa-Navarro}}, \bibinfo {author} {\bibfnamefont {D.}~\bibnamefont
  {Jamieson}}, \bibinfo {author} {\bibfnamefont {M.}~\bibnamefont {Baldi}},
  \bibinfo {author} {\bibfnamefont {G.}~\bibnamefont {Jung}}, \bibinfo {author}
  {\bibfnamefont {D.}~\bibnamefont {Karagiannis}}, \bibinfo {author}
  {\bibfnamefont {M.}~\bibnamefont {Liguori}}, \bibinfo {author} {\bibfnamefont
  {L.}~\bibnamefont {Verde}},\ \bibnamefont {and}\ \bibinfo {author}
  {\bibfnamefont {B.~D.}\ \bibnamefont {Wandelt}},\ }\bibfield  {title}
  {\bibinfo {title} {{Quijote-PNG: The Information Content of the Halo Power
  Spectrum and Bispectrum}},\ }\href {https://doi.org/10.3847/1538-4357/aca7c1}
  {\bibfield  {journal} {\bibinfo  {journal} {Astrophys. J.}\ }\textbf
  {\bibinfo {volume} {943}},\ \bibinfo {pages} {178} (\bibinfo {year}
  {2023})},\ \Eprint {https://arxiv.org/abs/2206.15450} {arXiv:2206.15450
  [astro-ph.CO]} \BibitemShut {NoStop}%
\bibitem [{\citenamefont {Coulton}\ and\ \citenamefont
  {Wandelt}(2023)\citenamefont {Coulton\ and\ Wandelt}}]{Coulton:2023sfu}%
  \BibitemOpen
  \bibfield  {author} {\bibinfo {author} {\bibfnamefont {W.~R.}\ \bibnamefont
  {Coulton}}\ \bibnamefont {and}\ \bibinfo {author} {\bibfnamefont {B.~D.}\
  \bibnamefont {Wandelt}},\ }\bibfield  {title} {\bibinfo {title} {{How to
  estimate Fisher information matrices from simulations}},\ }\href@noop {} {\
  (\bibinfo {year} {2023})},\ \Eprint {https://arxiv.org/abs/2305.08994}
  {arXiv:2305.08994 [stat.ME]} \BibitemShut {NoStop}%
\bibitem [{\citenamefont {Wilson}\ and\ \citenamefont {bean}(2024)\citenamefont
  {Wilson\ and\ bean}}]{Wilson:2024nhm}%
  \BibitemOpen
  \bibfield  {author} {\bibinfo {author} {\bibfnamefont {C.}~\bibnamefont
  {Wilson}}\ \bibnamefont {and}\ \bibinfo {author} {\bibfnamefont
  {R.}~\bibnamefont {bean}},\ }\bibfield  {title} {\bibinfo {title} {{Fisher's
  Mirage: Noise Tightening of Cosmological Constraints in Simulation-Based
  Inference}},\ }\href@noop {} {\  (\bibinfo {year} {2024})},\ \Eprint
  {https://arxiv.org/abs/2406.06067} {arXiv:2406.06067 [astro-ph.CO]}
  \BibitemShut {NoStop}%
\end{thebibliography}
%

\end{document}